\documentclass[amsmath,amssymb,rmp]{revtex4-2}

\usepackage[utf8]{inputenc}

\usepackage{graphicx}        
\usepackage{epstopdf}
\usepackage{float}

\usepackage{amssymb}
\usepackage{amsfonts}
\usepackage{mathtools}       
\usepackage{amsthm}
\usepackage{amsopn}
\usepackage{bm}              
\usepackage{bbm}
\usepackage{mathrsfs}
\usepackage{physics,braket}

\usepackage{dcolumn}         
\usepackage{booktabs}
\usepackage{multirow}
\usepackage{makecell,tabularx}

\usepackage{enumitem}

\usepackage{algorithmic}

\usepackage{xcolor}

\usepackage{tikz}
\usetikzlibrary{arrows.meta, positioning, calc, decorations.pathreplacing}

\usepackage{lipsum}

\usepackage{hyperref}
\usepackage{cleveref}

\ifpdf
  \DeclareGraphicsExtensions{.eps,.pdf,.png,.jpg}
\else
  \DeclareGraphicsExtensions{.eps}
\fi

\DeclareMathOperator*{\argmax}{arg\,max}

\newcommand{\U}{\hat{U}}
\newcommand{\A}{\mathcal{A}}

\newcommand{\twoline}[2]{\makecell{#1 \\ #2}}

\newcounter{algorithm}
\renewcommand{\thealgorithm}{\arabic{algorithm}}
\makeatletter
\newcommand{\alg@caption}[1]{%
  \noindent{\bfseries Algorithm \thealgorithm}\hspace{0.6em}#1\par
  \kern3pt\hrule\kern5pt}
\newenvironment{algorithm}{%
  \par\addvspace{\bigskipamount}%
  \refstepcounter{algorithm}%
  \let\caption\alg@caption
  \hrule height.8pt\kern4pt}{%
  \kern4pt\hrule height.8pt%
  \par\addvspace{\bigskipamount}}
\makeatother
\crefname{algorithm}{Algorithm}{Algorithms}
\Crefname{algorithm}{Algorithm}{Algorithms}

\begin{document}


\title{A Survey of Quantum Alternatives to Randomized Algorithms:\\
Monte Carlo Integration and Beyond}


\author{Philip Intallura}
\affiliation{Quantum Technologies Group, HSBC Holdings Plc., 8 Canada Square, London E14 5HQ, U.K.}

\author{Georgios Korpas}
\affiliation{%
Quantum Technologies Group, HSBC, 20 Pasir Panjang Road, 117439 Singapore.
}
\affiliation{Czech Technical University in Prague, Karlovo nam.\ 13, Prague 2, Czech Republic}

\author{Sudeepto Chakraborty}
\affiliation{Raytheon, Tuscon, AZ, U.S.A.}

\author{Rufus Lawrence}
\author{Ales Wodecki}
\author{Vyacheslav Kungurtsev}
\author{Jakub Marecek}
\affiliation{Czech Technical University in Prague, Karlovo nam.\ 13, Prague 2, Czech Republic}
 \email{jakub.marecek@fel.cvut.cz}

\begin{abstract}
Monte Carlo methods are a powerful class of numerical techniques widely used to tackle high-dimensional deterministic and probabilistic problems. In this paper, we survey the literature for implementing Monte Carlo procedures using quantum circuits, focusing on the potential to obtain a quantum computational advantage, with a particular focus on query complexity. We revisit the quantum analogues of classical Monte Carlo methods and then consider both existing quantum algorithms and their adaptive enhancements as alternatives to the classical approach.
\end{abstract}



\maketitle

\tableofcontents

\section{Introduction}

\subsection{Motivation}

Quantum computing promises to solve instances of certain problems currently intractable even with high-performance classical computers. See the surveys \cite{McKinsey2021},  \cite{Herman2022}, and \cite{9222275} discussing possible applications in chemistry, pharmaceuticals, and financial services, among other domains. 
Monte Carlo sampling~\cite[e.g.]{robert1999monte,kroese2013handbook} is a set of techniques that randomly generate numerical quantities for the purpose of simulating a statistical distribution and computing a moment thereof (e.g., mean, variance). It is prominent in many disciplines, including computational finance \cite{Savine2018-gt}, computational physics \cite{Thijssen2007}, artificial intelligence \cite{wang2012monte,llorente2021survey}, and various branches of engineering \cite{Shinozuka1972}. 
Although the concepts and ideas discussed in this paper readily generalize to other disciplines, we highlight applications in computational finance in particular. 

In computational finance, significant computational resources are deployed in the  risk management and asset pricing of, e.g., stocks, bonds, futures, and other exotic commodities such as derivatives. Classical and quasi-Monte Carlo methods feature prominently in such models, 
not least because their use is mandated by ever more stringent financial-markets regulations in most developed countries.
Consequently, there is a significant interest in improving the quality and efficiency of these methods.
Leading financial institutions, including HSBC \cite{Gmez2022,10250961,bohun2026entanglement}, Barclays \cite{braine2021quantum}, Fidelity Investments \cite{zhu2021generative,Iaconis2024}, Goldman Sachs \cite{giurgica2022low,Chakrabarti2021,stamatopoulos2022towards}, JPMorgan Chase \cite{Stamatopoulos2020optionpricingusing,Cherrat2023quantumdeephedging,herman2026quantum}, and Mitsubishi UFJ \cite{uno2021modified}, BBVA \cite{Alcazar2022}, and S\&P
\cite{Matsakos2024quantummontecarlo}, actively publish research in the field, while it is likely that there will be even more unpublished industrial research.
See, for example, \cite{rebentrost2018quantum,Stamatopoulos2020optionpricingusing,Chakrabarti2021,udvarnoki2023quantum,YuHan04032026,herman2026quantum} for work on option pricing and \cite{egger2020credit,stamatopoulos2024quantum,Matsakos2024quantummontecarlo,HAN2026109193} for work on risk assessment, including credit risk assessment \cite{egger2020credit,Matsakos2024quantummontecarlo}. 

The motivation for the search for quantum alternatives is rooted in the nature of these procedures, which are general and flexible enough to be effective for a wide array of possible real-life probability distributions but, in so doing, typically require a large number of samples to achieve good approximations. 
Once fully scalable fault-tolerant error corrected quantum
computers are available, quantum alternatives to Monte Carlo can potentially achieve an unconditional  advantage over the classical counterparts. This is because, as we will explain in the core of this survey, such tools have the potential to significantly improve the number of samples required. This could offer significant benefits over and beyond the immediate lowering of computational expense; fast computations allow greater potential for real-time decision making and thus more sophisticated trading strategies. 

Although quantum technology broadly holds tremendous potential, in this work, we show that the quantum alternatives and their adaptively tuned variants present a complicated picture regarding their potential to speed up the basic procedure. 
In this review, we shall also review the mentioned alternatives in order to present a comprehensive picture of sampling-based uncertainty quantification and the associated potential for quantum computing enhancements thereof.  

\subsection{Basic principles of classical Monte Carlo methods}

Let us review the basic principles of classical Monte Carlo methods in order to contrast them with the quantum alternatives. 
Let $f:[0,1]^d \to \mathbb{R}$ be a measurable function, and let $X \sim \mu$ be
a random variable taking values in $[0,1]^d$, where $\mu$ is a Borel
probability measure. A Monte Carlo method estimates the expectation
$\mathbb{E}[f(X)]$ through the estimator
\begin{align}
    I^n = \frac{1}{n}\sum_{j=1}^n f(X_j),
\end{align}
where $\{X_j\}_{j=1}^n$ are independent and identically distributed copies of
$X$. Provided $\mathbb{E}[|f(X)|] < \infty$, the strong law of large numbers
guarantees that the estimator converges to the desired expectation,
\begin{align}
    I^n \xrightarrow[n\to\infty]{} \mathbb{E}[f(X)]
    \qquad \text{(almost surely)}.
\end{align}
In practice, two complementary situations arise. In the first, the distribution
of $X$ is not known explicitly, but $X$ can be sampled efficiently. This occurs,
for instance, when $X$ is distributed according to the stationary distribution
of a Markov chain \cite{robert1999monte}. In the second, efficient sampling is
unavailable, yet the law of $X$ is known. When $\mu$ is absolutely continuous
with respect to the Lebesgue measure, with density $p_X$ (i.e.,
$\mathrm{d}\mu(x) = p_X(x)\,\mathrm{d}x$), the expectation may be written as the
integral
\begin{align}
    \mathbb{E}[f(X)] = \int_{[0,1]^d} f(x)\,p_X(x)\,\mathrm{d}x
    = \mathbb{E}_{U}\!\left[f(U)\,p_X(U)\right],
\end{align}
where $U$ is uniformly distributed on $[0,1]^d$. We can then sample points
uniformly on $[0,1]^d$ and average $f\,p_X$ over them to obtain an estimator of
$\mathbb{E}[f(X)]$. In particular, if $X$ itself is uniformly distributed on
$[0,1]^d$ (i.e., $p_X \equiv 1$), then $I^n$ estimates the deterministic
integral $\int_{[0,1]^d} f(x)\,\mathrm{d}x$, which may be analytically or
numerically intractable in high dimension. Historically, this has been one of
the principal applications of Monte Carlo.

A third scenario sits outside the dichotomy mentioned above, that is when the law of $X$ is specified only through some transform, for example through its characteristic function
$\varphi_X(t) = \mathbb{E}[e^{i\langle t, X\rangle}]$, with neither a tractable density nor an obvious direct sampler. A representative example is the class of $\alpha$-stable distributions, which (apart from a few special cases) possess no closed-form density. In this case, $p_X$ is defined implicitly through the Fourier inversion of $\varphi_X$, which is numerically delicate. In such settings we must either recover the density by numerical inversion, evaluate expectations directly in
the transform domain, or appeal to specialized samplers that bypass the density altogether.

The main advantage of Monte Carlo methods over other numerical methods lies in the independence of the query complexity (i.e., the number of evaluations of $f$) on the dimension of the domain, the shape of the domain and the measure. Other advantages include the ease of implementation, parallelization, and universality of error bounds \cite{robert1999monte, Fishman1996}. Despite these advantages, the variance of the statistical estimator $I^n$ given by the Monte Carlo method has an asymptotic convergence rate of

\begin{equation}\label{eq:standard-mc-conv}
\text{Var}(I^n)^{\frac{1}{2}}=\mathbb{E}\!\left[(I^n - I)^2\right]^{1/2} = \mathcal{O}(n^{-1/2}),
\end{equation}
where we assume $L^2$-regularity of the integrand, $n$ is the number of queries (or samples). 
The complexity result \eqref{eq:standard-mc-conv} can be reinterpreted using the Chebyshev inequality. More succinctly, for a given error tolerance $\varepsilon > 0$ and probability $1 - \delta > 0$ we require a sample size
\begin{equation}
n=\mathcal{O}\left(\frac{\sigma^{2}}{\varepsilon^{2}\delta}\right)
\end{equation}
to reach $\varepsilon$ precision with probability greater than $1 - \delta$.

Although the complexity bound \eqref{eq:standard-mc-conv} is independent of dimension, the convergence rate $n^{-\frac{1}{2}}$ is still slow.  This issue is typically addressed by the use of high-performance computing (HPC), which can improve wall-clock times. However, even the use of HPC can not always guarantee sufficient accuracy in time-sensitive applications, for example intraday expected shortfall computations in risk analysis, and is always associated with large computational costs. 

This has motivated the development of quantum generalizations of classical Monte Carlo methods. Once fully scalable fault-tolerant error corrected quantum computers are available, quantum alternatives to Monte Carlo can achieve a competitive advantage over their classical counterparts. This is because quantum variants of Monte Carlo have provably better asymptotic convergence rates in terms of query complexity under a variety of assumptions (cf. Table \ref{tab:det-ran-quantum} on p. \pageref{tab:det-ran-quantum}). These advantages could then be leveraged in situations where classical Monte Carlo methods find utility.

In the absence of any additional assumptions, classical Monte Carlo methods are optimal. However, there exist variants, such as quasi-Monte Carlo methods, which achieve strictly faster convergence rates under additional assumptions. Nevertheless, these assumptions are relatively strong and often not satisfied in practice. Section \ref{sec:alt} provides more details on the benefits and shortcomings of such methods. Hence, the speedup achieved by quantum Monte Carlo over classical algorithms should be thought of as an advantage over all possible classical randomized algorithms.

In this survey, we describe the origins and the state-of-the-art of quantum Monte Carlo methods. In addition to explaining the standard quantum Monte Carlo algorithms and their variants, we present an end-to-end perspective on the application of quantum Monte Carlo methods to real world problems. In particular, we discuss how aspects of the upstream problem (e.g., the regularity of the integrand) inform the (quantum) implementation of the algorithm (see Figure \ref{fig:hourglass}).

Throughout this work we use the term \emph{Quantum Monte Carlo} exclusively to
denote quantum algorithms that serve as an alternative to classical Monte Carlo
for estimating expectations and integrals. We caution the reader that the same
term is also used for a family of \emph{classical} algorithms for simulating
quantum systems~\cite{gubernatis2016quantum}, such as diffusion Monte
Carlo~\cite{Reynolds1990}. These, however, are unrelated to the methods considered here and lie beyond our scope.


\section{Quantum amplitude estimation and quantum approximate counting}\label{sec:quantum_alternatives}

Quantum variants of classical Monte Carlo algorithms have been considered since the late 1990s \cite{abrams1999fast, Brassard2002,Brassard1998}, first appeared as generalizations of Grover's quantum search algorithm in terms of the subroutines utilized. Like Grover's algorithm for unstructured search \cite{Grover1997}, they yield a quadratic speedup over the best classical randomized algorithms. There are two main approaches, one based on the procedure of Quantum Amplitude Estimation (QAE) and one referred to as Quantum Approximate Counting (QAC). Of the two, QAE and its variants have in recent years become the dominant paradigm, but QAC also remains an active research area. In the remainder of the introduction, we survey the ``state of play'' in the early 2000s, when the field began to mature.

\begin{table}[tbh]
  \centering
  \renewcommand{\arraystretch}{1.3}
  \begin{tabular}{ll}
    \toprule[2pt]
    \textbf{Symbol} & \textbf{Meaning} \\
    \hline
    $f$ & Function to be summed/integrated (e.g.\ $f:\{0,\dots,N-1\}\to\{0,1\}$ or $f:[0,1]^d\to[0,1]$) \\
    $n$ & Number of qubits in the index register \\
    $N$ & Size of the domain (number of basis states), $N=2^n$ \\
    $S_N(f)$ & Boolean-function average $\frac{1}{N}\sum_{j=0}^{N-1}f(j)$ \\
    $U_f$ & Phase oracle, $U_f\ket{j}=(-1)^{f(j)}\ket{j}$ \\
    $\mathcal{A}$ & Arbitrary state-preparation algorithm/subroutine \\
    $\mathcal{A}^{-1}$ & Inverse (adjoint) of $\mathcal{A}$ \\
    $\ket{s}$ & Equal superposition state $\frac{1}{\sqrt{N}}\sum_{j=0}^{N-1}\ket{j}$ \\
    $\ket{w}$ & \emph{Good} state (superposition over $f(j)=1$ outcomes) \\
    $\ket{w^\perp}$ & \emph{Bad} state (superposition over $f(j)=0$ outcomes) \\
    $\ket{\Psi}=\ket{w}+\ket{w^\perp}$ & State $\mathcal{A}\ket{0}$ decomposed into bad ($\ket{w^\perp}$) and good ($\ket{w}$) parts \\
    $M$ & Number of marked elements, $M=|\{j:f(j)=1\}|$ \\
    $a$ & Amplitude/probability of measuring a good state, $a=\sin^2\theta_a$ \\
    $\theta,\ \theta_a$ & Rotation angle of the Grover operator; $\sin^2(\theta/2)=M/N$ \\
    $G$ & (Generalized) Grover operator: $(2\ket{s}\!\bra{s}-\mathbbm{1})U_f$ in the Boolean case, $-\mathcal{A}\,S_0\,\mathcal{A}^{-1}\,S_f$ in general \\
    $S_0,\ S_f$ & Reflections about the zero state and the good subspace (the latter defined by $f$) \\
    $\lambda_\pm=e^{\pm 2i\theta_a}$ & Eigenvalues of $G$ on the good--bad subspace \\
    $r$ & Number of Grover-operator applications (iterations); in Sec.~\ref{sec:apps}\,ff.\ also the smoothness order of a function class \\
    $T$ & Dimension of the counting register in amplitude estimation, $T=\mathcal{O}(1/\varepsilon)$ \\
    $F_T$ & Quantum Fourier transform on the $T$-dimensional counting register \\
    $\Lambda_T(G)$ & Controlled application of powers of $G$, $\ket{j}\ket{y}\mapsto\ket{j}G^j\ket{y}$ \\
    $\varepsilon$ & Additive error tolerance \\
    $\delta,\ \alpha$ & Failure probability \\
    $\mathcal{P}$ & State-preparation oracle loading a distribution, $\ket{0}^n\mapsto\sum_j\sqrt{p_j}\ket{j}$ \\
    $p_j,\ p_X(x)$ & Probability (mass/density) of the random variable $X$ \\
    $\mathcal{R}$ & Controlled $Y$-rotation encoding $f_j$ onto the ancilla \\
    $f_j$ & Discretized function value $f(x_j)$ \\
    $\mathbb{E}[f(X)]$ & Target expectation, $=\sum_j p_j f_j=a$ \\
    $\mathcal{S}=\langle b\rangle$ & Approximate-counting estimate, $\mathcal{S}=K/(NQ)$ \\
    $K$ & Number of marked pairs in approximate counting, $K=|b^{-1}(1)|$ \\
    $b$ & Boolean discretization of a real-valued $f$ via threshold parameter $q\in[1,Q]$ \\
    $Q$ & Range-discretization resolution in quantum approximate counting (threshold $q\in[1,Q]$) \\
    $A$ & Size of the QFT register in approximate counting (sets accuracy $1/A$) \\
    $Q_f$ & Quantum query oracle of \cite{heinrich2002quantum} (function-class complexity, Sec.~\ref{sec:apps}) \\
    $D\subset\mathbb{R}^d$ & Spatial domain of an elliptic boundary-value problem (Sec.~\ref{sec:apps}) \\
    $\pi,\ \pi_i$ & Probability distribution(s) of the random variable(s) \\
    $\ket{\pi_i}$ & Coherent quantum-sample state $\sum_x\sqrt{\pi_i(x)}\ket{x}$ \\
    $\bigl((\pi_i)_{i\in I},(A_j)_{j\in J}\bigr)$ & Abstract QMC problem specification (the ``hourglass waist'') \\
    \bottomrule[2pt]
  \end{tabular}
  \caption{Notation used throughout Sections~\ref{sec:quantum_alternatives} and~\ref{sec:apps}.}
  \label{tab:notation}
\end{table}

\subsection{Summation of Boolean functions}

One of the simplest implementations of quantum amplitude estimation is a procedure for computing the average of a Boolean function $f: \{0, \dots, N-1\} \rightarrow \{0,1\}$,

\begin{align}\label{eq:SN}
    S_N(f) = \frac{1}{N}\sum_{j=0}^{N-1} f(j),
\end{align}
assuming we have access to a (phase-flip) oracle for the function $f$, i.e. a unitary $U_f$ such that 

\begin{align}
U_f(|j\rangle) = (-1)^{f(j)}|j\rangle.
\end{align}

In this context, the \textit{quantum query complexity} \cite{ambainis2018understanding} of an algorithm is defined as the number of times the oracle is applied during the computation. The quantum query complexity of a quantum algorithm is a good proxy for the overall complexity, and is hardware agnostic, i.e. does not depend on the choice of universal gate set used in the quantum computer, unlike, for instance, the gate complexity or the qubit complexity of a quantum algorithm. Moreover, in most situations we encounter, applying the oracle $U_f$ will in general be more costly than applying ``generic'' quantum gates such as CNOT, Hadamard or SWAP gates. We should remark at this point that constructing $U_f$ is not necessarily straightforward, this is discussed in more detail in \ref{subseq:oraclecontstruction}.

The basic algorithm begins by initializing an $n$-qubit register in the state $|0\rangle^{\otimes n} \equiv \ket{0}_n$. Next, a Hadamard transform is applied to place the system in an equal superposition

\begin{align}
    |s\rangle = \frac{1}{\sqrt{N}}\sum_{j=0}^{N-1} |j\rangle.
\end{align}

Following this, we introduce the (generalized) Grover operator 

\begin{align}
    G = (2|s\rangle\langle s| - \mathbbm{1})U_f.
\end{align}

Now, if $M$ denotes $|\{j\,|\,f(j)=1\}|\ $, then $S_N = \frac{M}{N}$. Next, we introduce the states

\begin{align}
    |w^\perp\rangle = \sqrt{\frac{1}{N-M}}\sum_{f(j)=0}|j\rangle, \quad
    |w\rangle = \sqrt{\frac{1}{M}}\sum_{f(j)=1}|j\rangle.
\end{align}
These states satisfy the relation

\begin{align}
    |s\rangle = \cos(\theta/2) |w^\perp\rangle + \sin(\theta/2)|w\rangle,
\end{align}
where $\theta/2$ is the angle between $|s\rangle$ and $|w^\perp\rangle$, and so $\sin^2(\theta/2) = M/N$. Hence, computing the sum reduces to estimating $\theta$. The most common way of estimating $\theta$ is to proceed via quantum phase estimation, as in \cite{Brassard2002}. 

The phase estimation procedure can be described as follows: let $\mathcal{H}$ denote the Hilbert space spanned by $|w^\perp\rangle$ and $|w\rangle$. This space is $G$-invariant, and the restriction of $G$ to $\mathcal{H}$ has matrix representation

\begin{align}
    \begin{pmatrix}
        \cos \theta & -\sin \theta \\
        \sin \theta & \cos \theta
    \end{pmatrix}
\end{align}
The eigenvalues of $G$ are then $\exp(\pm i\theta)$ with corresponding eigenvectors $|\xi_\pm\rangle$. 

Now, we can estimate $\theta$ by phase estimation, as in \cite{Brassard2002, Papageorgiou2009}. In order to obtain an error of order $\mathcal{O}(\varepsilon)$, choose $t = \mathcal{O}(-\log \varepsilon)$. Performing phase estimation with initial state $|0\rangle^{\otimes t}|s\rangle$ yields the result, with $2^t = \mathcal{O}(\frac{1}{\varepsilon})$ oracle calls. Note that an approximate phase estimation is possible even though $|s\rangle$ is not an eigenvector of $G$. Of course, it is also possible to write the error as a function of the number of queries: in this case $\varepsilon = \mathcal{O}(\frac{1}{n})$. Note that in utilizing quantum phase estimation, we assume not only that we have access to the oracle $U_f$, but also that we are able to implement controlled-$G$ operations. Moreover, phase estimation involves applying the quantum Fourier transform, which is notoriously hard to implement on NISQ-era devices. 

An alternative approach, following \cite{abrams1999fast}, forgoes phase estimation (and hence the necessity to implement controlled-$G$ operations) and instead infers $\theta$ directly from repeated computational-basis measurements. Up to a factor logarithmic in $1/\varepsilon$, it attains the same query complexity $\mathcal{O}(1/\varepsilon)$.

Applying the Grover operator $r$ times to $\ket{s}$ gives
\begin{align}
    G^r\ket{s} = \cos\!\big(\tfrac{(2r+1)\theta}{2}\big)\ket{w^\perp} + \sin\!\big(\tfrac{(2r+1)\theta}{2}\big)\ket{w},
\end{align}
so the probability of finding the system in the good subspace $\ket{w}$ is $p_r = \sin^2\!\big(\tfrac{(2r+1)\theta}{2}\big)$. Repeating the experiment $k$ times and recording the number $s_g$ of good outcomes, the data are described by the binomial likelihood
\begin{align}\label{eq:goodbad}
    L(\theta) = \binom{k}{s_g}\, p_r^{\,s_g}\,(1-p_r)^{\,k-s_g},
\end{align}
from which $\theta$, and hence $S_N = \sin^2(\theta/2)$, is inferred by maximization. Note that some form of Grover amplification is needed to obtain a quantum speedup: with $r=0$ (no amplification), each shot is a Bernoulli draw with probability $p_0 = \sin^2(\theta/2) = S_N$, so inferring $\theta$ to accuracy $\varepsilon$ requires resolving this probability to $\mathcal{O}(\varepsilon)$, which by the variance of a binomial estimator demands $k = \mathcal{O}(1/\varepsilon^2)$ shots (by Chebyshev), just as in the classical randomized setting. 

The obstacle is that $p_r$ is periodic in $\theta$, so a single deep circuit leaves $\theta$ ambiguous among $\mathcal{O}(r)$ candidates. Abrams and Williams \cite{abrams1999fast} resolve this through a sequence of refinements. Given the current estimate $E$ of the mean, they re-center the integrand to $f' = f - E$ and amplify only the residual $D = S - E$; because $D$ is small it can be amplified by a large factor without overshooting, and measuring the amplified residual yields a sharper estimate. This refined estimate defines a new $f'$, and the process repeats, the resolution improving geometrically. The essential point is that, since the residual is re-centered at every step, the operator that is amplified -- and hence the circuit that is run -- changes from one iteration to the next: it is \emph{not} a single fixed operator raised to successively higher powers, but a new operator built from the updated $f'$. Each round uses only a constant number of measurements, so the total number of shots is logarithmic in $1/\varepsilon$, while the query cost is dominated by the final, deepest round, giving $\mathcal{O}(1/\varepsilon)$ overall.

A more recent variant \cite{suzuki2020amplitude} keeps the measure-and-infer idea but removes both the phase estimation and the adaptivity. Instead of re-centring, the authors fix a schedule of amplification depths $\{m_k\}$ in advance -- for instance $m_k = 2^{k-1}$ -- applies powers $G^{m_k}$ of the \emph{same} Grover operator, and combines the resulting good-counts into a single likelihood $\prod_k L_k(\theta)$ whose maximizer is the estimate. The shallow circuits fix the period while the deep ones supply the resolution, so the combined likelihood has a unique maximum with no adaptive feedback. Because the schedule is predetermined, the circuits are independent and may be run in parallel, and only powers of one fixed operator are needed; an exponential schedule recovers the optimal $\mathcal{O}(1/\varepsilon)$ query complexity. The absence of phase estimation -- no quantum Fourier transform and no controlled-$G$ operations -- together with the constant qubit count makes this approach well suited to near-term devices.

\begin{figure}[H]
    \centering
    \begin{tikzpicture}[>={Stealth[length=2.4mm]},line cap=round]
  \def\h{13}

  \draw[->] (0,0) -- (6.4,0) node[right] {$|w^\perp\rangle$};
  \draw[->] (0,0) -- (0,6.4) node[above] {$|w\rangle$};

  \draw[->] (0,0) -- (\h:5.9)    node[right]       {$|s\rangle$};
  \draw[->] (0,0) -- (3*\h:5.9)  node[right]       {$G|s\rangle$};
  \draw[->] (0,0) -- (5*\h:5.5)  node[above right] {$G^2|s\rangle$};
  \draw[->] (0,0) -- (82:5.9)    node[above right] {$G^r|s\rangle$};

  \draw[gray,dashed] (3.3,0) arc (0:5*\h:3.3);
  \draw[gray,dashed] (2.5,0) arc (0:3*\h:2.5);
  \draw[gray,dashed] (1.15,0) arc (0:\h:1.15);

  \node at (0.55*\h:1.65) {$\theta/2$};
  \node at (2.35*\h:2.15) {$\dfrac{3\theta}{2}$};
  \node at (3.95*\h:2.95) {$\dfrac{5\theta}{2}$};
\end{tikzpicture}
    \caption{The action of powers of the Grover operator $G$ on the state $\ket{s}$.}
    \label{fig:groverfig}
\end{figure}

\subsection{Amplitude estimation with an arbitrary subroutine}
\label{sec:brassard-ae}

The Boolean-summation routine above is a special case of a much more general construction. Rather than presenting the input as a phase oracle $U_f$ for a Boolean function, we now allow access to an arbitrary quantum subroutine $\A$ and its inverse, and we seek to estimate the probability that measuring the state it prepares yields a designated ``good'' outcome. This is the setting of Brassard, H{\o}yer, Mosca and Tapp \cite{Brassard2002}, which we describe next.
 
Concretely, let $\A$ be any quantum algorithm acting on $\mathcal{H}$ that uses no measurements, and as before let $f : \mathbb{Z} \to \{0,1\}$ be a Boolean function partitioning the basis states into \emph{good} ($f = 1$) and \emph{bad} ($f = 0$) ones. Writing $\ket{\Psi} = \A\ket{0}$ and decomposing it into its good and bad parts $\ket{\Psi} = \ket{w} + \ket{w^\perp}$, the quantity of interest is the probability
\begin{equation}
  a = \langle w | w \rangle = \sin^2(\theta_a), 
\end{equation}
where $0 \le \theta_a \le \tfrac{\pi}{2}$, that a measurement of $\ket{\Psi}$ returns a good state. 

Returning to QAE, $\A$ enters the algorithm only as a black box: it is enough to be able to run $\A$ and its inverse $\A^{-1}$ (the latter obtained simply by running the circuit for $\A$ in reverse), together with the reflections $S_0$ and $S_f$ that flip the sign of the zero state and of the good subspace respectively. From these ingredients one forms the \emph{Grover operator}
\begin{equation}
  G = -\,\A\, S_0\, \A^{-1}\, S_f .
  \label{eq:grover}
\end{equation}
 
The key structural fact (see~\cite[Section~2]{Brassard2002}) is that $G$ acts as a rotation by angle $2\theta_a$ on the two-dimensional subspace spanned by $\ket{w}$ and $\ket{w^\perp}$. Equivalently, restricted to this subspace $G$ has the two eigenvalues
\begin{equation}
  \lambda_{\pm} = e^{\pm 2 i \theta_a}.
\end{equation}
Estimating the amplitude $a$ therefore reduces to estimating the eigenphase $\theta_a$ of a unitary, which can be done using the classical quantum phase estimation algorithm, as above. This is the content of~\cite[Section~4]{Brassard2002}.
 
Brassard et al.'s algorithm, $\textsc{Est\_Amp}(\A,f,T)$, uses two registers: a first ``counting'' register of dimension $T$, and a second register holding the state produced by $\A$. With $F_T$ the quantum Fourier transform on the first register and $\Lambda_T(G)$ the controlled application of powers of the Grover operator, $\ket{j}\ket{y} \mapsto \ket{j}\,(G^{j}\ket{y})$, the algorithm is:
 
\begin{algorithm}
\caption{$\textsc{Est\_Amp}(\A,f,T)$}
\label{alg:estamp}
\begin{algorithmic}[1]
  \STATE Initialize two registers to $\ket{0}\,\A\ket{0}$.
  \STATE Apply $F_T$ to the first register.
  \STATE Apply $\Lambda_T(G)$, with $G = -\A S_0 \A^{-1} S_f$.
  \STATE Apply $F_T^{-1}$ to the first register.
  \STATE Measure the first register; call the outcome $\ket{y}$.
  \RETURN $\tilde{a} = \sin^2\!\left(\pi \tfrac{y}{T}\right)$.
\end{algorithmic}
\end{algorithm}

The whole circuit is thus $(F_T^{-1} \otimes I)\,\Lambda_T(G)\,(F_T \otimes I)$ applied to $\ket{0}\,\A\ket{0}$, followed by a measurement and the classical post-processing $y \mapsto \sin^2(\pi y / T)$. The accuracy is governed by the main theorem of~\cite{Brassard2002}: for any $k \ge 1$, the estimate satisfies
\begin{equation}
  \lvert \tilde{a} - a \rvert \;\le\; 2\pi k \frac{\sqrt{a(1-a)}}{T} + \frac{k^2 \pi^2}{T^2},
  \label{eq:ae-bound}
\end{equation}
with probability at least $8/\pi^2$ when $k=1$ (and rapidly approaching $1$ as $k$ grows), using \emph{exactly} $T$ applications of $\A$. In other words, to estimate $a$ to additive error $\varepsilon$ it suffices to take $T = \mathcal{O}(1/\varepsilon)$, as against the $\mathcal{O}(1/\varepsilon^2)$ samples a classical Monte Carlo estimate of the same probability would require, demonstrating the quadratic speedup.

\subsection{Weighting by a probability distribution}\label{subsubsection:load_dist}

The Boolean-summation and Brassard constructions above estimate the probability of a ``good'' outcome. With a small modification they also compute the expectation of a function of a random variable,
\begin{align}
    \mathbb{E}[f(X)] = \int_{[0,1]^d} f(x)\,p_X(x)\,dx,
\end{align}
where $f:[0,1]^d\to[0,1]$ and $p_X$ is the density of $X$. Discretizing the domain onto a grid of $N=2^n$ points $\{x_j\}$ and writing $p_j \approx p_X(x_j)$, $f_j = f(x_j)$, the target becomes the weighted sum $\mathbb{E}[f(X)] \approx \sum_{j=0}^{N-1} p_j f_j$.

Rather than building a single oracle for the combined integrand $f\cdot p_X$, we encode the distribution and the function \emph{separately}, through two independent subroutines. The first oracle is a state-preparation oracle $\mathcal{P}$ that loads the distribution,
\begin{align}\label{circ:p}
    \mathcal{P}: \ket{0}^n \mapsto \sum_{j=0}^{N-1}\sqrt{p_j}\,\ket{j},
\end{align}
the canonical construction being that of Grover and Rudolph~\cite{grover2002creating} (loading a distribution is a non-trivial problem in its own right and can dominate the cost; cf.\ Sec.~\ref{challenge:stateprep}). 

To encode the values of $f$, rather than using a phase flip-oracle as before, we instead use an ancilla register and construct the oracle controlled $Y$-rotation $\mathcal{R}$,
\begin{align}\label{eq:R}
    \mathcal{R}: \ket{j}\ket{0} \mapsto \ket{j}\!\left(\sqrt{1-f_j}\,\ket{0} + \sqrt{f_j}\,\ket{1}\right), \qquad f_j\in[0,1],
\end{align}
so that the ancilla's $\ket{1}$-amplitude carries $\sqrt{f_j}$. The idea of encoding function values by controlled rotation goes back at least to \cite{grover1998framework}, although here we use the more modern formalism of \cite{rebentrost2018derivative}.  This separable encoding underlies the amplitude-estimation approach to expectation values~\cite{Montanaro_2015} and is the standard template in quantitative-finance applications~\cite{rebentrost2018quantum,Woerner2019,Stamatopoulos2020optionpricingusing}.

Composing the two and writing $\mathcal{A} = \mathcal{R}(\mathcal{P}\otimes\mathbbm{1})$ prepares
\begin{align}\label{eq:sprime}
    \ket{\Psi} = \mathcal{A}\ket{0}^n\ket{0} = \sqrt{1-a}\,\ket{w^\perp} + \sqrt{a}\,\ket{w},
    \qquad
    \ket{w} = \sum_{j=0}^{N-1}\sqrt{p_j f_j}\,\ket{j}\ket{1},
\end{align}
with $\ket{w^\perp}$ the orthogonal (ancilla-$\ket{0}$) component. The good-state weight is now exactly the quantity of interest,
\begin{align}\label{good}
    a = \langle w | w\rangle = \sum_{j=0}^{N-1} p_j f_j = \mathbb{E}[f(X)].
\end{align}
It is worth pausing on what has changed. In the Boolean case $a=M/N$ was a counting probability; here, because $f_j\in[0,1]$ is real-valued, $a$ is a \emph{weighted mean}. Encoding $f$ as a rotation angle, rather than as a marked/unmarked label, is precisely what turns ``probability of a good outcome'' into ``expectation of $f$''.

The connection to the general algorithm of Sec.~\ref{sec:brassard-ae} is now immediate: the composite subroutine $\mathcal{A}=\mathcal{R}(\mathcal{P}\otimes\mathbbm{1})$ \emph{is} the arbitrary state-preparation algorithm that $\textsc{Est\_Amp}$ takes as input, and $\ket{\Psi}=\mathcal{A}\ket{0}$ is already in the good/bad form~\eqref{eq:sprime} that the algorithm requires. Running amplitude estimation (Algorithm~\ref{alg:estamp}) on this $\mathcal{A}$ therefore returns $\mathbb{E}[f(X)]$ to additive error $\varepsilon$ using $\mathcal{O}(1/\varepsilon)$ applications of $\mathcal{A}$ -- the same quadratic speedup over the $\mathcal{O}(1/\varepsilon^2)$ classical samples, now for a general expectation rather than a Boolean average. The one caveat is that every query now invokes $\mathcal{P}$, so the speedup is contingent on the distribution being efficiently loadable; when it is not, state preparation can erode the advantage~\cite{herbert2021quantum}, a point we return to in Sec.~\ref{challenge:stateprep}.

A number of variants of amplitude estimation algorithms have recently been proposed, which we examine in the next section. See Table~\ref{QAEvariants} for an analytical overview of a few. In particular, some of them, for example \cite{grinko2021iterative,suzuki2020amplitude,aaronson2020quantum,uno2021modified} have removed the 
need to terminate the circuit with the Quantum Fourier transform (QFT) \cite{Nielsen2012}. Although the QFT adds only $\mathcal{O}( \log \log (1/\varepsilon))$ 
to the depth of the circuit when applied to $\mathcal{O}(\log(1/\varepsilon))$ qubits and various controlled gates are allowed, 
it tends to yield very deep circuits after compiling with respect to commonly used gate sets. 
Practically speaking, \cite{grinko2021iterative} often performs the best on \emph{noisy intermediate-scale quantum} (NISQ) devices due to the absence of Quantum Phase Estimation (QPE) \cite{Nielsen2012} and the fact that the required circuit depth is smaller. Note that a comparison of various QAE variants was presented in \cite{yu2020comparison}.

\subsection{Quantum approximate counting}\label{sec:QAC}

\emph{Quantum approximate counting} \cite{Brassard1998, abrams1999fast} is an alternative approach to estimating the sum \eqref{eq:SN}. Assume $f:\mathbb{R}^d\to [0,1]$. This method works by converting the real-valued function $f(\vb*{x})$ to a Boolean function. This can be accomplished by introducing a new parameter $q \in [1,Q]$ that determines the value of the Boolean function as follows:
\begin{align}
    b\left(a_{1}, a_{2}, \ldots a_{d}, q\right):=\begin{cases}
1 \text { if } q \leq f\left(a_{1}, a_{2}, \ldots a_{d}\right) Q \\
0 \text { if } q>f\left(a_{1}, a_{2}, \ldots a_{d}\right) Q.
\end{cases}
\end{align}
Therefore, for a given $\{\boldsymbol{a}_j\}_{j=1}^N$, the fraction of values $q$ for which
$b(a_1,\ldots ,a_d, q) = 1$ is the best approximation to $f(a_1,\ldots ,a_d)$, and as a result, the average value of $b$ is identical to the average value of $f$. 

Since $b$ is a Boolean-valued function, its average value can be estimated by approximate counting
\begin{align}
    \mathcal{S} = \braket{b} = \mathbb{E}[b] = \frac{K}{NQ},
\end{align}
where $K$ corresponds to the counts of $b$ evaluating to one, $K = |b^{-1}(1)|$, for a given $\{\boldsymbol{a}_j\}_{j=1}^N$. Recall from the previous discussion on the principle of QAE/QAA that, counting the number of solutions $K$, the state of the system rotates by applying the Grover operator $G$ on $\ket{s} = \sqrt{(1-a)} \ket{w^\perp} + \sqrt{a}\ket{w}$, where $\ket{w^\perp},\ket{w} \in \mathcal{H}$ are the \emph{bad} and \emph{good} states, respectively, as in Eq. \eqref{eq:sprime}\footnote{The correspondence between \cite{abrams1999fast} and our article is $\alpha \ket{s} + \beta U^{-1}\ket{t} \equiv \sqrt{1-a}\ket{w^\perp}+\sqrt{a}\ket{w}$.}. However, similarly to the previous subsection, if the good state is a superposition over all basis vectors $\ket{i}$ for which $b(i) = 1$, that is, $\ket{w} = \sum_{i:b(i)=1}\ket{i}$, then its amplitude is $\sqrt{K/(NQ)}$. After a threshold number of applications of $G$, the amplitudes of $\ket{w^\perp}$ and $\ket{w}$ oscillate depending on $K$. Therefore, we can create and measure the following superposition:
\begin{align}
    \ket{\psi} = \frac{1}{A}\sum_{i=0}^{A-1}\ket{i}G^i\ket{w^\perp}.
\end{align}
Finally, the value of $K$ can be computed by performing the QFT on the first
register; the accuracy $1/A$ is controlled by the size $A$ of the QFT register, which in turn fixes the number of logic operations \cite{abrams1999fast}. The sum $\mathcal{S}$ can thus be estimated to $\varepsilon$ accuracy with $\mathcal{O}(1/\varepsilon)$ operations, as is the case with the QAA algorithm. Also, as in the case of QAE, the number of operations does not depend on the size of the
domain of the integral $\mathcal{S}$ approximates, but only on the desired value of $\varepsilon$.

In \cite{Gall2022}, quantum approximate counting was revisited in a generalized setting where it estimates the number of marked states of a Markov chain, making it possible to construct quantum approximate counting algorithms from \emph{quantum-walk}-based search algorithms.

\subsection{The end-to-end perspective}\label{subsection:taxonomy}

So far, we have seen Brassard et al.'s general procedure for quantum amplitude estimation \cite{Brassard1998} as well as some variants and alternatives for the problems of Boolean summation and mean estimation. In each case, we begin by assuming access to some oracle $U_f$, and perhaps also a general quantum subroutine $\mathcal{A}$ encoding the input of the algorithm. However, it is instructive to consider the entire procedure of solving a problem using a quantum Monte Carlo method.

Figure \ref{fig:hourglass} summarizes the \textit{end-to-end} implementation of quantum Monte Carlo methods at the highest level of abstraction. Like most quantum algorithms, the general procedure first involves some classical reduction, followed by the application of a quantum algorithm, potentially followed by further classical post-processing.

The first step is to reduce the problem of interest to the problem of estimating the expectation of a function of a discrete random variable. The second step is to construct quantum subroutines (e.g. oracles or controlled oracles) that grant access to the values of the function, and to prepare states whose amplitudes encode the probability distribution of the random variable. The third step is to implement one of the quantum algorithms described in the introduction, or one of the more modern variants described in Section \ref{sec:quantum_opp}.

\begin{figure}[h!]
    \centering
    \begin{tikzpicture}[
    every node/.style={align=center, inner sep=3pt},
    arrow/.style={-{Stealth[length=5pt]}, thick},
  ]

  \node[text width=3cm, font=\small] (ex1) at (-5, 4)
    {Path integration\\(e.g.\ Feynman--Kac,\\Traub et al.)};
  \node[text width=3.5cm, font=\small] (ex2) at (0, 4)
    {Option pricing \&\\estimation of risk\\measures (e.g.\ VaR, CVaR)};
  \node[text width=3cm, font=\small] (ex3) at (5, 4)
    {Estimation of partition\\functions\\(e.g.\ Ising model, graph coloring...)};

  \draw[decorate, decoration={brace, amplitude=8pt}]
    (-7, 5.5) -- (7, 5.5)
    node[midway, above=10pt, font=\itshape] {Problems amenable to (quantum) Monte Carlo methods};

  \node[text width=3.5cm] (app1) at (-5, 0)
    {Approximating $\int_{[0,1]^d} f(x)\,d\mu$ \\ $f$ a member of some \\
    function class $F$};
  \node[text width=4.5cm] (app2) at (0, 0)
    {Estimating $\mathbb{E}[f(X)]$
     for $X$ a random variable and
     $f$ an arbitrary measurable function};
  \node[text width=4cm] (app3) at (5, 0)
    {Approximating a partition function
     $Z(\beta)=\sum_{x\in\Omega}e^{-\beta H(x)}$};
  \draw[arrow, dashed] (app1) -- (app2);

  \draw[arrow] (ex1) -- (app1);
  \draw[arrow] (ex2) -- (app2);
  \draw[arrow] (ex3) -- (app3);

  \node[draw] (estimate) at (0, -4)
    {Estimate $\mathbb{E}_{\pi_i}\bigl(f_j(X_i)\bigr)$\\$X_i$ discrete, $i \in I, j\in J$ finite};
  \node[anchor=west] (classical) at (4, -4) {Classical \\ randomized \\algorithms};
  \draw[arrow, dashed] (estimate) -- (classical);

  \draw[arrow] (app1) -- (estimate);
  \draw[arrow] (app2) -- (estimate);
  \draw[arrow] (app3) -- (estimate);

  \node[draw] (PA) at (0, -8.5)
    {$\left(\pi_{i}\right)_{i\in I},\,\left(U_{f_j}\right)_{j \in J}, \,\left(\mathcal{A}_{k}\right)_{k\in K}$};
  \draw[arrow] (estimate) -- (PA);

  \node[text width=3cm, font=\small, anchor=east] (amplitudes) at (-2.5, -8.5)
    {amplitudes of\\state(s) encoding\\distributions of $X_j$};
  \draw[arrow] (amplitudes) -- (PA);

  \node[text width=3cm, font=\small, anchor=west] (oracle) at (2.5, -8.5)
    {quantum subroutine(s)\\(e.g.\ oracle(s)) giving access to \\
    values of function $f$, \\ or more general subroutines for \\ state preparation};
  \draw[arrow] (oracle) -- (PA);

  \node (qae) at (-5, -13) {Quantum amplitude \\ estimation via \\ phase estimation};
  \node (qsp) at (-1.7, -13) {Quantum signal \\ processing};
  \node (mle) at (1.7, -13) {Measurement-based \\ methods};
  \node (qac) at (5, -13) {Quantum approximate \\ counting};

  \draw[arrow] (PA) -- (qae);
  \draw[arrow] (PA) -- (qsp);
  \draw[arrow] (PA) -- (mle);
  \draw[arrow] (PA) -- (qac);

  \draw[decorate, decoration={brace, mirror, amplitude=8pt}]
    (-6.2, -13.8) -- (6.2, -13.8)
    node[midway, below=10pt, font=\itshape] {Quantum algorithms estimating $\mathbb{E}_{\pi_j}\bigl(f_i(X_j)\bigr)$, \\ using quantum computational resources $\left(\left(\pi_{i}\right)_{i\in I},\left(A_{j}\right)_{j\in J}\right)$};

  \draw[decorate, decoration={brace, mirror, amplitude=8pt}]
    (-7.5, 4.9) -- (-7.5, -4.5)
    node[midway, left=10pt, font=\itshape] {Step 1};
  \draw[decorate, decoration={brace, mirror, amplitude=8pt}]
    (-7.5, -5) -- (-7.5, -9)
    node[midway, left=10pt, font=\itshape] {Step 2};
  \draw[decorate, decoration={brace, mirror, amplitude=8pt}]
    (-7.5, -9.5) -- (-7.5, -14)
    node[midway, left=10pt, font=\itshape] {Step 3};

\end{tikzpicture}
    \caption{A schematic, hourglass-shaped view of the end-to-end procedure for quantum Monte Carlo methods. Figure~\ref{fig:QAEvars} gives an expanded view of the bottom half of the diagram.}
    \label{fig:hourglass}
\end{figure}

The top half of the hourglass is essentially classical in nature: it captures the procedure of reducing a problem to the problem of estimating $\mathbb{E}_{\pi_{j}}[f_{i}(X_{j})]$. As the dashed arrow suggests, once the
discrete families $\{f_{i}\}$ and $\{X_{j}\}$ have been constructed, we could in principle dispense with a quantum computer altogether and pass to a classical randomized algorithm. The bottom half of the diagram comprises the
genuinely quantum parts of the procedure: the preparation of the amplitude-encoding states $(\ket{\pi_{i}})$, the realization of the oracles $(\mathcal{A}_{j})$, and the QAE-style algorithms that operate on them.

Although every problem class passes through the abstraction
$((\pi_{i})_{i\in I},(A_{j})_{j\in J})$, the (optimal) way to construct this pair can depend on the upstream problem. We refer to this phenomenon as the \emph{path-dependence} of
the hourglass. 

A point that is easy to miss when working at the level of the abstract template is that the upstream problem almost always supplies additional structure that can be exploited when constructing $(\ket{\pi_{i}})$ and $(\mathcal{A}_{j})$. One rarely meets
the pair as a genuine black box: in practice, some physical or mathematical characteristic of the original problem will inform the choice of $(\pi_{i})$ and $(\A_{j})$, as well as the appropriate choice of quantum Monte Carlo method (the bottom row of Figure \ref{fig:hourglass}). Several representative cases
(developed in detail in \ref{sec:apps}) illustrate this:
\begin{itemize}
    \item For the integration problem $\int_{[0,1]^{d}} f\,\mathrm{d}\mu$, the
    regularity class of $f$ (e.g.\ membership of an $L_{p}$ or Sobolev space
    $W^{r}_{p}$) is typically known a priori. This regularity informs both the
    optimal choice of discretization $\Pi_{i}$, as well as the oracle $\mathcal{A}_{i}$
    and the best achievable convergence rate~\cite{novak2001quantum}.
    \item As an application of the previous example, we show how these regularity-based quantum complexity results inform the design of algorithms for path integration \cite{traub2002path} and for the solution of elliptic PDE \cite{heinrich2006elliptic}. 
    \item For the Ising-model partition function, each $\pi_{i}$ is a Gibbs
    measure associated with a \emph{known} classical Hamiltonian. The state
    $\lvert\pi_{i}\rangle$ can therefore be prepared via a quantum walk built
    from a Markov chain (e.g.\ the Glauber dynamics) whose mixing time and
    spectral gap can be analyzed directly from the interaction graph of the
    model.

\end{itemize}
Moreover, in section \ref{sec:finance}, we discuss the application of quantum Monte Carlo methods to problems in mathematical finance.

\section{Reduction to a quantum amplitude estimation/quantum counting problem}\label{sec:apps}

As mentioned in Subsection \ref{subsection:taxonomy}, all possible applications reduce to estimating expectations of the form $\mathbb{E}_{\pi}\left[f\left(X\right)\right]$, where $f$ is a function and $\pi$ is the distribution of the random variable $X$. Given an application, one makes choices that ultimately lead to a well-defined problem

\begin{equation}\label{eq:qmc-problem-definition}
\left(\left(\pi_{i}\right)_{i\in I}, \left(U_{f_j}\right),\left(\mathcal{A}_{k}\right)_{k\in K}\right),    
\end{equation}
where $I, J$ are finite index sets for the distributions and oracles, respectively. This data then serves as the input for one of the quantum Monte Carlo algorithms discussed in the introduction, or one of the more modern variants surveyed in Section \ref{sec:quantum_opp}. In this section, we discuss how problems of interest are reduced to the form (\ref{eq:qmc-problem-definition}) by considering three applications: (i) the integration of H\"older and Sobolev functions, (ii) the estimation of the partition function of an Ising model, and (iii) path integration. The discussion of applications to problems in quantitative finance is postponed to Section \ref{sec:finance}.

\subsection{Query complexity of integration of Hölder and Sobolev functions}\label{subseq:holdersobolev}

In Section \ref{sec:quantum_alternatives}, we have treated the oracle $U_{f}$ as a given, abstract object: a ``black-box'' unitary allowing us to access $f$ through its values. In practice, however, building such an oracle requires choosing a discretization of the domain and range. These choices affect the precision or equivalently the query complexity of the resulting quantum algorithm. This is another instance where a priori knowledge of the problem (in this case the regularity of the integrand) influences the optimal choice of discretization and hence, oracle.

To make this dependence precise, we specialize the framework of Heinrich \cite{heinrich2002quantum, heinrich2003monte} in order to discuss the integration operator $S$ of functions over a hypercube. Let $F$  be a set of functions from $D = [0,1]^d$ to $\mathbb{R}$, and let $S \colon F \to K$ be defined as
\[
  S(f) \;=\; \int_{D} f(x)\, dx.
\]

Using a quantum computer to compute $S$ requires two discretizations: one of the domain and one of the range. To discretize the domain we choose a non-empty index set $Z \subseteq \{0, \ldots, 2^{m'}-1\}$
together with a mapping
\[
  \tau \colon Z \to D
\]
that assigns to each bit-string label $i \in Z$ a point in the domain $\tau(i) \in D$ at which $f$ may be evaluated. 

The range is discretized by function values $f(\tau(i)) \in K \subseteq \mathbb{R}$, which are encoded into
an $m''$-qubit register by a mapping
\[
  \beta \colon K \to \{0, \ldots, 2^{m''}-1\}.
\]
If $f$ takes values in $[a,b]$ a standard choice is to pick $\beta$ representing a uniform partition of $[a,b]$ into $2^{m''}$ subintervals.
 
With these choices fixed, the tuple $(m, m', m'', Z, \tau, \beta)$ defines, for each $f \in F$, a unitary \emph{quantum query} $Q_{f} \in U(H_{m})$, where $m \ge m' + m''$. On the computational basis state $|i\rangle\,|x\rangle\,|y\rangle$, with $|i\rangle \in H_{m'}$, $|x\rangle \in H_{m''}$ and $|y\rangle \in H_{m - m' - m''}$ representing the index, value and workspace registers respectively, $Q_f$ acts as
\begin{equation}
  Q_{f}\, |i\rangle\,|x\rangle\,|y\rangle
  \;=\;
  \begin{cases}
    |i\rangle\,|x \oplus \beta(f(\tau(i)))\rangle\,|y\rangle
      & \text{if } i \in Z, \\[4pt]
    |i\rangle\,|x\rangle\,|y\rangle
      & \text{otherwise,}
  \end{cases}
  \label{eq:heinrich-query}
\end{equation}
where $\oplus$ denotes addition modulo~$2^{m''}$. When the value register is initialized to $|0\rangle$, the action $|i\rangle\,|0\rangle \mapsto |i\rangle\,|\beta(f(\tau(i)))\rangle$ recovers the familiar picture of the oracle ``writing the function value into the second register''.

A quantum algorithm is a sequence of unitaries $U_{0}, U_{1}, \ldots, U_{n}$ (independent of $f$) interleaved with applications of $Q_{f}$:
\[
  B_{f} \;=\; U_{n}\, Q_{f}\, U_{n-1}\, Q_{f}\, \cdots\, U_{1}\, Q_{f}\, U_{0},
\]
applied to a fixed initial state followed by a measurement and a classical post-processing step that returns the output in $\mathbb{R}$. The algorithm is said to compute $S(f)$ with error $\varepsilon$ if the output lies within $\varepsilon$ of $S(f)$ with probability at least $3/4$. The \emph{query complexity} of the problem is then the smallest number of applications of $Q_{f}$ required to achieve error $\varepsilon$, uniformly over $f \in F$, and serves as the basic measure against which classical deterministic and randomized complexities are compared.

With the query model in place, we can make precise how the regularity of the integrand controls the achievable quantum query complexity. Introducing the multi-index notation $\nu = (\nu_1,\dots,\nu_d) \in \mathbb{N}_0^{d}$ we set $|\nu| = \nu_1 + \cdots + \nu_d$ and $\partial^{\nu} = \partial_{x_1}^{\nu_1}\cdots\partial_{x_d}^{\nu_d}$, we define the following two function classes. The (unit norm) H\"older class consists of $r$-times continuously differentiable functions whose derivatives of order $r$ are $\alpha$-H\"older continuous,
\begin{equation}
    F^{r,\alpha}_{d} \;=\; \bigl\{ f \in C^{r}([0,1]^{d}) \;:\;
    \|f\|_{\infty} \le 1,\;
    |\partial^{\nu} f(x) - \partial^{\nu} f(y)| \le |x-y|^{\alpha} \ \text{whenever}\ |\nu| = r
  \bigr\},
\end{equation}
and the (unit norm) Sobolev class is
\begin{equation}
W^{r}_{p,d} \;=\; \bigl\{ f \in L^{p}([0,1]^{d}) \;:\;
    \|\partial^{\nu} f\|_{p} \le 1 \ \text{for all}\ |\nu| \le r
  \bigr\},
\end{equation}
where in the Sobolev case $\partial^{\nu}$ denotes the weak partial derivative. Heinrich and Novak \cite{heinrich2002quantum, novak2001quantum, heinrich2002optimal} use a classical preprocessing step to obtain the optimal quantum query complexity bound. The authors choose $n$ deterministic sample points and a corresponding interpolation (or quadrature) operator $P_{n}$ adapted to the class $F$, and writes
\begin{equation}
  \int_{D} f(x)\, dx
  \;=\; \int_{D} P_{n} f(x)\, dx \;+\; \int_{D} \bigl(f(x) - P_{n} f(x)\bigr)\, dx.
  \label{eq:separation}
\end{equation}
The first integral is a finite linear combination of function values and may be evaluated exactly by a classical  quadrature at cost $\mathcal{O}(n)$. The remainder $g \coloneqq f - P_{n} f$ has small sup-norm: classical results in approximation theory \cite{novak1988} show that $P_{n}$ may be chosen so that
\begin{equation}
  \|g\|_{\infty} \;=\; \|P_{n} f - f\|_{\infty} \;=\; \mathcal{O}(n^{-\gamma}),
  \label{eq:approx-rate}
\end{equation}
where the exponent $\gamma$ depends on the class: $\gamma = (r+\alpha)/d$ for $f \in F^{r,\alpha}_{d}$, and $\gamma = r/d$ for $f \in W^{r}_{p,d}$ \cite{novak1988}. The ratio of smoothness to dimension -- $(r+\alpha)/d$ or $r/d$ -- is therefore intrinsic to $F$, and the appearance of $d$ in the denominator is the familiar curse of dimensionality.

It remains to integrate the remainder $g$. Since $\|g\|_{\infty} = \mathcal{O}(n^{-\gamma})$, the rescaled function $\tilde{g} \coloneqq g/\|g\|_{\infty}$ can be integrated to precision $\varepsilon$ with query complexity $\mathcal{O}(n^{-1})$ by the quantum mean-estimation result of Brassard, H{\o}yer, Mosca and Tapp \cite{Brassard2002} (see also Lemma~7 of Heinrich \cite{heinrich2002quantum}). Applying this bound to $\tilde{g}$ and rescaling restores the factor $\|g\|_{\infty} = \mathcal{O}(n^{-\gamma})$, so that the remainder integral is approximated with error
\begin{equation}
    \varepsilon
  \;=\; \mathcal{O}\bigl(\|g\|_{\infty} \cdot n^{-1}\bigr)
  \;=\; \mathcal{O}\bigl(n^{-\gamma - 1}\bigr).
\end{equation}
The total error then decomposes neatly as the product
\begin{equation}
  \underbrace{n^{-\gamma}}_{\text{separation of main part}}
  \;\cdot\;
  \underbrace{n^{-1}}_{\text{quantum summation}}
  \;=\; n^{-\gamma - 1},
  \label{eq:product-rate}
\end{equation}
in which the two stages contribute independently: the classical quadrature exploits the regularity of $f$, while the quantum subroutine delivers the quadratic quantum speed-up on the residual. Inverting this rate yields a quantum query complexity of $O\bigl(\varepsilon^{-1/(1+\gamma)}\bigr)$ to approximate the integral of $f$ to precision $\varepsilon$. The resulting rates are summarized in Table~\ref{table:functionclasses} below, and match the optimal quantum rates established (up to logarithmic factors) by Heinrich and Novak \cite{heinrich2002quantum, novak2001quantum}.
 
\textbf{Remark}: There are two ways to fit this decomposition into the hourglass framework. The simplest is to consider the replacement of $f$ with $f-P_nf$ as a classical preprocessing step. However, Heinrich also shows, in Lemmas 4 and 5 of \cite{heinrich2002quantum} how to coherently exchange an oracle (in his language, a quantum query) for $f$ with an oracle for $f-P_n$. This still requires some classical preprocessing in order to compute $P_n$, however, these results effectively allow us to start in the middle of the hourglass, with an oracle for $f$. This is another illustration of how a priori knowledge of some aspect of the problem can influence the construction of the optimal quantum algorithm.

\subsection{Extension to other function classes}

In the following, we discuss the possible extension of the results above to Besov functions. Let
\[
  \mathbb B^d=\{x\in\mathbb R^d: |x|\le 1\},
\]
and let
\[
  w_\mu(x)=b_d^\mu (1-|x|^2)^{\mu-1/2},\qquad \mu\ge 0,
\]
be the normalized Jacobi weight on $\mathbb B^d$(where $b_d^\mu$ is chosen so that $\int_{\mathbb{B}^{d}}\left|w_{\mu}(x)\right|dx=1$).
Define
\[
  \|f\|_{p,\mu}
  =
  \left(\int_{\mathbb B^d}|f(x)|^p w_\mu(x)\,dx\right)^{1/p},
\]
when $1\le p<\infty$ and let $\|f\|_{\infty,\mu}$ be the uniform norm (which covers the case $p=\infty$).

Let $\Pi_L^d$ denote the space of polynomials of degree at most
$L$ restricted to $\mathbb B^d$, and define the best weighted polynomial approximation
error by
\[
  E_L(f)_{p,\mu}
  =
  \inf_{P\in \Pi_L^d}\|f-P\|_{p,\mu}.
\]
For $r>0$, $1\le p\le\infty$, and $0<\tau\le\infty$, the weighted Besov
space $B^r_\tau(L_{p,\mu})$ is defined by the quasi-norm
\[
  \|f\|_{B^r_\tau(L_{p,\mu})}
  =
  \|f\|_{p,\mu}
  +
  \left(
    \sum_{j=0}^{\infty}
    2^{jr\tau} E_{2^j}(f)_{p,\mu}^{\tau}
  \right)^{1/\tau},
  \qquad 0<\tau<\infty,
\]
for $1 \leq 0 < \infty$ and 
\[
  \|f\|_{B^r_\infty(L_{p,\mu})}
  =
  \|f\|_{p,\mu}
  +
  \sup_{j\ge 0} 2^{jr}E_{2^j}(f)_{p,\mu}
\]
for $p = \infty$. The corresponding Besov class $BB^r_\tau(L_{p,\mu})$ is the unit ball of
$B^r_\tau(L_{p,\mu})$.

This definition is equivalent to other standard definitions of weighted
Besov spaces on the ball.  Moreover, if
\[
  r>\frac{d+2\mu}{p},
\]
then $B^r_\tau(L_{p,\mu})$ is compactly embedded into $C(\mathbb B^d)$,
so point evaluations and numerical integration are well-defined.

The state of the art result is the independence of the quadrature error of
\[
  \int_{\mathbb B^d} f(x)w_\mu(x)\,dx,
\]
on $\tau$, resulting in the optimal quadrature error \cite{li2022besovball}
\[
  n^{-r/d-1/2+(1/p-1/2)_+},
\]
where $+$ denotes the positive part, which holds
in the Monte Carlo setting.
An analogous result holds in the case of Sobolev
spaces, which is levereged by Heinrich \cite{heinrich2002optimal, heinrich2003monte}, resulting in the improvement of the quantum query complexity bounds described in the previous section. We conjecture that repeating Heinrich's analysis using the quadrature of ~\cite{li2022besovball} should yield an analogous optimal quantum query complexity result for the integration of Besov functions. 

At present, however, the aforementioned weighted Besov results apply only to classical Monte Carlo, which is reflected by Table~\ref{tab:det-ran-quantum}, where we gather the results on Besov, Sobolev and Hölder spaces. Note that we also include the regime $\tau=\infty$ which is identical to the anisotropic Hölder results
treated in \cite{hu2005anisotropic}.
 
\begin{table}[H]\label{table:functionclasses}
  \centering
  \scriptsize
  \setlength{\tabcolsep}{2pt}
  \renewcommand{\arraystretch}{1.4}
  \begin{tabular}{llllll}
    \hline
    Problem & Class & Determ. & Random. & Quantum & Ref. \\
    \hline
    \multicolumn{6}{l}{\emph{Mean of $N$ numbers (range $n\le cN$)}} \\
    $S_N$ & $\mathcal{B}(L_\infty^N)$
      & $1$ & $n^{-1/2}$ & $n^{-1}$
      & \cite{Brassard2002} \\
    $S_N$ & $\mathcal{B}(L_p^N)$, $p\ge 2$
      & $1$ & $n^{-1/2}$ & $n^{-1}$
      & \cite{heinrich2002quantum} \\
    $S_N$ & $\mathcal{B}(L_p^N)$, $1\le p<2$
      & $1$ & $n^{-2(1-1/p)}$
      & \begin{tabular}[t]{@{}l@{}}$\min(n^{-2(1-1/p)},$\\ $\quad n^{-2/p}N^{2/p-1})$\end{tabular}
      & \cite{heinrich2003problem} \\
    \hline
    \multicolumn{6}{l}{\emph{Integration on $D=[0,1]^d$}} \\
    $I_d$ & $F^{r,\alpha}_d$
      & $n^{-(r+\alpha)/d}$ & $n^{-(r+\alpha)/d-1/2}$
      & $n^{-(r+\alpha)/d-1}$
      & \cite{novak2001quantum} \\
    $I_d$ & $\mathcal{B}(W^r_p)$, $p\ge 2$
      & $n^{-r/d}$ & $n^{-r/d-1/2}$ & $n^{-r/d-1}$
      & \cite{heinrich2003sobolev} \\
    $I_d$ & $\mathcal{B}(W^r_p)$, $1<p<2$
      & $n^{-r/d}$ & $n^{-r/d-1+1/p}$ & $n^{-r/d-1}$
      & \cite{heinrich2003sobolev} \\
    $I_d$ & $\mathcal{B}(W^r_1)$
      & $n^{-r/d}$ & $n^{-r/d}$ & $n^{-r/d-1}$
      & \cite{heinrich2003sobolev} \\
    \hline
    \multicolumn{6}{l}{\emph{Integration: Besov classes,
      $r/d>1/p$, any $\tau$}} \\
    $I_d$ & $\mathcal{B}(B^r_{\tau})$
      & $n^{-r/d}$ & $  n^{-r/d-1/2+(1/p-1/2)_+}$ (a) & open (b)
      & \cite{li2022besovball}  \\
    $I_d$ & $H^{r}_p$ anis.\ ($\theta{=}\infty$)
      & $n^{-g(r)}$ & $n^{-g(r)}$ & $n^{-g(r)-1}$
      & \cite{ye2008anisotropic} \\
    \hline
  \end{tabular}
  \caption{The minimal errors for Monte Carlo integration dependent of function classes in the deterministic, randomized, and
    quantum settings; where logarithmic factors are suppressed.
    }
  \label{tab:det-ran-quantum}
\end{table}
 
\subsection{Concrete applications of regularity-based complexity bounds}\label{subseq:regularityapp}

To a less mathematically inclined reader, the regularity-based quantum query complexity discussed in the previous section may appear as little more than a (highly technical) curiosity. However, in many physically interesting situations, we do have a priori knowledge about the regularity of the integrand. Moreover, the methods of Heinrich \cite{heinrich2002quantum} extend to problems more general than integration over a hypercube. Table~\ref{tab:op-eq} collects the rates for a representative selection of such problems, including path integration on a Gaussian-measure space \cite{traub2002path}, the related Feynman--Kac integration of expectation functionals of a Brownian motion \cite{kwas2006feynmankac}, parametric integration whose output is itself a function of an external parameter \cite{wiegand2006parametric}, the initial-value problem for ordinary differential equations \cite{kacewicz2004ivp,kacewicz2005ivp}, the Sturm--Liouville eigenvalue problem \cite{papageorgiou2005sturm}, elliptic boundary-value problems with the solution sampled on a submanifold \cite{heinrich2006elliptic}, and $L_q$-approximation of Sobolev functions \cite{heinrich2004approximation}. 

In this subsection, we single out two illustrative examples: path integration \cite{traub2002path} and the solution of elliptic PDE \cite{heinrich2006elliptic}.

\begin{table}\label{table:regularityapplications}[H]
  \centering
  \renewcommand{\arraystretch}{1.9}
  \setlength{\tabcolsep}{6pt}
  \resizebox{\textwidth}{!}{%
  \footnotesize
  \begin{tabular}{llllll}
    \hline\hline
    \textbf{Problem} & \textbf{Class}
      & \textbf{Deterministic} & \textbf{Randomized}
      & \textbf{Quantum} & \textbf{Ref.} \\
    \hline
    \multicolumn{6}{l}{\emph{Operator equations}} \\[3pt]
    Path integration
      & Integrand in $FL_r$, Gaussian measure
      & $\mathrm{e}^{\,\Theta(1/\varepsilon)}$
      & $\varepsilon^{-2}$
      & $4.46\,\varepsilon^{-1}$
      & \cite{traub2002path} \\[4pt]
    Feynman--Kac path integration
      & smooth payoff and potential
      & $\varepsilon^{-\Theta(\log \varepsilon^{-1})}$
      & $\varepsilon^{-2}$
      & $\varepsilon^{-1}$
      & \cite{kwas2006feynmankac} \\[4pt]
    Parametric integration
      & \twoline{$C^r([0,1]^{d_1+d_2})$,}{output a function of $s\in[0,1]^{d_1}$}
      & $n^{-r/(d_1+d_2)}$
      & $n^{-r/(d_1+d_2)-1/2}$
      & $n^{-\min\!\left(\frac{r}{d_2},\,\frac{r}{d_1+d_2}+1\right)}$
      & \cite{wiegand2006parametric} \\[4pt]
    IVP (ODE)
      & $C^r$ right-hand side
      & $n^{-r}$
      & $n^{-r-1/2}\;{}^{(f)}$
      & $n^{-r-1}\;{}^{(f)}$
      & \cite{kacewicz2005ivp} \\[4pt]
    Elliptic PDE, order $2m$
      & \twoline{$C^r(D)$ data,}{output on dim-$d_1$ submanifold}
      & $n^{-r/d}$
      & $n^{-\min\!\left(\frac{r+2m}{d_1},\,\frac{r}{d}+\frac{1}{2}\right)}$
      & $n^{-\min\!\left(\frac{r+2m}{d_1},\,\frac{r}{d}+1\right)}$
      & \cite{heinrich2006elliptic} \\[4pt]
    Sturm--Liouville (smallest eigenvalue)
      & $q\in C^r([0,1])$, $\|q\|_\infty\le 1$
      & $\varepsilon^{-1/r}$
      & $\varepsilon^{-1/(r+1/2)}$
      & $\varepsilon^{-1/(r+1)}$
      & \cite{papageorgiou2005sturm} \\[4pt]
    $L_q$-approximation
      & $\mathcal{B}(W^r_p([0,1]^d))$, $p\ge q$
      & $n^{-r/d}$
      & $n^{-r/d}\;{}^{(g)}$
      & $n^{-r/d}\;{}^{(g)}$
      & \cite{heinrich2004approximation} \\[4pt]
    \hline\hline
  \end{tabular}}
  \caption{The (deterministic, randomized and quantum) query complexity of various algorithms based around QAE/QAC in terms of the integrand. Quantities in terms of $\varepsilon$ are asymptotic query complexities, while quantities in terms of $n$ denote asymptotic error obtained using $n$ queries.}
  \label{tab:op-eq}
\end{table}

\subsubsection*{Path integration via QAE}

\emph{Path integration} is the problem of computing $\int_X f(x)\,\mu(dx)$, where $X$ is an infinite-dimensional separable Banach space and $\mu$ is a zero-mean Gaussian measure on $X$. The canonical example is the Wiener measure on $X = C([0,1])$, which assigns probability to continuous trajectories of Brownian motion and underlies, among other things, the Black--Scholes--Merton family of derivative-pricing models. Up to an embedding into $L_2([0,1])$, the measure $\mu$ is characterized by the eigenvalues $\lambda_j$ of its covariance operator, and we assume the standard decay $\lambda_j = \Theta(j^{-k})$ with $k>1$ (the Wiener case is $k=2$). Integrands are assumed to lie in the class $FL_r$ of functions with bounded Fréchet derivatives up to order $r-1$ and Lipschitz $r^{th}$ derivative. In the worst-case deterministic setting this problem is \emph{intractable}: its $\varepsilon$-complexity grows exponentially in $\varepsilon^{-1}$ \cite{traub2002path}. Monte Carlo restores tractability, achieving complexity $\varepsilon^{-2(1+o(1))}$ but no better.

The route to a quantum algorithm proposed by \cite{traub2002path} is essentially a quantum variant of a deterministic quadrature due to Curbera \cite{curbera1998path}. The first step is to truncate the infinite-dimensional integral to a $d$-dimensional Gaussian integral, where $d = \Theta(\varepsilon^{-c/(k-1)})$ active variables suffice to capture the integrand to accuracy $\varepsilon/2$ (with $c=2$ for $r=1$ and $c=1$ for $r\ge 2$). Curbera then approximates the $d$-dimensional Gaussian integral by an equal-weight quadrature on a tensor-product grid of $m^d$ nodes, with $m = \Theta(\varepsilon^{-1})$ chosen large enough to control the remaining error. The output is an algorithm of the form
\begin{equation}\label{eq:curberasum}
    S_n(f) \;=\; \frac{1}{n}\sum_{\vec j} f\bigl(x_{\vec j}\bigr),
  \qquad
  n \;=\; m^d,
\end{equation}
which already has the sum-of-function-values structure that quantum mean estimation operates on. The number of summands $n$ is exponential in $\varepsilon^{-1}$ and this is exactly what makes the deterministic complexity exponential, since each summand must be evaluated.

Traub and Wo\'zniakowski's approach is simple: they propose to evaluate the Curbera sum \ref{eq:curberasum} using QAE or one of its variants. Recall that QAE computes a mean of $n$ bounded values to additive error $\varepsilon$ using $O(\varepsilon^{-1})$ quantum queries, while the qubit complexity depends only logarithmically on $n$. The exponential summand count is therefore no longer a barrier: $\log n$ is polynomial in $\varepsilon^{-1}$, and the resulting algorithm has polynomial total cost. Quantitatively, it achieves error $\varepsilon$ with probability $\ge 3/4$ using at most $4.46\,\varepsilon^{-1}$ quantum queries together with $O\!\bigl(\varepsilon^{-(1+\gamma(r))/(k-1)} \log \varepsilon^{-1}\bigr)$ qubits, where $\gamma(1)=1$ and $\gamma(r)=0$ for $r\ge 2$.

\subsubsection*{Elliptic boundary-value problems on a submanifold}

Another application of the regularity-based complexity theory pertains to the problem of solving an elliptic boundary-value problem and reading off the solution on a chosen subset of the domain \cite{heinrich2006elliptic}. Concretely, we fix a general elliptic partial differential operator $L$ of order $2m$ with smooth coefficients on a smooth bounded domain $D\subset\mathbb{R}^d$, together with homogeneous boundary conditions, and consider the equation $Lu = f$ for right-hand sides $f\in C^r(D)$. The output is not the function $u$ in its entirety but its restriction to a smooth $d_1$-dimensional submanifold $M\subseteq D$, with error measured in the $L_\infty(M)$ norm. We will see that restricting the solution to a submanifold influences (favorably) the query complexity: $d_1$ appears in the final complexity bound \ref{eq:elliptic-rate}. Of course, if we wish, we can set $D = \{x\},\;d_1 = 0$, corresponding to evaluating $u$ at a single point $x \in M$, or we can set $D = M, \;d_1 = d$ to obtain a global solution $u$ on all of $M$. 

The algorithm rests on the representation of the solution as an integral. Standard elliptic theory (see e.g. \cite{krasovskij1967greens}) gives a kernel $k(x,y)$, the Green's function of $L$, such that
\begin{equation}\label{eq:kernelconvolution}
  u(x) \;=\; \int_Q k(x,y)\,f(y)\,dy,
\end{equation}
so evaluating $u$ at any point $x$ is itself an integration problem in the input $f$. Heinrich considers integral kernels $k(x,y)$ such that the integral operator constructed by integrating a test function against $k$ is $\textit{weakly singular}$, see Section 5 of \cite{heinrich2006elliptic}.

The algorithm builds an approximate $u$ on $M$ by sampling values $u(x_i)$ at a grid of evaluation points covering the submanifold, with each $u(x_i)$ computed as a weighted quantum mean estimate of $f$, with $k(x_i, \cdot)$ serving as a known weight. $k$ is decomposed into a multilevel sum of pieces of controlled support, and each piece is integrated by an instance of a \emph{weighted} quantum mean-estimation primitive, which achieves error $\mathcal{O}(n^{-1}\|g\|_{L_1})$ for weighted means $N^{-1}\sum_i g(i) f(i)$. The overall query budget is then allocated across grid points and decomposition levels via a multiplicativity lemma for $n$-th minimal errors, which separates the cost of representing $u$ on $M$ from the cost of the kernel-integration itself.

Heinrich proves that the $n$-th minimal quantum query error is, up to logarithmic factors,
\begin{equation}
  e_n^{q}(S, F) \;\asymp\;
  n^{-\min\!\bigl(\tfrac{r+2m}{d_1},\,\tfrac{r}{d}+1\bigr)}.
  \label{eq:elliptic-rate}
\end{equation}
For comparison, the classical deterministic and randomized rates are $n^{-r/d}$ and $n^{-\min((r+2m)/d_1,\,r/d+1/2)}$ respectively. Two features are worth reading off this result. The deterministic rate $n^{-r/d}$ ignores the submanifold structure: whether one wants the solution at a single point or on the entire domain, the cost is the same, since deterministic algorithms cannot exploit the localization. The randomized and quantum settings, by contrast, both see $d_1$, and the quantum gain over Monte Carlo appears as the familiar upgrade from ``$+1/2$'' to ``$+1$'' in the second argument of the minimum.

It should be noted that the construction of the kernel/Green's function $k$ is not part of the problem specification: we assume that we are given $k$, or can construct it. In practice, this is not a problem, indeed, for many common elliptic partial differential operators (for instance, the Laplacian), we even have an analytic representation for the kernel $k$, and computing the convolution \eqref{eq:kernelconvolution} to obtain the solution to the inhomogeneous problem $Lu = f$ is the computationally expensive step. 

\subsubsection*{Summary}

The common thread throughout the entries of Table \ref{tab:op-eq} is simple. First, we take a continuous, potentially infinite dimensional problem, and reduce this to the problem of computing a finite dimensional integral. Next, using a classical quadrature method, approximate the integral by a finite sum: it is at this stage that the regularity of the problem enters into consideration. Finally compute this sum using QAE or QAC, or one of their variants (see Section \ref{sec:quantum_alternatives}). 

Let us briefly discuss the remaining entries in the table. Feynman--Kac path integration \cite{kwas2006feynmankac} generalizes the Wiener-measure case by allowing a drift and potential, and inherits the same exponential-to-polynomial improvement. The initial-value problem for ordinary differential equations \cite{kacewicz2004ivp,kacewicz2005ivp} reduces to integration via a Picard iteration. The Sturm--Liouville eigenvalue problem \cite{papageorgiou2005sturm} fits the template via a power method whose inner products are computed by quantum mean estimation. Parametric integration \cite{wiegand2006parametric} exhibits a subtler phenomenon worth singling out: the quantum subroutine, in its favorable regime, sees only the dimension being integrated over and is insensitive to the dimension of the parameter, so the speed-up can be substantial when most of the dimensionality lives in the parameter rather than the integration variables. In every case the structure of the speed-up is the same as in (\ref{eq:product-rate}): a regularity-and-dimension exponent set by the problem, with the additive constant upgraded from $1/2$ (randomized) to $1$ (quantum). The deterministic setting either matches the regularity exponent or, as for path and Feynman--Kac integration, fails tractability altogether.

It is important to note that we do not always obtain a quantum advantage, illustrated by the $L_q$-approximation entry of Table~\ref{tab:op-eq} \cite{heinrich2004approximation}. There the \emph{output} of the problem is itself a function rather than a single number, and the quantum advantage disappears: deterministic, randomized, and quantum methods all achieve the same rate. The speed-up that runs through this section is fundamentally a speed-up for problems whose answer is a scalar obtained by averaging; it is the contraction onto that scalar that quantum mean estimation accelerates, not the underlying function-space operation.

\subsection{Estimating partition functions}\label{subseq:partitionfunctions}

The estimation of partition functions plays a key role in classical statistical
physics \cite{ValleauCard1972}, graph coloring \cite{FriezeVigoda2007}, combinatorics \cite{DyerFrieze1992}
and other areas. Given an exponentially large state space $\Omega$, the problem is
to estimate the sum
\begin{equation}\label{eq:partition-function}
Z(\beta)=\sum_{x\in\Omega}e^{-\beta H(x)},
\end{equation}
where $\beta = \tfrac{1}{k_B T}$ ($k_B$ is Boltzmann's constant) is the inverse
temperature and $H:\Omega \rightarrow \mathbb{R}^+_0$ is the Hamiltonian, expressing
the energy of the system in state $x$. In non-physical (combinatorial) applications Boltzmann's
constant is absorbed into a dimensionless parameter and $H$ becomes a dimensionless
function.

Rather than evaluating the sum in~\eqref{eq:partition-function} directly, the
standard approach reduces the problem to sampling from a sequence of carefully
chosen probability distributions. Introduce a sequence of inverse temperatures
\begin{equation}
0=\beta_{0}<\beta_{1}<\dots<\beta_{\ell}=\beta
\end{equation}
and write $Z(\beta)$ as the telescoping product
\begin{equation}
Z(\beta)=Z\left(\beta_{0}\right)\frac{Z\left(\beta_{1}\right)}{Z\left(\beta_{0}\right)}\cdots\frac{Z\left(\beta_{\ell}\right)}{Z\left(\beta_{\ell-1}\right)}=\left|\Omega\right|\prod_{i=0}^{\ell-1}\alpha_{i},
\end{equation}
where $\alpha_{i}=\dfrac{Z\left(\beta_{i+1}\right)}{Z\left(\beta_{i}\right)}$ for each
$i=0,1,\dots,\ell-1$ and we have used $Z(\beta_0)=Z(0)=|\Omega|$. To express each
ratio as an expectation, consider the Gibbs (Boltzmann) probability mass function
\begin{equation}
\pi_{i}\left(x\right)=\frac{e^{-\beta_{i}H(x)}}{Z\left(\beta_{i}\right)}
\end{equation}
together with the random variable
\begin{equation}
Y_{i}\left(x\right)=e^{-(\beta_{i+1}-\beta_{i})H(x)}.
\end{equation}
A short calculation then shows that each factor in the telescoping product is an
expectation,
\begin{equation}
\mathbb{E}_{\pi_{i}}\left[Y_{i}\right]
=\frac{1}{Z(\beta_i)}\sum_{x\in\Omega}e^{-\beta_i H(x)}\,e^{-(\beta_{i+1}-\beta_i)H(x)}
=\frac{Z(\beta_{i+1})}{Z(\beta_i)}=\alpha_{i},
\end{equation}
so the problem fits the general quantum Monte Carlo template
of~(\ref{eq:qmc-problem-definition}). Classically, $\alpha_i$ is estimated by drawing
samples $x_{1},\dots,x_{k}$ from $\pi_{i}$ and forming
\begin{equation}
\hat{\alpha}_{i}=\frac{1}{k}\sum_{j=1}^{k}e^{-(\beta_{i+1}-\beta_{i})H(x_{j})},
\end{equation}
where $H$, $\beta_{i+1}$ and $\beta_i$ are known.

In the quantum Monte Carlo setting we instead require that:
\begin{itemize}
    \item Each distribution $\pi_i$, $i=0,1,\dots,\ell-1$, is loaded into a coherent
    quantum-sample state $\left|\pi_{i}\right\rangle =\sum_{x\in\Omega}\sqrt{\pi_{i}(x)}\left|x\right\rangle$,
    where $\left|x\right\rangle$ is a qubit representation of a configuration in
    $\Omega$;
    \item There exist $\ell$ quantum oracles $\mathcal{A}_i$, $i=0,1,\dots,\ell-1$, acting as
    \begin{equation}
        \left|x\right\rangle \left|0\right\rangle \overset{\A_{i}}{\longrightarrow}\left|x\right\rangle \left|\Pi_{i}\!\left(e^{-(\beta_{i+1}-\beta_{i})H(x)}\right)\right\rangle,
    \end{equation}
    where $\Pi_i$ is the discretization operator that maps the value
    $e^{-(\beta_{i+1}-\beta_{i})H(x)}$ onto a finite register.
\end{itemize}
The resulting problem is therefore specified by finitely many probability mass
functions and finitely many oracles, i.e.\ the tuple
$\left(\left(\pi_{i}\right)_{i=0}^{\ell-1},\left(\A_{i}\right)_{i=0}^{\ell-1}\right)$.

To run the quantum mean-estimation algorithm we must replace classical sampling from $\pi_i$ with the preparation of the corresponding \emph{coherent Gibbs state}  $\ket{\pi_i}=\sum_{x\in\Omega}\sqrt{\pi_i(x)}\,\ket{x}$, and two broad strategies are available. Due to the physical nature of the distributions $\pi_i$ one can achieve this by using a class of initial state preparation algorithms that are typically used to supply a ground state of a given Hamiltonian \cite{Wocjan_2008, Montanaro_2015}.


\section{Modern variants of quantum amplitude estimation}\label{sec:quantum_opp}

\begin{figure}[h!]
    \centering
    \resizebox{\textwidth}{!}{%
    \begin{tikzpicture}[
    every node/.style={align=center, inner sep=3pt},
    arrow/.style={-{Stealth[length=4pt]}, semithick},
    concept/.style={font=\bfseries\small, text width=4.8cm},
    detail/.style={font=\footnotesize, text width=2.1cm},
    cite/.style={font=\tiny\ttfamily, text=gray, text width=2.4cm},
  ]
  \node[concept] (qae) at (0, 0)
    {Quantum Amplitude\\Estimation};
  \node[concept] (qpe) at (-7, -2)
    {Phase estimation};
  \node[concept] (qsp) at (0, -2)
    {Quantum signal processing / \\Quantum singular value transformation};
  \node[concept] (qpefree) at (7, -2)
    {Direct measurement\\+ classical\\post-processing};
  \draw[arrow] (qae) -- (qpe);
  \draw[arrow] (qae) -- (qsp);
  \draw[arrow] (qae) -- (qpefree);
  \node[detail] (qft) at (-9.2, -4.3)
    {QFT on ancilla\\register};
  \node[detail] (parallel) at (-7, -4.3)
    {Parallelization\\(qubits for depth)};
  \node[detail] (coherent) at (-4.8, -4.3)
    {Bit-by-bit via\\block-encodings};
  \draw[arrow] (qpe) -- (qft);
  \draw[arrow] (qpe) -- (parallel);
  \draw[arrow] (qpe) -- (coherent);
  \node[cite] at (-9.2, -5.7) {\cite{Brassard2002}};
  \node[cite] at (-7, -5.7) {\cite{Braun2022}};
  \node[cite] at (-4.8, -5.7) {\cite{Rall2021fastercoherent}};
  \node[detail] (qsp_poly) at (-1.4, -4.3)
    {Polynomial\\transformations};
  \node[detail] (gen_qubit) at (1.4, -4.3)
    {Generalized\\qubitization};
  \draw[arrow] (qsp) -- (qsp_poly);
  \draw[arrow] (qsp) -- (gen_qubit);
  \node[cite] at (-1.4, -6.2)
    {\cite{low2017optimal} \cite{gilyen2019quantum} \cite{martyn2021grand} \cite{Rall2023amplitudeestimation}};
  \node[cite] at (1.4, -5.7) {\cite{lu2023quantum}};
  \node[detail] (mle) at (4.8, -4.3)
    {MLE (various\\schedules)};
  \node[detail] (ci) at (7, -4.3)
    {Iterative\\confidence intervals};
  \node[detail] (elf) at (9.2, -4.3)
    {Engineered\\likelihood functions};
  \draw[arrow] (qpefree) -- (mle);
  \draw[arrow] (qpefree) -- (ci);
  \draw[arrow] (qpefree) -- (elf);
  \node[cite] at (4.8, -6.0)
    {\cite{suzuki2020amplitude} \cite{giurgica2022low}};
  \node[cite] at (7, -6.5)
    {\cite{grinko2021iterative} \cite{Zhao2022} \cite{callison2022improved} \cite{Fukuzawa2023}};
  \node[cite] at (9.2, -5.7) {\cite{PRXQuantum.2.010346}};
  \node[concept] (fourier) at (-4, -8.5)
    {Fourier decomposition\\of integrand};
  \node[detail] at (-4, -9.6)
    {uses QAE variant as subroutine};
  \node[cite] at (-4, -10.6) {\cite{herbert2021quantum}};
  \node[concept] (var) at (4, -8.5)
    {Variational approximation\\of Grover iterates};
  \node[detail] at (4, -9.6)
    {feeds MLE post-processing};
  \node[cite] at (4, -10.6) {\cite{plekhanov2022variational}};
  \draw[arrow] (qae.south west) .. controls (-2.9,-2) and (-3.1,-6.2) .. (fourier.north);
  \draw[arrow] (qae.south east) .. controls ( 2.9,-2) and ( 3.1,-6.2) .. (var.north);
\end{tikzpicture}%
    }

    \vspace{2em}

    \resizebox{\textwidth}{!}{%
    \begin{tikzpicture}[
    every node/.style={align=center, inner sep=3pt},
    arrow/.style={-{Stealth[length=4pt]}, semithick},
    barrow/.style={-{Stealth[length=4pt]}, semithick, blue},
    paper/.style={align=center, text width=2.7cm},
  ]
  \node[paper] (grover) at (0, 0)
    {{\bfseries\small Grover's search operator $G$}\\[1pt]{\tiny\ttfamily\color{gray}\cite{Grover1997}}};
  \node[paper] (qae) at (-1.5, -2.4)
    {{\bfseries\small Brassard \emph{et al.}}~{\tiny\ttfamily\color{gray}\cite{Brassard2002}}\\[1pt]{\footnotesize Quantum Amplitude Estimation (QAE)}};
  \node[paper] (count) at (6.5, -2.0)
    {{\bfseries\small Brassard \emph{et al.}}~{\tiny\ttfamily\color{gray}\cite{Brassard2002}}\\[1pt]{\footnotesize Quantum Approximate Counting}};
  \draw[arrow]  (grover) -- (qae);
  \draw[barrow] (grover) -- (count);
  \node[paper] (uno) at (2.3, -3.5)
    {{\bfseries\small Uno \emph{et al.}}~{\tiny\ttfamily\color{gray}\cite{uno2021modified}}\\[1pt]{\footnotesize Modified Grover operator (no QFT)}};
  \node[paper] (rall) at (-9, -3.7)
    {{\bfseries\small Rall}~{\tiny\ttfamily\color{gray}\cite{Rall2021fastercoherent}}\\[1pt]{\footnotesize Faster QAE}};
  \node[paper] (brau) at (-9.3, -5.0)
    {{\bfseries\small Braun \emph{et al.}}~{\tiny\ttfamily\color{gray}\cite{Braun2022}}\\[1pt]{\footnotesize Parallelized QAE}};
  \node[paper] (mont) at (-6.5, -5.4)
    {{\bfseries\small Montanaro}~{\tiny\ttfamily\color{gray}\cite{Montanaro_2015}}\\[1pt]{\footnotesize Quantum speedup of Monte Carlo methods}};
  \node[paper] (suz) at (-3.4, -5.4)
    {{\bfseries\small Suzuki \emph{et al.}}~{\tiny\ttfamily\color{gray}\cite{suzuki2020amplitude}}\\[1pt]{\footnotesize QAE without QPE (MLE heuristics)}};
  \node[paper] (herb) at (-0.3, -5.4)
    {{\bfseries\small Herbert}~{\tiny\ttfamily\color{gray}\cite{herbert2021quantum}}\\[1pt]{\footnotesize Quantum Monte Carlo integration (no QFT, uses Fourier decomposition)}};
  \node[paper] (ham) at (3.2, -5.4)
    {{\bfseries\small Hamoudi \& Magniez}~{\tiny\ttfamily\color{gray}\cite{Hamoudi2019}}\\[1pt]{\footnotesize Quantum Chebyshev's inequality and applications}};
  \draw[arrow] (qae) -- (rall);
  \draw[arrow] (qae) -- (brau);
  \draw[arrow] (qae) -- (mont);
  \draw[arrow] (qae) -- (suz);
  \draw[arrow] (qae) -- (herb);
  \draw[arrow] (qae) -- (ham);
  \draw[arrow] (qae) -- (uno);
  \node[paper] (wie) at (10, -3.7)
    {{\bfseries\small Wie}~{\tiny\ttfamily\color{gray}\cite{wie2019simpler}}\\[1pt]{\footnotesize Simpler quantum counting}};
  \node[paper] (aar) at (8, -5.4)
    {{\bfseries\small Aaronson \& Rall}~{\tiny\ttfamily\color{gray}\cite{aaronson2020quantum}}\\[1pt]{\footnotesize Quantum approximate counting revisited}};
  \draw[barrow] (count) -- (wie);
  \draw[barrow] (count) -- (aar);
  \node[paper] (gt) at (-6.5, -8.4)
    {{\bfseries\small Giurgica-Tiron, Kerenidis \emph{et al.}}~{\tiny\ttfamily\color{gray}\cite{giurgica2022low}}\\[1pt]{\footnotesize Low depth algorithms for QAE}};
  \node[paper] (cb1) at (-3.4, -8.4)
    {{\bfseries\small Callison \& Browne}~{\tiny\ttfamily\color{gray}\cite{callison2022improved}}\\[1pt]{\footnotesize Improved maximum likelihood QAE}};
  \node[paper] (zhao) at (-0.3, -8.4)
    {{\bfseries\small Zhao \emph{et al.}}~{\tiny\ttfamily\color{gray}\cite{Zhao2022}}\\[1pt]{\footnotesize Adaptive QAE}};
  \node[paper] (grinko) at (3.4, -8.9)
    {{\bfseries\small Grinko \emph{et al.}}~{\tiny\ttfamily\color{gray}\cite{grinko2021iterative}}\\[1pt]{\footnotesize Iterative QAE}};
  \draw[arrow] (suz) -- (gt);
  \draw[arrow] (suz) -- (cb1);
  \draw[arrow] (suz) -- (zhao);
  \draw[arrow] (suz) -- (grinko);
  \draw[barrow] (count.south) .. controls (6.3,-5) and (5.6,-7.6) .. (grinko.east);
\end{tikzpicture}%
    }

    \caption{An expanded view of the bottom half of the hourglass diagram in Figure~\ref{fig:hourglass}. The top diagram outlines the dependencies between the different conceptual approaches while the bottom diagram shows the relationship between the key papers.}
    \label{fig:QAEvars}
\end{figure}

Section \ref{sec:apps} detailed the top part of the hourglass (Figure \ref{fig:hourglass}) which covered the reduction of numerical problems to the triple $\left(\left(\pi_{i}\right)_{i\in I}, \left(U_{f_j}\right),\left(\mathcal{A}_{k}\right)_{k\in K}\right)$. In the present section, we address the bottom part of Figure \ref{fig:hourglass} and detail specific state-of-the-art quantum algorithms applied to the triple. We warn the reader that while many of the papers referenced may appear to trace a path from the top to the bottom of Figure \ref{fig:hourglass}, the main contribution in each case is situated at the bottom of the hourglass. In particular, while the content of the previous section was largely classical in nature, it is in this section, where we survey the quantum algorithms.

Indeed, we have already seen several such algorithms in \ref{sec:quantum_alternatives}, where we surveyed the state of the art around the year 2000. The years since have seen a variety of improvements. Some, such as Montanaro's work \cite{Montanaro_2015}, are refinements of the approaches introduced in the introduction (QAE, QAC, Grover-Rudolph inspired methods). Others, such as the proposal of Suzuki \emph{et al.}\ \cite{suzuki2020amplitude}, are modifications of the algorithms presented in the introduction so as to be better suited to NISQ-era hardware. Notably, several of the approaches we present are concerned with bypassing the need to implement the (notoriously unstable) quantum Fourier transform as part of the phase estimation procedure. Similarly, \cite{giurgica2022low} explore variants of QAE with lower circuit depth (at the cost of a higher qubit complexity), partially addressing the challenges posed by NISQ hardware, where deep circuits are particularly sensitive to noise. These variants can be roughly classified into three types, each of which reduce the problem to learning the angle $\theta$ of Eq.~\eqref{eq:goodbad}, with the quantity of interest recovered through $\theta=\arcsin\sqrt{a}$:

\begin{enumerate}
    \item \textbf{Phase estimation based}: these methods refine the original approach \cite{Brassard2002}, in which $\theta$ is obtained as the eigenphase of $G$ using quantum phase estimation.
    
    \item \textbf{Quantum signal processing / quantum singular value transformation}: $\theta$ is extracted by applying an engineered polynomial transformation to $G$ via block encodings, replacing phase estimation with a structured, robust subroutine.
    
    \item \textbf{Measurement and classical post-processing}: phase estimation is avoided entirely. The angle $\theta$ is infered from measurement and grover iterates $G$ are used to improve the accuracy at a fixed number of queries. This approach was first suggested by ~\cite{abrams1999fast}.
\end{enumerate}
The conceptual and bibliographic dependencies are presented in Figure \ref{fig:QAEvars}. In the following sections, we categorize several recent landmark contributions according to this scheme. We also discuss two recent papers that do not fit into the characterization above.

\subsection{Phase-estimation based approaches}

\subsubsection{Montanaro's algorithm}
 
Working in the same generalized setting of \cite{Brassard2002}, \cite{Montanaro_2015} provided a quantum algorithm that can accelerate Monte Carlo methods estimating the expected output value of an arbitrary randomized or quantum subroutine with bounded variance, achieving a near-quadratic speedup over the best possible classical algorithm. Furthermore, he showed that combining the algorithm with quantum walks gives a quantum speedup over the fastest known classical algorithms with rigorous performance bounds for computing partition functions using multiple-stage Markov chain Monte Carlo techniques. The same quantum algorithm can also be used to efficiently estimate the total variation distance between probability distributions. The key distinction is that Montanaro's algorithm no longer requires the random variable $\nu(\mathcal{A})$ to be bounded: it is sufficient for $\nu(\mathcal{A})$ to exhibit bounded variance.
 
Specifically, Montanaro considers a randomized quantum algorithm $\mathcal{A}$ whose output is $v(\mathcal{A})$, and the aim is to find a way to compute its expectation value, that is, $ \mathbb{E}[v(\mathcal{A})]\eqqcolon \mu$.
 
First, he considers the special case $0\leq v(\mathcal{A})\leq 1$, and repeats the analysis of Brassard \emph{et al.}\ \cite{Brassard2002} to present a procedure that can compute an estimate $\hat{\mu}$ such that $|\mu-\hat{\mu}| =\mathcal{O}(\sqrt{\mu}/t+1/t^2)$ (where $t$ is the number of iterations) with high probability. The procedure is based on familiar principles from above: apply an operator $W$ that takes the basis state $\ket{j}\ket{0}$ to $W\ket{j}\ket{0}=\ket{j}\left(\sqrt{1-f_j}\ket{0}+\sqrt{f_j}\ket{1}\right)$, where $f_j$ is the value of $v(\mathcal{A})$ for the measurement outcome $j$; then apply amplitude estimation with target state $\ket{\psi}=(I\otimes W)(\mathcal{A}\otimes I)\ket{0}^{n+1}$ for a number of repetitions and output the median value.
 
Subsequently, Montanaro generalizes to the following scenario: $0 \leq v(\mathcal{A}) \leq \gamma$, where $\gamma \in \mathbb{R}_{> 0}$ is finite but not necessarily one. Essentially, this case corresponds to algorithms where $\| v(\mathcal{A})\|_2$ is controlled and its computation amounts to a generalization of \cite{heinrich2002quantum}: by considering a truncated version $\mathcal{A}_{p,q}$ such that the output is 
\begin{itemize}
    \item $v(\mathcal{A})$ if  $p\leq v(\mathcal{A})< q$,
    \item $0$ otherwise.
\end{itemize}
Then $\mu\coloneqq\mathbb{E}[v(\mathcal{A}_{p,q})]$ can be estimated by sequentially applying and summing the ``standard'' Brassard \emph{et al.}\ algorithm over $\mathcal{O}(1/\varepsilon)$ increments. Montanaro further suggests that when $|\mu| = \mathcal{O}(1)$, with high probability it holds that $|\mu-\hat{\mu}| = \mathcal{O}(\varepsilon)$ for appropriate parameters.
 
Generalizing further, Montanaro considers probabilistic or quantum algorithms $\mathcal{A}$ with bounded variance $\sigma^2$, such that $v(\mathcal{A})$ need not be non-negative or $\ell_2$-bounded, and by finding an appropriate transformation of $\mathcal{A}$ to an algorithm $\mathcal{B}$, an approximation of $\mu$ is determined. This can be achieved as follows: assume that $v(\mathcal{A})$ has mean $\mu$ and that the variance is bounded above by $\sigma^2$. The transformation required can be given by considering $(v(\mathcal{A})-\mu)/\sigma$, which has $\ell_2$ norm bounded by one. In this way, estimating it to additive error $\varepsilon / \sigma$ yields an estimate of $\mu$ to additive error $\varepsilon$. The problem of not knowing the value of $\mathbb{E}[v(\mathcal{A})] = \mu$ can be overcome by first running $\mathcal{A}$ once to obtain a value $\widetilde{\mu}$, and setting this as an approximate proxy for $\mu$. As explained in \cite{Montanaro_2015}, $\widetilde{\mu}$ is expected to be $\mathcal{O}(\sigma)$ away from $\mu$, and as a result, there is a high probability that $\mathcal{B}$ is bounded by a constant. This makes it possible to estimate the positive and negative parts of $\mathbb{E}[v(\mathcal{B})]$ separately and then combine and rescale them with accuracy $\varepsilon $ in $\tilde{\mathcal{O}}(\sigma/\varepsilon)$ time.

\subsubsection{Parallelized QPE and QAE}
\cite{Braun2022} showed how QPE and QAE algorithms can be parallelized, reducing the gate depth of the quantum circuits to that of a single Grover operator, with small overhead. 
 
\begin{figure}[!tbh]
    \centering
    \includegraphics[scale=0.7]{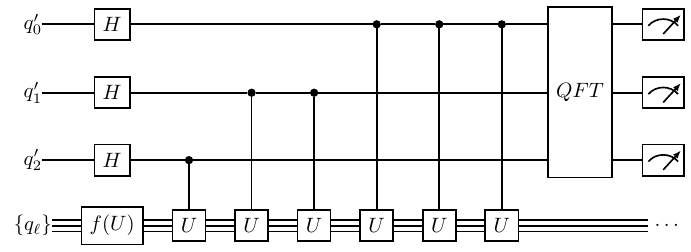}
    \caption{A unitary $f(U)$ prepares an eigenstate of $U$ on the $\ell$ registers whose eigenvalue is computed at 3-bit precision using QPE.}
    \label{fig:qpe}
\end{figure}

Recall that QPE computes the eigenvalues of unitaries $U \in U(n)$. For such a unitary $U$ on $\ell$ qubits, QPE requires $2^b-1$ executions of $U$ to estimate the eigenvalues with $b$-bit precision (see Fig.~\ref{fig:qpe}).  The proposed parallelization refers to including several copies of the registers $\{ 
q_\ell \}$ such that each $U$ gate required by QPE is controlled uniquely. 
 
For each of the new copies of the $\{q_\ell\}$ register, additional $f(U)$ gates need to be applied so as to generate the same eigenstate of $U$. As a result of these extra ancilla registers, the phase kickbacks are naturally parallelized with the $q_0$ qubit obtaining $2^{b-1}$ kickbacks.  We can also refer to this type of parallelization as ``vertical'' parallelization.
 
\begin{figure}[!htb]
    \centering
    \includegraphics[scale=0.8]{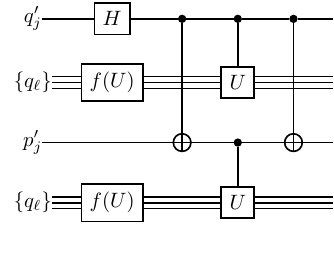}
    \caption{An extra entangling qubit $p_j$ achieves a horizontal compression in the gate execution, as compared to the standard and directly parallelized QPE.}
    \label{fig:parqpe}
\end{figure}

Indirect parallelization requires adding an extra qubit $p_j$ for each $U$, and entangling them such that sequential application is eliminated (see Fig.~\ref{fig:parqpe}). In this way, indirect parallelization amounts to vertical parallelization with horizontal compression at the expense of additional entangling gates.
 
The result of the parallelization described above is reduced gate execution time and, consequently, longer coherence times. Another benefit of this procedure is the reduction of the effect of unitary errors, as compared to the standard QPE algorithm, on the qubit registers where Grover operators are applied (see Fig.~\ref{fig:parqpe}). In general, the depth ratio of serial over parallelized QPE, after neglecting overhead relative to the controlled-unitary depth, is approximately $2^{1-b}$~\cite[Sec. 3.2]{Braun2022}. The effectiveness of this procedure can be realized even for low bit resolutions. For example, the circuit depth can be reduced by a factor of 128 for 8-bit resolution. This follows simply from parallel-to-serial arguments~\cite[Sec. 3.2]{Braun2022}. 
 
Despite the additional qubit cost of parallelization with approximately the same factor of depth improvement, there is tremendous potential in this algorithm for application in near-term computers where the qubit coherence times are relatively low. 
 
The authors of \cite{Braun2022} propose a method of parallelization by reinitialization. This method essentially resets the $\{ q_\ell\}$ qubit registers after each phase kickback. However, in this approach, the unitary $f(U)$ needs to be repeatedly applied in serial, which essentially eliminates the low-depth advantage of the previously mentioned parallelization methods. Nevertheless, the main advantage of this approach is that the reduced number of qubits (which, once again, favors current near-term devices) allows the execution of QPE and QAE with a higher level of precision. 
 
Note that by setting $U$ as the Grover operator, we obtain the corresponding parallelized QAE schemes. The proposed parallelized schemes for QAE are indeed better amplitude estimators
than the standard serial version. Benefits include, among others, the need for fewer circuit evaluations to achieve the same accuracy. 
 
To summarize this section, we highlight the three main advantages of the parallelization schemes for QPE and QAE: 
\begin{itemize}
\item[(1)] reduction of the computation time, which may be useful for NISQ devices that suffer from short
decoherence times. 
\item [(2)] full QAE estimation may be possible even in the presence of large amounts of decoherence noise; a feature ideal for NISQ devices, and 
\item[(3)] easy implementation of error-correcting techniques.
\end{itemize}
 
\begin{center}
    \begin{table}[htb!]
\begin{tabular}{cccc} \toprule[2pt]
Task & Description & Reference & Type\\ \hline
\multirow{2}{*}{\begin{tabular}[c]{@{}c@{}}Classical MC\\ algorithms\end{tabular}} & \multirow{2}{*}{\begin{tabular}[c]{@{}c@{}}Metropolis–Hastings, Gibbs,\\ Solovay–Strassen to name a few\end{tabular}} & \multirow{2}{*}{\cite{robert1999monte}} & \multirow{2}{*}{Classical} \\ &&&\\
\multirow{2}{*}{\begin{tabular}[c]{@{}c@{}}QME\end{tabular}} & \multirow{2}{*}{\begin{tabular}[c]{@{}c@{}}Solving the mean
estimation  \\ problem with relative error $\varepsilon$ \end{tabular}} & \multirow{2}{*}{\cite{Brassard2002,Wocjan2009}} & \multirow{2}{*}{Quantum} \\ &&&\\    
\multirow{2}{*}{\begin{tabular}[c]{@{}c@{}}QME\end{tabular}} & \multirow{2}{*}{\begin{tabular}[c]{@{}c@{}}Alternative approach to the
  \\ mean estimation problem \end{tabular}} & \multirow{2}{*}{\cite{Brassard1998, abrams1999fast}} & \multirow{2}{*}{Quantum} \\ &&&\\ 
\multirow{2}{*}{\begin{tabular}[c]{@{}c@{}}QME\end{tabular}} & \multirow{2}{*}{\begin{tabular}[c]{@{}c@{}}Solving the previous 
problem \\ for arbitrary distributions \end{tabular}} & \multirow{2}{*}{\cite{montanari2015computational}} & \multirow{2}{*}{Quantum} \\ &&&\\     
\multirow{2}{*}{\begin{tabular}[c]{@{}c@{}}QME\end{tabular}} & \multirow{2}{*}{\begin{tabular}[c]{@{}c@{}}Improvement over the
 \\ Montanari paper \cite{montanari2015computational} \end{tabular}} & \multirow{2}{*}{\cite{Hamoudi2019}} & \multirow{2}{*}{Quantum} \\ &&&\\   \hline  
\end{tabular}
\caption{Summary of various Quantum Mean Estimation (QME) algorithms with references.}
\label{tab:qme}
\end{table}
\end{center}

\subsection{Quantum signal processing / quantum singular value transformation}

\subsubsection{Faster coherent QPE and QAE}
\cite{Rall2021fastercoherent} presented an approach to QPE and QAE under the assumptions that (1) only one copy of the input state is given, (2) the input state is not an eigenstate of the unitary, and (3) the state need not be measured. 
This algorithm relies on the notion of \emph{block encoding} \cite{Low2017,gilyen2019quantum,martyn2021grand}. Block encoding tackles the problem of implementing non-unitary matrices with bounded spectral norms within quantum circuits. For such a matrix $A$, block encoding amounts to constructing a unitary matrix $U_A \in U(n)$, for some $n$, such that
\begin{align}
U_A = \begin{pmatrix}
A & * \\
* & *
\end{pmatrix}.
\end{align}
Another important tool of \cite{Rall2021fastercoherent} is the \emph{singular value transformation} (SVT): given a block encoding of
\begin{align*}
    A = \sum_{i}\lambda_i \ketbra{\psi_i},
\end{align*}
and a polynomial $p$ in one variable, we can construct the block encoding
\begin{align*}
    p(A) = \sum_{i}p(\lambda_i) \ketbra{\psi_i}.
\end{align*}
 
Using block encoding and SVT, as well as the ``block-measurement lemma'' \cite[see Eq. 193]{Rall2021fastercoherent}, Rall is able to compute an $n$-bit approximation of  $\lambda_i$ in a non-destructive manner (rounding promise), and to construct a non-destructive energy estimation algorithm. Eventually, the latter is used in a novel amplitude
estimation algorithm which has several performance advantages over the algorithm by Brassard \emph{et al.}\ \cite{Brassard2002}, and even the non-destructive variant of \cite{harrow2020adaptive}. For example, this algorithm requires far fewer ancilla qubits, and it runs in a fixed amount of time with just a single application of the previously mentioned energy
estimation with a constant-factor speedup.
 
\subsubsection{Quantum signal processing}\label{sec:qsp}
 
Low and Chuang \cite{low2017optimal} introduced the concept of \emph{quantum signal processing} (QSP). The authors argue that, ideally, we should apply the physical dynamics directly to a quantum computer without relying on the established robust time-optimal control methods implemented currently. The parametrized discrete single-qubit rotation gates, such as 
\begin{equation*}
    R_\phi(\theta)\coloneqq e^{-i\tfrac{\theta}{2} (\sigma_x \cos \phi + \sigma_y \sin \phi) },
\end{equation*}
should be understood as (quantum) programs that output the values of certain functions. In the case of the $R_\phi(\theta)$ gate, we can encode the $\sin$ function by noting that $\braket{1|R^N_{\pi/2}(\theta)|0}=\sin(N\theta/2)$. More generally, given that any single-qubit gate can be written as a superposition of the identity and the three Pauli matrices, for $\phi_k \in \mathbb{R}^N$, we can consider discrete gates of the form 
\begin{equation}\label{qsp}
\begin{aligned}
V(\theta) &= \prod_{k=0}^{N-1}R_{\phi_{N-k}}(\theta) \\
          &=  A_0(\theta)\pmb{1} + i \sum_{\ell\in \{x,y,z\}}A_\ell(\theta)\sigma_\ell,
\end{aligned}
\end{equation}
and compute more general functions of $\theta$ given by the real-valued functions $\{A_0,A_x,A_y,A_z \}$ for certain choices of $\phi$. Due to its inherent discreteness, controlling gate $V(\theta)$ or a similar function is quite suitable for the digital fault-tolerant model of quantum computation. 
 
With these considerations, it is proven that QSP can approximate $V_{\rm ideal}$ with very high success probability bounded by $1-16\varepsilon$ in the best-case scenario, and by $6\varepsilon$ when conditions (1) and (2) are not satisfied, thus allowing one to apply this method to find a Fourier approximation to unitaries that implement Hamiltonian evolution.
See also \cite{Rall2023amplitudeestimation, Mitarai2023perturbationtheory} and \cite{Stamatopoulos2020optionpricingusing} for applications in finance.

\subsubsection{Generalized qubitization}
In a spirit similar to quantum signal processing in Sec.~\ref{sec:qsp}, \cite{lu2023quantum} introduced the notion of ``generalized qubitization'' for QAE. Qubitization \cite{low2017hamiltonian,lu2023quantum,Gilyn2019} refers to a technique used to construct block encodings for functions of matrices. 
 
Essentially, generalized qubitization of \cite{lu2023quantum} refers to a subroutine that replaces QPE within QAE and, despite requiring no prior information on the quantum amplitude, the number of queries to the oracle is improved to $\pi/(\sqrt{6}\,\varepsilon)$, which is half of that required by QPE.

\subsection{Measurement and classical post-processing approaches}

\subsubsection{MLE-QAE}\label{first2} 
 
As briefly mentioned in a previous section, \cite{suzuki2020amplitude} proposed a QAE variant in which quantum phase estimation (QPE) is eliminated in favor of classical post-processing techniques, specifically, maximum likelihood estimation (MLE). The motivation behind this idea is related to the fact that controlled 2-qubit gates are quite difficult to implement in NISQ devices \cite{Leymann2020}, rendering amplitude estimation via phase estimation delicate. Avoiding phase estimation and inferring $\theta$ from measurement had already been proposed in \cite{abrams1999fast}, however, this approach requires the construction of a new oracle for the residual at each measurment step. MLE-QAE is a modification of this approach that uses only the original oracle. 

In \cite{suzuki2020amplitude}, a quantum speedup on NISQ devices has been demonstrated numerically, despite utilizing the classical MLE technique. The main idea is to estimate the eigenvalues required for QAE using a likelihood function that takes as input the ``good-bad basis'' states of repeated amplitude amplification processes. Specifically, the role of MLE is to provide a good estimate for the angle $\theta \in [0,2\pi]$ in the Grover operator,
\begin{align}
    G^{r}\ket{s}=\cos \left((2 r+1) \theta_{}\right)\ket{w^\perp} +\sin \left((2 r+1) \theta_{}\right)\ket{w}.
\end{align}
The likelihood function $L_s(s,\theta)$ is essentially the expression \eqref{eq:goodbad}. After preparing the states $G^{r}\ket{s}$, a discrimination procedure follows with the aim of constructing and then classically maximizing $L_s(s,\theta)$. Repetition of the experiment yields a vector of likelihood functions $L_{}(\boldsymbol s, \theta)$ parametrized by $\boldsymbol s = (s_1, \ldots , s_m)^\top$. Therefore, the objective to be maximized is
\begin{align}
    L_{}(\boldsymbol s, \theta) = \prod_{i=1}^m L_{s_i}(s_i,\theta),
\end{align}
and the problem of estimating the desired parameter $\theta$ reduces to 
\begin{align}
    \hat\theta \coloneqq \argmax_{\theta} L(\boldsymbol s, \theta). 
\end{align}
Generally, the resulting probabilities depend on the number $r$ of Grover operators applied. Moreover, the outcomes are not i.i.d. Specifically, while they are independent due to the fact that in each amplification process there is only one free parameter $r$, they are not identically distributed since the probability distribution for the measurement outcomes is different for each $r$. Finally, MLE-QAE may be biased, since it is a MLE. The issue of bias was very recently addressed in the context of Random-depth QAE \cite{lu2023adaptive}.
 
Despite all of the above, the proposed MLE-QAE algorithm was able to estimate the target values with lower query complexity than the traditional QAE. A lower bound on the estimation error was derived in terms of Fisher information, and it was observed that the accuracy of the results of the numerical experiment came sufficiently close to the theoretical Heisenberg limit. Furthermore, this semi-classical QAE algorithm required fewer CNOT gates and qubit registers than the standard QAE, which is a good indicator for this algorithm as an efficient NISQ-compatible quantum alternative to classical Monte Carlo simulations.
 
Several works followed \cite{suzuki2020amplitude}, where the efforts were concentrated on either reducing the number of quantum arithmetic operations or reducing the number of (CNOT) gates and qubits, while maintaining some form of quantum advantage, often interpolating between the classical regime and the quadratic speedup, similar to \cite{Woerner2019,Stamatopoulos2020optionpricingusing}. In a similar approach, \cite{giurgica2022low} proposed a low-depth quantum algorithm for QAE that sacrifices some of the quantum speedup in favor of a shallower circuit. This algorithm, along with a few others \cite{Gosset2026}, 
uses ancilla qubits. They are further discussed in Sec.~\ref{sec:challenges}.
Additionally, \cite{callison2022improved} provide subtle new insights and improvements that expand the work of Suzuki \emph{et al.}\ \cite{suzuki2020amplitude}.
It is worth mentioning \cite{nakaji2020faster} as a further QFT-free approach. Moreover, \cite{Ram_a_2025, li2025harnessingbayesianstatisticsaccelerate} tackle the problem of avoiding the QFT via a Bayesian approach.
 
\subsubsection{Iterative QAE}
 
\cite{grinko2021iterative} introduce Iterative QAE, which is a variant of QAE that does not rely on QPE but is based only on the amplitude amplification subroutine. Similarly to MLE-QAE, this algorithm reduces the depth of the quantum circuit, but it does not rely on heuristic classical post-processing techniques. Interestingly, IQAE maintains the (quantum) quadratic speedup over classical Monte Carlo integration (up to a double-logarithmic term $\log \left(2 / \alpha \log_2(\pi / 4 \varepsilon)\right)$ where $\varepsilon$ corresponds to the accuracy) while requiring a small constant overhead. Similarly to all QAE algorithms, IQAE approximates the probability of measuring the good state, that is, the sine term of expression \eqref{eq:goodbad} for different $r$, or alternatively, for different numbers of iterations of the Grover amplification operator. Doing so, the authors obtain an approximation of the angle $\theta$ that is directly related to the integral we are interested in computing via $\theta = \arcsin{\sqrt{a}}$. IQAE is made up of two nested loops, the inner and outer loops, which narrow down the confidence interval $[\theta_{\rm low}, \theta_{\rm up}]$ in which $\theta$ lies. The role of the outer loop is to increase the value of $r$ at each iteration so that the computation is performed for $r_1,\ldots, r_m$, i.e.\ $m$ times. Let $R_i = 2r_i + 1$, $i=1,\ldots, m$. Then it is shown that when the amplification circuit is applied to the power $r_i$, it returns the good state with probability $\sin^2(R_i \theta)$ and, by repeated measurements, for fixed $r_i$, the authors obtain a confidence interval $[\theta_{\rm low}, \theta_{\rm up}]$ where both bounds are functions of $R_i$. Due to a certain invariance of IQAE, the outer loop is allowed to increase and provide another estimate for $[\theta_{\rm low}, \theta_{\rm up}]$, eventually converging to a \emph{sufficiently accurate} estimate.
 
\cite{Fukuzawa2023} proposed a variation of IQAE that achieves a better query complexity of $O\left(\frac{1}{\varepsilon} \log \frac{1}{\alpha}\right)$ while managing to retain only small constant factors.

\subsubsection{Robust amplitude estimation}
 
\cite{PRXQuantum.2.010346} suggested a quantum-enhanced algorithm
to estimate expectation values, called the \emph{robust amplitude estimation} (RAE) algorithm, suitable for NISQ and ``early fault-tolerant'' (EFT) devices. Consider an $n$-level system: the idea is that for the expectation value of an observable $\braket{O} = \braket{\psi | O | \psi}$, with $\ket{\psi} = A \ket{0}^{\otimes n}$ and $A \in U(n)$ (the group of unitary $n\times n$ matrices), RAE uses the measurements extracted from enhanced sampling circuits \cite{PRXQuantum.2.010346}, and performs classical post-processing using heuristics, specifically, maximum likelihood estimation (MLE) \cite{katabarwa2021reducing}. The connection to amplitude estimation comes from the fact that each of the $L$ layers of the noisy enhanced sampling circuits essentially corresponds to Grover routines for QAE. Considering enhanced sampling circuits with different likelihood functions for post-processing, the authors obtain an empirical estimate of $\braket{O}$. 
 
The main advantage of using RAE, as opposed to the usual sampling techniques, is that it can provide higher accuracy and precision \cite{PRXQuantum.2.010346,katabarwa2021reducing}. Furthermore, \cite{dalal2022noise} showed that randomized compiling (RC) \cite{PhysRevA.94.052325} of RAE (a set of methods that efficiently introduces random single-qubit gates into a logical quantum circuit without altering the logical circuit) converts coherent errors into stochastic errors. This makes RAE a candidate amplitude-estimation algorithm for NISQ devices. 
 
While RAE is not expected to substitute fault-tolerant algorithms as replacements for MC, we can implement this algorithm in conjunction with algorithms mentioned previously, to reduce the high cost of estimating expectation values of observables, with desired precision and accuracy.

\subsubsection{Adaptive QAE}
 
\cite{Zhao2022} proposed a Grover-only-based QAE. Specifically, the authors designed an \emph{adaptive QAE} (AQAE) algorithm that uses only Grover-like subroutines to estimate the intervals of probability amplitudes. The adaptive nature arises from an adaptive factor that adjusts the amplitude of good states so
that the amplitude after the adjustment and the original amplitude can be estimated without
ambiguity in the subsequent step. 
 
One problem with QAE and its Grover-operator dependence is that it is impossible to uniquely estimate the desired amplitude based on the measurements of Grover operators applied to a single circuit. This problem is often termed ``period ambiguity'' in the literature. The architecture of AQAE begins with Eq.~\eqref{eq:goodbad}. As discussed in detail previously, usually in QAE one applies the Grover operator $G$, $r$ times, to increase the amplitude $\sqrt{p}$ ($\sqrt{a}$ in Eq.~\eqref{eq:sprime}) linearly for (not necessarily small) $p$, while a classical brute-force search algorithm would increase the probability $p$ instead. Due to the periodicity of the sine term in Eq.~\eqref{eq:goodbad}, the confidence interval $[L,U]$ is equivalent to the union of $2r+1$ intervals for $\theta$:
\begin{equation}
    \begin{aligned}
&I^{+,(j)}=\left[\frac{\arcsin \sqrt{L}+j \pi}{2 r+1}, \frac{\arcsin \sqrt{U}+j \pi}{2 r+1}\right], \\
&I^{-,(l)}=\left[\frac{-\arcsin \sqrt{U}+l \pi}{2 r+1}, \frac{-\arcsin \sqrt{L}+l \pi}{2 r+1}\right],
\end{aligned}
\end{equation}
where $ j=0,1, \ldots, r$ and $l=1, \ldots,r$. Each of those intervals, in turn, is contained in one of the ``periods''
\begin{align}
    [0,\tfrac{1}{2r+1}\tfrac{\pi}{2}], \, [\tfrac{1}{2r+1}\tfrac{\pi}{2},\tfrac{2}{2r+1}\tfrac{\pi}{2}],\ldots, [\tfrac{2r}{2r+1}\tfrac{\pi}{2},\tfrac{\pi}{2}].
\end{align}
For the correct period, the estimation error for both $\theta$ and $p$ is given as $\mathcal{O}(1/r)$. 
 
The algorithm of \cite{Zhao2022} aims to determine this period as follows. First, for $\ket{\psi}$ and $r=0$, the measurements are used to construct an initial confidence interval $\mathcal{I}$ for $\theta$, which does not (yet) have period ambiguity. Then, $\mathcal{I}$ is used to determine the period of $\theta$. Assuming that $r$ grows at a geometric rate $\sim K^r$, for $K\in 2\mathbb{Z}+1$, the number of oracle queries is $\mathcal{O}(K^{r_{max}})$, where $r_{max}$ is the maximum number of Grover iterations. As a result, the estimation error is reduced to $\mathcal{O}(1/N_{\rm oracle})$.
 
Let $p\in [0,\tfrac{1}{2}]$ and let $\{r_0,r_1,\ldots, r_T\}$ be the schedule of Grover-iteration counts, where $r_T$ is the largest allowed value and $T$ denotes the largest index, such that the confidence interval for $\theta$ is less than $\tfrac{2}{2r+1}\tfrac{\pi}{2}$. Furthermore, for each different $r_t$, the algorithm increases the sample size $N_t$ in step $t$. The authors then show that their algorithm satisfies $\mathbb{P}\Big[p \in [p^L,p^U]\Big] \geq 1-\alpha$, $|p^L-p^U|\leq \varepsilon$, where $\varepsilon$ is a small positive number and 
\begin{equation}
    N_{\rm oracle} = \mathcal{O}\left( \log \Big( \frac{\pi^2(T+1)}{3\alpha} \Big)\frac{1}{\varepsilon} \right),
\end{equation}
where $T = \log \tfrac{\pi}{K\varepsilon}/\log K$. The authors rigorously prove these statements, which translate into a bound $\mathcal{O}(1/\varepsilon)$ on the number of oracle queries and a classical complexity of $\mathcal{O}(\log(1/\varepsilon))$. This is a significant improvement over the MLE approach of Suzuki \emph{et al.}\ and the iterative QAE of Grinko \emph{et al.}\ \cite{grinko2021iterative}.

\subsubsection{Power-law and QoPrime QAE}
 Power-law QAE and QoPrime QAE are two variations of QAE introduced in \cite{giurgica2022low}. Their convergence rate can be viewed as interpolating between the classical Monte Carlo one and the full QAE one, making them very interesting for practical implementation in NISQ devices. 
  The first one, Power-Law QAE (also known as the ``Kerenidis-Prakash'' approach to QAE), is a generalization of the MLE-QAE \cite{suzuki2020amplitude} algorithm, where power-law schedules are utilized. For the Power-Law QAE we follow the exposition presented in \cite{bouland2020prospects}. This variant of QAE is parametrized by a number $\beta \in [0,1]$ that controls the interpolation between the classical algorithm (for $\beta=1$, with $\mathcal{O}\left(1 / \varepsilon^2\right)$ repetitions) and the vanilla QAE \cite{Brassard1998} (for $\beta=0$, with $\mathcal{O}(1 / \varepsilon)$ repetitions). The ability to choose values in the range $[0,1]$ makes it possible to construct QAE circuits of lower depth at the expense of a larger number of oracle calls to the quantum circuit. Although this increases the convergence speed, the asymptotic advantage over classical Monte Carlo remains.
 
The QoPrime QAE is a number-theoretic variant of QAE. Let $U$ be the unitary whose amplitude, $\braket{U} \coloneqq \braket{0|U|0}$, we wish to estimate
with $n$ qubits, using the QoPrime QAE of \cite{giurgica2022low}. This algorithm is a number-theoretic variant of the standard QAE algorithm in which the authors choose $k$ distinct co-prime moduli $\{n_i\}_{i=1}^k$, each satisfying $n_i \sim \mathcal{O}(1 / \varepsilon^{1/k})$ such that $\prod_{i=1}^k n_i \sim \mathcal{O}(1 / \varepsilon)$. The true value of the amplitude is assumed to be $\braket{U} = \tfrac{\pi M}{2N}$,  $M \in[0, N]$. Then, QoPrime QAE estimates $M \pmod{N_{i}}$, where $N_i$ is the product of $l < k$ moduli, using $N/N_i$ sequential calls to the oracle $O_f$, which are then followed by measurements in the measurement and Hadamard bases. Then, using the Chinese remainder theorem, we can obtain the congruence $M \pmod {N}$ (recall, this amounts to summing the products $b_ix_iN_i$ for all $i\in [k]$), while the fractional part of $M$ is computed separately. This algorithm requires a circuit depth of
$d \cdot \left(\frac{1}{\varepsilon}\right)^{1-l/k}$, where
$d$ is the circuit depth for a single application of $U$, $\varepsilon$ is the additive error, $\beta \in (0,1]$, $k\geq 2$, $l \in [k-1]$.

\begin{table}[!htb]
    \centering
    \begin{tabular}{lcccc}\toprule[2pt]
   Algorithm & Ref.  & Qubits & Depth & Query complexity  \\
    \hline
    Original QAE & \cite{Brassard2002} & $n + \log(1/\varepsilon)$ & $ \frac{d}{\varepsilon}+\log\log(1/\varepsilon)$ & $\frac{1}{\varepsilon}$ \\
    QFT-free QAE &\cite{uno2021modified,suzuki2020amplitude,aaronson2020quantum}  & $n$ & $ \frac{d}{\varepsilon}$ & $\frac{1}{\varepsilon}$ \\
     Burchard's &\cite{burchard2019lower}   & $n$ &  $ \left(\frac{d}{\varepsilon}\right)^{1-\beta} \log(1/\varepsilon)$ & $ \left( \frac{1}{\varepsilon} \right)^{1+\beta}\log(1/\varepsilon)$ \\
     Iterative QAE &\cite{grinko2021iterative} & $n$ & $\frac{d}{\varepsilon}$ & ${1}/{\varepsilon}$  \\
    Power-law QAE &\cite{giurgica2022low}  & $n$ &  $\left(\frac{d}{\varepsilon}\right)^{1-\beta}  $ & $ \left( \frac{1}{\varepsilon} \right)^{1+\beta}$  \\
      QoPrime QAE &\cite{giurgica2022low}   & $n$ &  $ \left(\frac{d}{\varepsilon}\right)^{1-q/k}$ & $ \left( \frac{1}{\varepsilon} \right)^{1+q/k}$  \\
    \hline
    \end{tabular}
    \caption{Asymptotic trade-offs of amplitude estimation algorithms from \cite{giurgica2022low}. Parameters:  $n$ is the number of qubits and $d$ is the circuit depth for a single application of $U$, $\varepsilon$ is the additive error, $\beta \in (0,1]$, $k\geq 2$, $q \in [k-1]$.  More QAE-like algorithms may be available, for example, \cite{Nagaj2009} considers a QAE-like algorithm by utilizing the connection between a product of two reflections and a generalized quantum walk.}
    \label{QAEvariants}
\end{table}

\subsubsection{Optimal mean estimation}
 
Recently,~\cite{kothari2023mean} introduced a new algorithm, based on Grover's algorithm, but with complex phases. This algorithm is suitable for computing $\mathbb{E}[v(\mathcal{A})]=\mu$, the ``mean'' of interest, without any assumptions on the random variable $v$, by accessing a randomized circuit $C$. Specifically, under a certain assumption -- knowledge of the ``code'' for a random variable $v$ -- their algorithm requires $\mathcal{O}(n)$ samples and outputs an estimate $\hat \mu$ with 
\begin{align*}
    \mathbb{P}\left( |\mu-\hat \mu|>\sigma/n\right)\leq 1/3.
\end{align*}
It should be noted that this algorithm requires the use of a classical randomized circuit (without input) that generates a sample from $v$.
 
In a broader sense, it is possible to achieve the same outcome of obtaining a sample from $v$ by using a unitary quantum circuit with a predetermined input $|\psi \rangle$. After measuring the output and discarding some bits, the circuit provides a sample from $v$. It is important to note that a quantum circuit with intermediate measurements can be modified to meet this requirement. Finally, and less broadly, we can use a unitary quantum circuit that produces $\frac{1}{\sqrt{N}} \sum_{j=1}^N|j\rangle\left|\psi_j\right\rangle$, where $v$ is the uniform distribution on a multiset of real numbers $\left\{\psi_1, \psi_2, \ldots, \psi_N\right\}$. This model enables Grover's algorithm to function effectively.
 
This work provides near-optimal extensions of a variety of algorithms discussed earlier, notably \cite{hamoudi2021quantum} and \cite{cornelissen2022near}.

\subsection{Other Approaches} 
 
\subsubsection{Full advantage with minimal depth}
 
\cite{herbert2021quantum} proposes a quantum algorithm for MC integration that avoids the quantum Fourier transform (QFT), as well as most quantum arithmetic. Specifically, this approach considers the decomposition of the sum $\mathcal{S}$ (which approximates the integral $\mathcal{I}$) in the Fourier basis where each component is estimated using an instance of QAE. The reasoning behind this goes as follows: vanilla QAE as in Sec.~\ref{sec:brassard-ae} is known to have ``simple enough'' circuits for easy distributions and for trigonometric functions $f$. Essentially, Herbert proposes the Fourier decomposition of an arbitrary function $f$,
\begin{align}
    f \sim \sum_{n=1}^\infty \Big(\tilde a_n \cos(n\omega x) + \tilde b_n \sin(n\omega x) \Big),
\end{align}
(see \cite{herbert2021quantum} for details of the notation) and shows that $\mathcal{I} = \mathbb{E}[\mathcal{S}]$ can be estimated with a mean square error that scales as $\Theta(q^{-\lambda})$, where $q$ is the number of runs of the circuit that prepares the distribution $p(\boldsymbol x)$ (circuit $\mathcal{P}$ in Eq.~\eqref{circ:p}) and $\lambda$ is the convergence rate of the QAE subroutine utilized. Crucially, this method retains the full quadratic speedup while it is claimed that the relevance of this algorithm is not to be reserved for the fault-tolerant, error-corrected era only (see the interesting discussion in the last section of \cite{herbert2021quantum}).
 
\subsubsection{Variational quantum amplitude estimation}
 
\cite{plekhanov2022variational} introduced the notion of \emph{variational quantum amplitude estimation} (VQAE). This approach is an amalgam of standard variational quantum algorithm (VQA) techniques \cite{cerezo2021variational} and the Suzuki \emph{et al.}\ approach \cite{suzuki2020amplitude}, where QFT is eliminated. VQAE is an interesting approach to perform Monte Carlo integration using NISQ devices \cite{preskill2018quantum} with complexity $\mathcal{O}(1/\varepsilon^{1+\beta})$, $\beta \in [0,1]$. However, its double reliance on heuristics may obscure its potential advantages, as the authors observe that these heuristics are not only not guaranteed to work but also yield a
larger approximation error \cite{guo2026settinganglesquantumapproximate}.

\cite{plekhanov2022variational} also propose an adaptive-VQAE. VQAE reduces the time-consuming Grover iterations by implementing variational steps. It prepares the state $\ket{s'}$ as in Eq.~\eqref{eq:sprime}. While in the standard QAE we arrive at the desired state by applying the Grover operator $r$ times as $G^r\ket{s}$ (see Eq.~\eqref{eq:sprime}), in VQAE we perform the following:
\begin{enumerate}
    \item First, we find a parametrized state $\ket{\phi_{i=0}} = \ket{s}$ that encodes $f(x)$ and $p(x)$, and we pick a $k\in\mathbb{Z}$.
    \item While $0\leq m \leq r$ we set $i = \lfloor m/k \rfloor$ and $j = m \% k$ (modulo operation) and:
    \begin{itemize}
        \item sample the circuit $G^j\ket{\phi_i}$ 
        \item record the frequency $h_w^{(m)}$ with which its ancilla qubit is measured in the $\ket{1}$ state, where $h^{(m)}$ is the total number of measurements.
    \end{itemize}
    \item If $j=k-1$ then variationally approximate $\ket{\phi_{i+1}} \approx G^k \ket{\phi_i}$.
    \item Collect the set $h_w^{(m)}$ for all $m$ iterations of the algorithm and perform the maximum likelihood estimation (MLE), that is, maximize $\prod_m L_m(h_w^{(m)},x)$ where $L_m(h_w^{(m)},x) \coloneqq \sin^2((2m+1)x)^{h_w^{(m)}} \cos^2((2m+1)x)^{h^{(m)}-h_w^{(m)}}$.
\end{enumerate}
In the end, the authors obtain an approximation of $G^m\ket{s}$ by applying the Grover operator to the parametrized state only $j$ times, $G^j\ket{\phi_i}$ with $i,j$ defined above. The details of the MLE procedure are given in \cite{plekhanov2022variational}.
 
The authors of \cite{plekhanov2022variational} claim that VQAE can be more efficient than classical MC sampling. Achieving this means finding a variational circuit such that $\beta <1$. On a negative note, VQAE suffers from all the usual problems of VQA, such as the barren plateau problem, the fact they are \textbf{NP}-hard to train \cite{bittel2021training}. Later, it was shown that to reach a prescribed target expectation value is \textbf{QCMA}-hard \cite{bittel2022optimizing}, and later it has been conjectured that VQAs are undecidable. Furthermore, in the presence of biased noise \cite{kungurtsev2024iteration}, a characteristic evident in near-term quantum devices, the asymptotic rate of convergence of VQAs is not affected, but unfortunately, the level of bias negatively affects the constant term therein as well as the asymptotic distance to stationarity. This raises concerns about the applicability of VQAE for obtaining a practical (quantum) advantage in Monte Carlo simulations.

\section{Applications in Finance}\label{sec:finance}

It is well known that finance provides a particularly suitable application domain of the general quantum Monte Carlo
framework described in the previous sections.  The central theme that underlies this connection is a family of problems and questions whose answers rely, for the most part, in resource intensive Monte Carlo simulations. Such problems include the determination of the future price of a derivative, the tail risk of a portfolio of assets under management, the credit
exposure of a bank to a counterparty, or the sensitivity of such quantities to market
inputs, to name a few.  Typically, after the classical modeling step with some underlying probability distribution $\pi$, each of these problems is reduced to the
estimation of one or more expectations of the form (cf. Alg. \ref{alg:estamp})
\begin{align}
    \mathbb{E}_{\pi}[f(X)] ,
\end{align}
where $X \sim \pi$ is a random vector of discretized risk factors and $f$ is a reward, a
loss, an exposure or some other generic sensitivity functional of interest.  Finally, quantumly, an
algorithm $\mathcal{A}$ must prepare amplitudes proportional to $\sqrt{\pi}$, implement a reversible
version of $f$, and estimate the resulting marked amplitude.   

For all matters and purposes, therefore, quantum Monte Carlo applications in finance is a rich, often
high-dimensional, source of problems materialized exactly as the expectation-estimation problems for which
quantum amplitude estimation can be useful.

\subsection{Derivative pricing}

The basic object in modern derivative pricing is a discounted expectation. A derivative is a contract whose payoff depends on the future behavior of some underlying quantity, the risk factors (stock prices, interest rates, foreign exchange rates, valitilities, credit intensities and such).  Let
$S_t\in\mathbb{R}^d$ denote a vector of risk factors. The contract's payoff is usually denoted by $\Phi$.  Under the usual
no-arbitrage formulation\footnote{ Essentially, \emph{no-arbitrage} means it is not possible to make a guaranteed profit from a zero investment. In turn, this implies the existence of a probability measure $\mathbb{Q}$, under which today's price for any asset equals its expected discounted future value~\cite{harrison1979martingales}. This is why prices are modeled expectations, as in Eq.~\eqref{eq:V0}. Note that $\mathbb{Q}$ should not be confused with the real-world probability of market outcomes. Instead it is really a reweighting of paths chosen to make pricing consistent and arbitrage-free. \cite{harrison1979martingales}.}, the value at time zero of a contract paying
$\Phi((S_t)_{0\leq t\leq T})$ at maturity $T$ is
\begin{align}\label{eq:V0}
            V_0 =
        \mathbb{E}_{\mathbb{Q}}\!\left[
        D(0,T)\,
        \Phi\!\left((S_t)_{0\leq t\leq T}\right)
        \right],
\end{align}
\begin{align}\label{eq:discount}
    D(0,T)=\exp\!\left(-\int_0^T r_s\,ds\right),
\end{align}

where $\pi = \mathbb{Q}$ is the risk-neutral pricing measure and $D(0,T)$ is the discount
factor.  Here, $\mathbb{Q}$ should simply be
viewed as the probability measure under which the discounted traded assets have the
martingale property required by the absence of arbitrage.  

Here, $\mathbb{Q}$ is the probability measure used to price the contract, and is 
fixed by a single requirement. Generally, along with considering risky assets, there is also an
idealized reference asset, the \emph{risk-free bank account}. This is a hypothetical
investment that grows \emph{deterministically} at the prevailing risk-free interest
rate $r_s$, with no randomness whatsoever. Its value is
\begin{align}
    B_t = \exp\!\left(\int_0^t r_s\,ds\right),
        \qquad B_0 = 1 .
\end{align}
This may be understood simply as the continuously
compounded value of one unit of cash invested at the risk-free rate, and 
serves as the benchmark against which every risky return is compared.

Typically, instead of tracking an asset by its raw price $S_t$, we track the ratio
$S_t / B_t$, i.e., a normalized version. Dividing by $B_t$ removes the guaranteed deterministic growth of cash and expresses the price essentially in present-value terms, and this is what in finance is referred to as
\emph{discounting}.

Researchers in quantitative finance model the risk factors obey a stochastic process, the Itô diffusion process (essentially a Langevin process{\footnote{%
The term \emph{Langevin equation} here is the same as the one used in statistical physics where the drift $\mu(S_t,t)$ denotes a
deterministic force and $\sigma(S_t,t)\,dW_t$ is the stochastic noise term. The
density of $S_t$ then evolves according to the associated Fokker--Planck (forward
Kolmogorov) equation, while the backward version is the Feynman--Kac equation used
for pricing.}}) given by the following a stochastic differential equation (SDE):
\begin{align}\label{eq:Ito}
    dS_t = \mu(S_t,t)\,dt + \sigma(S_t,t)\,dW_t.
\end{align}

We see therefore, that the dynamics of the any such price or ratio split into a sum two terms: a
predictable trend (the drift, the $dt$ term) plus a random Brownian fluctuation (the
$dW_t$ term). Under the measure $\mathbb{Q}$, the discounted price $S_t/B_t$ of every traded asset has vanishing drift. That means that while the discounted price fluctuates, it
has no predictable direction of motion and as a result its expected future value under
$\mathbb{Q}$ equals its present value. A process with this property is a
\emph{martingale} \cite{Musiela2005}. This property is foundamental since if a discounted price had a predictable trend (amounting to a nonzero drift relative to the risk-free account) that trend could be harvested by a self-financing trading strategy
for riskless profit which is not rational. Therefore, the existence of such probability measure $\mathbb{Q}$, also called the
\emph{risk-neutral measure}, is required by the martingale property.

Returning to the pricing problem, the expectation in Eq.~\eqref{eq:V0} is rarely
available in a closed form. Therefore, the best we can typically do is to evaluate it numerically. As expected, first we may
discretize time into $m$ steps, $0 = t_0 < t_1 < \cdots < t_m = T$, and replace
the continuous trajectory by the simulated path
\begin{align}
    x = (S_{t_1}, \ldots, S_{t_m}) \in \mathbb{R}^{D},
        \qquad D = d\,m,
\end{align}
a single point in a space whose dimension $D$ is the number of risk factors $d$
times the number of time steps $m$. Each such path carries a probability density
$p(x)$ induced by the dynamics under $\mathbb{Q}$, and the discounted payoff
becomes a function $\varphi(x)$ of the path. The price in Eq.~\eqref{eq:V0} is then
the high-dimensional integral
\begin{align}
    V_0 \simeq
    \int_{\mathbb{R}^{D}} \varphi(x)\, p(x)\, dx
    = \mathbb{E}_{\mathbb{Q}}\!\left[\varphi(X)\right],
    \label{eq:V0discrete}
\end{align}
where the expectation is taken over the paths $X$ drawn from $p$. The right hand side looks oddly familiar now and this is where Monte Carlo methods are employed: the price is an average of a payoff over a large number of
simulated paths.

There is an equivalent partial-differential-equation view.  For Markovian diffusion
models, the Feynman--Kac theorem \cite{Kac1949} identifies the same price with the solution of a
backward parabolic equation,
\begin{align}\label{eq:feynman-kac}
    \partial_t u
        + \frac12 \mathrm{Tr}\!\left(\sigma\sigma^{\top}\nabla^2 u\right)
        + \mu^{\top}\nabla u
        - r u =0,
\end{align}
where $u(s,T)=\Phi(s)$. The dimension of this PDE equals the number of factors $d$, that is, the number
of independent stochastic risk variables used in the model. Now, when $d$ is small as, for example, for a single asset, or an asset together with a
stochastic volatility or a stochastic interest rate, giving $d=1$ or $d=2$, then the
equation can be solved directly on a grid using finite-difference or
finite-element methods. However, scaling this is quite hard since, as is well-known, grid-based solvers scale exponentially in $d$ with $n^{d}$ nodes required, that is exponential in the problem dimension. Already at $d \sim 5-10$
this is at the edge of tractability, and many problems of interest such as derivative
written on a basket of many underlying assets, have $d$ in the tens or hundreds. There, the grid is hopelessly large due to the 
\emph{curse of dimensionality} and actually it is the central
obstacle that Monte Carlo is designed to avoid.  If $X^{(1)},\ldots,X^{(N)}$ are independent
samples of $p$, the density associated with the risk measure $\mathbb{Q}$, the estimator
\begin{align}
    \widehat V_N = \frac1N\sum_{j=1}^N \varphi(X^{(j)})
\end{align}
has root-mean-square error
\begin{align}\label{eq:error}
    \left(\mathbb{E}_Q[\widehat V_N-V_0]^2\right)^{1/2}
        =
        \frac{\sigma_\varphi}{\sqrt{N}},
\end{align}
where $\sigma_\varphi^2=\mathrm{Var}_p[\varphi(X)]$. In Eq.~\eqref{eq:error} the exponent $1/2$ is independent of the dimension $D$ and, as a matter of fact, only the variance constant depends on the underlying model and the payoff function.  For this reason, Monte Carlo methods are quite an indispensable tool in quantitative finance. And while they indeed do avoid the curse of dimensionality, their drawback is also clear from the practitioner point of view: having one more
decimal digit of statistical accuracy costs roughly $100$ times more samples. 
Variance-reduction methods, quasi-Monte Carlo, and multilevel Monte Carlo can greatly
improve constants and discretization overheads, but for generic nonsmooth payoffs the leading statistical scaling remains governed by Eq.~\eqref{eq:error}. It is this factor, the sample size that practitioners, are interested in optimizing with the quantum Monte Carlo methods.

\subsection{Risk estimations}
\label{sec:risk_estimations}

Derivative pricing is a very good example where we choose a stochastic model, generates
sample paths, evaluates a payoff, discounts it, and averages.  Standard examples
include European options\footnote{Strictly speaking, most trading desks will not utilize Monte Carlo methods for European options, but use the Heston model of stochastic volatility instead.}, whose payoff depends only on $S_T$, basket options, whose
payoff depends on several assets, and path-dependent contracts, so-called Asian options, whose payoff may
depend on maxima, minima, barrier crossings, or accumulated quantities along the path. The case is that the quantity of interest is always the
expectation of some observable over a probability law on paths.

Risk measurement is the second major application in finance, where quantum Monte Carlo methods can massively enhance the state of the art. Under the post-2008 crisis regulatory frameworks implemented worldwide, a bank, holding a portfolio of assets must be in the position to regularly estimate rare but severe losses.  If $L$ is
the loss over a fixed horizon, the value-at-risk at the confidence level $\alpha$ is the
$\alpha$-quantile
\begin{align}
    \mathrm{VaR}_\alpha(L)
        =
        \inf\{\ell:\mathbb{P}(L\leq \ell)\geq \alpha\},
\end{align}
while the expected shortfall (ES), also called conditional value-at-risk (CVaR), is
\begin{align}
     \mathrm{ES}_{\alpha}(L)
        =
        \frac{1}{1-\alpha}\int_\alpha^1
        \mathrm{VaR}_u(L)\,du .
\end{align}
For continuous loss distributions, this is equivalently
\begin{align}
    \mathrm{ES}_{\alpha}(L)
        =
        \mathbb{E}\!\left[L\,\middle|\,L\geq
        \mathrm{VaR}_{\alpha}(L)\right].
\end{align}
To compute these, one needs to be able to perform arithmetics  \cite{wang2025comprehensive} on the samples obtained from Monte Carlo.  

We realize therefore that the expected shortfall is an observable that admits information from the tail of the tail of a distribution.  Estimating it is  much more challenging than estimating
a mean because the relevant samples are rare.  In regulatory market-risk calculations, such tail estimators must be repeated across risk classes, liquidity horizons, and
regulatory stress scenarios taking place continuously.  Therefore, we realize that the computational burden goes from a single but potentially large Monte Carlo calculation towards a large family of related tail Monte Carlo calculations.

\subsection{Nested simulations}

Generally, the most expensive workload within a financial institution amounts to nested simulations. CVA analysis tries to understand the counterparty risk's probability of default before a given contract's maturity date and computes the value loss, in expectation, because of that risk, as well as what needs change to compensate at any given time.

Such counterparty-risk
and valuation-adjustment computations actually require quantities such as credit valuation
adjustment,
\begin{align}
    \mathrm{CVA}
        =
        (1-R)\int_0^T
        \mathbb{E}_{\mathbb{Q}}\!\left[
        D(0,t)\,\mathrm{EE}(t)
        \right]\,d\mathrm{PD}(t),
\end{align}
where $R$ is a recovery rate, $\mathrm{PD}(t)$ is a default-probability curve, and
\begin{align}
    \mathrm{EE}(t)
        =
        \mathbb{E}_{\mathbb{Q}}\!\left[
        (V_t)^+\,\middle|\,\mathcal{F}_t
        \right]
\end{align}
is the expected positive exposure at time $t$ where $V_t$ is the contract value and $V_t^+ \coloneqq \max \left(V_t, 0\right)$. The inner expectation prices the institution's portfolio conditional on an outer market scenario.  A direct nested Monte Carlo therefore has an outer loop over scenarios and an inner loop over conditional portfolio valuations.  Bias-variance analyses of such nested estimators lead to
costs that scale worse than ordinary Monte Carlo, often as $\varepsilon^{-3}$ or
$\varepsilon^{-4}$ for root-mean-square error $\varepsilon$, depending on the
allocation of inner and outer samples. As such, quantum Monte Carlo methods can provide singificant accuracy improvements in this family of computations as well. For example, \cite{YuHan04032026} study exactly how one can approach multi-option portfolio pricing and valuation adjustments

\subsection{Greeks}

Financial institutions require knowledge of sensitivities, the so-called \emph{Greeks}, namely
derivatives of prices and risks with respect to model parameters and market input \cite[Sec. 2.1]{stamatopoulos2022towards}. For example, if a portfolio depends on $k$ risk factors, finite-difference bump-and-revalue
estimation requires $\mathcal{O}(k)$ additional pricing calculations for first derivatives and $\mathcal{O}(k^2)$ calculations for second-order cross sensitivities.  Adjoint algorithmic differentiation reduces the cost of many first-order sensitivities, but the large second-order sensitivity sets remain computationally quite demanding.

As an example, let us consider the first-order sensitivity of a price
$V_0=\mathbb{E}_{\mathbb{Q}}\!\left[D(0,T)\,\Phi\!\left((S_t)_{0\le t\le T}\right)\right]$ (c.f. Eq. \eqref{eq:V0})
to the initial level $S_0^{j}$ of the $j$-th risk factor, the so-called \emph{Delta}:
\begin{align}\label{eq:Delta}
    \Delta^{j}
        =
        \frac{\partial V_0}{\partial S_0^{j}}
        =
        \mathbb{E}_{\mathbb{Q}}\!\left[
        D(0,T)\,
        \frac{\partial \Phi}{\partial S_0^{j}}
        \right],
        \qquad j=1,\dots,k,
\end{align}
where, for sufficiently smooth payoffs, the derivative is moved inside the expectation and applied to the payoff path by path. Here, $D(0,T)$ is the discount factor as in Eq.~\eqref{eq:discount}, and $\partial\Phi/\partial S_0^{j}$ is the pathwise sensitivity of the
payoff to the $j$-th factor. The full first-order set $\{\Delta^{j}\}_{j=1}^{k}$ thus
comprises $k$ such expectations, and the second-order set
$\Gamma^{jl}=\partial^2 V_0/\partial S_0^{j}\partial S_0^{l}$ comprises $\mathcal{O}(k^2)$,
which is the source of the computational burden noted previously.

\vspace{1em}


Crucially, the structure of the greeks problem exposes two
distinct aspects along which a quantum advantage would be highly desired. The first is the statistical error
$\varepsilon$ on each individual sensitivity. Since every $\Delta^{j}$ in Eq.~\eqref{eq:Delta} is itself an expectation of the form~\eqref{eq:V0}, it inherits directly the quadratic amplitude-estimation speedup already discussed for pricing. The cost of estimating a single greek to error $\varepsilon$ scales as $\mathcal{O}(\varepsilon^{-1})$ quantumly versus $\mathcal{O}(\varepsilon^{-2})$ classically. The second, and probably more
interesting, aspect is the number of sensitivities $k$.


Classically, the first-order set $\{\Delta^{j}\}_{j=1}^{k}$ is usually
assembled component by component. A finite-difference \emph{bump \& revalue scheme} reprices the portfolio at $\mathcal{O}(k)$ shifted parameter configurations, and therefore obtaining all $k$ Greeks to error $\varepsilon$ incurs an $\mathcal{O}(k)$ multiplicative overhead relative to the cost of one
price estimate~\cite{giles2007mc,cathcart2011calculating}. The second-order set $\{\Gamma^{jl}\}$ is larger still, with $\mathcal{O}(k^2)$ distinct entries. Adjoint algorithmic differentiation can remove the linear-in-$k$ overhead for first-order
sensitivities in classical Monte Carlo pricing,~\cite{giles2006smoking,capriotti2010fast} but second-order Greeks remain substantially more challenging in practice.

The quantum improvement here is of a different nature. Instead of
estimating each component separately, one may encode the price as a
function of all $k$ inputs and try to apply a quantum gradient estimation
algorithm in the style \cite{jordan2005fast}, with the high-accuracy formulation of \cite{gilyen2019optimizing}. The gradient
$\nabla V_0 = (\Delta^{1},\dots,\Delta^{k})$ is then recovered with a
number of queries to the (reversible) pricing oracle that scales as
$\widetilde{\mathcal{O}}(\sqrt{k}/\varepsilon)$, a quadratic improvement
in the dimension of the sensitivity vector on top of the quadratic
improvement in $\varepsilon$. \cite{stamatopoulos2022towards} followed this approach for
pricing derivatives of practical interest and managed to demonstrate numerically
that the empirical resource requirements can sit well below the
worst-case bounds. As such, and that the additional advantage in computing market risk can materially lower the logical clock rate required to reach a
quantum advantage for such a financial exercise relative to the estimate of \cite{Chakrabarti2021}. We note, however, that realizing the $\sqrt{k}$ scaling presupposes a coherent pricing oracle that is
differentiable to the required precision and is queried in superposition
across the parameter space and, unfortunatelly, constructing such an oracle, together with
the associated state-preparation cost, remains for now the dominant practical
obstruction, mirroring the oracle-construction caveat that pervades the
entire amplitude-estimation programme. Worth noting that a more modern approach for computing gradients, still requiring a heavy Tofolli count in the digital realm, which have not yet been tested in this domain is that of the Quantum Hamiltonian Descent \cite{leng2023quantum}.

\subsection{Going beyond Black-Scholes}
\label{sec:beyond-bs}
The end-to-end speedups surveyed so far are, almost without exception,
established for the geometric Brownian motion (GBM) dynamics of the
Black-Scholes model, where the terminal law of the risk factors is an
explicitly known log-normal distribution that can be loaded
directly~\cite{stamatopoulos2022towards,chakrabarti2020quantum}. However, real
pricing pipelines rely on quite much richer underlying dynamics such as mean-reverting short-rate models, stochastic-volatility models, and their
multi-dimensional correlated extensions. In general, for all these model no closed-form terminal density is available and the path itself must be simulated.
Whether the usual quadratic Monte Carlo advantage survives in this context is a
genuinely open question, since naively simulating a discretized stochastic differential equation (SDE) inside the amplitude-estimation oracle may inflate the circuit depth enough to erase any asymptotic gain.

Recent work of Herman et al.~\cite{herman2026beyond} aims to address this
gap exactly. There they identify a structural property of the underlying
SDE, which they term \emph{fast-forwardability}, which allows the law of
$S_T$ to be prepared at a cost independent of the number of time steps. Subsequently they show that two of the models of actual practical relevance, the
Cox-Ingersoll-Ross (CIR) short-rate model \cite{cox1985cir} and a variant of the Heston
stochastic-volatility model \cite{heston1993}, possess it. For these two models they establish novel end-to-end quadratic speedups of the same
character as in the GBM case, thereby making some progress beyond the Black-Scholes setting. 

For general models lacking the fast-forwardable structure, they instead
attack the path-simulation layer itself, introducing a \emph{quantum
Milstein sampler} built on a new quantum subroutine for sampling
L\'evy areas. Embedding this sampler within a quantum multi-level Monte
Carlo (qMLMC) framework allows for maintaining the quadratic speedups for multi-dimensional processes exhibiting certain correlation structures, extending the
quantum-accelerated MLMC programme of An et
al.~\cite{an2021quantum} to the higher-order strong-convergence regime
required when the diffusion coefficients do not commute. Furthermore,
the authors give an improved analysis of the numerical-integration step
underlying derivative pricing, and manage to offer constant-factor
reductions in the resource estimates for both the GBM and CIR models.

In direct relevance to the present survey's framing \cite{herman2026beyond} offers a critique of the alternative route in which one
prices derivatives by solving the associated Fokker-Planck or
Feynman-Kac PDE~\eqref{eq:feynman-kac} with a quantumly, e.g., with a quantum PDE solver, and then reading off the answer by QAE. They identify
several theoretical barriers to obtaining a genuine speedup along this
path. For example such issues arise in the form of the insufficiency of history states for path-dependent payoffs, the integration overhead associated with the
curse of dimensionality, and also any runtime obstructions intrinsic to quantum
PDE solvers. This critique is a useful counterpoint to the popular PDE-based pricing analyses that have appeared in the literature~\cite{chen2026quantum}.

\subsection{Exotic and path-dependent payoffs}
\label{sec:exotics}
Orthogonal to the choice of model dynamics is the structure of the
payoff functional $\Phi$ itself. For European contracts, often used as running
examples in the literature, it is well-known that they depend on the terminal state $S_T$ only. However, when considering a large fraction
of traded volume, the problem sits in \emph{path-dependent} and \emph{multi-asset}
contracts, that is Asian options, briefly mentioned in Sec. \ref{sec:risk_estimations}, whose payoff depends on a time-average of the
underlying. Such options are the barrier and lookback options, which depend on the running maximum or minimum. Also the so-called basket, best-of, and call-on-min options, whose payoff couples several assets at maturity fall in the same family. Each of these requires the reversible payoff circuit $f$ of Algorithm~\ref{alg:estamp} to compute not a simple function of a single register but an accumulated or comparative quantity over the entire simulated path $x=(S_{t_1},\dots,S_{t_m})$.

Quantumly the additional cost of these payoffs is constrained to the
construction of the marking oracle, while the amplitude-estimation outer
loop is unchanged. The relevant arithmetic (running sums for
averages, comparison chains for maxima and minima, piecewise-affine
evaluation for the resulting kinks) may be assembled from standard
reversible building blocks used by practitioners, and gate-level
constructions for the continuous piecewise-affine payoff class (basket, spread, call-on-min, and best-of-call options) have been proposed with full error and complexity analysis~\cite{chen2026quantum}. However, a recurring subtlety in this topic is that such payoffs are typically unbounded and only linearly growing, which forces the rotation that encodes the payoff into an amplitude to
use a vanishing scaling parameter. The resulting loss in the
inverse-precision exponent, that is an $\mathcal{O}(\varepsilon^{-3})$ rather
than $\mathcal{O}(\varepsilon^{-1})$ query complexity, and this is really a
feature of every end-to-end construction in which the oracle is built
and not assumed. As such, the full quadratic advantage is recovered only
for bounded payoffs~\cite{chen2026quantum,herbert2022quantum},
unfortunately. Genuinely path-dependent contracts, such as Asian,
barrier, and lookback options, raise the further difficulty that the
marking oracle must act on the entire simulated path rather than its
terminal state alone. And American-style contracts, whose holder may
exercise early, fall outside the plain expectation-estimation template
altogether, requiring instead a quantum least-squares Monte Carlo
treatment of the associated optimal-stopping
problem~\cite{doriguello2022optimalstopping}.

\subsection{Quantum Monte Carlo for simulation-based optimisation}
\label{sec:sbo}
A theme cutting across the applications above is that the estimated
expectation is rarely the final answer since in practice it is the inner
objective of an outer optimisation. Portfolio construction under a tail
constraint (Mean-CVaR), risky Mean-Variance allocation, and
distributionally robust pricing all share the structure
\begin{align}
    \min_{\theta\in\Theta}\;
        \mathcal{R}\!\left(
        \mathbb{E}_{\pi_\theta}[f_\theta(X)]
        \right),
\end{align}
in which amplitude estimation supplies the inner expectation
$\mathbb{E}_{\pi_\theta}[f_\theta(X)]$ for each candidate $\theta$ and an
outer routine searches over $\theta$. Because QAE returns the inner
quantity only up to a confidence interval, the interaction between
estimation error and optimisation accuracy must be controlled jointly.
Cui et al.~\cite{cui2024quantum} carry out a systematic study of this
regime where they account for all possible systematic errors in the QMC integration of risk functionals such as VaR and CVaR, analyse the resources needed to encode the relevant distributions, and then apply the machinery to
Mean-CVaR and Mean-Variance optimisation problems, including a
Mean-Variance experiment executed under hardware noise with a dedicated
error-mitigation scheme for QAE. Their analysis serves as a 
useful template for the resource accounting that any QAE-in-the-loop
optimisation will require, and it exposes the central tension that the
outer loop multiplies the already substantial per-query cost of the
inner amplitude-estimation routine.

\section{Discussion: Current challenges to quantum Monte Carlo}
\label{sec:challenges}

In this section, we list some of the most significant sticking points in applying quantum alternatives to classical Monte Carlo.
In general, there are three major considerations in the context of QAE that are important to address in any reasonable implementation:
\begin{enumerate}
    \item[(A)] The difficulty of constructing an oracle (of either the phase-flip, rotation, or quantum query variety) without evaluating $f$ at each point in the domain,
    \item[(B)] the depth of the QAE (and variants) circuit,
    \item[(C)] the complexity of state preparation.
\end{enumerate}
These challenges can be addressed by the careful use of ancilla qubits, albeit at a cost of (some of) the quantum speedup, which depends on the number of ancilla qubits used.
For example, the state preparation can be performed in a linear-depth circuit \cite{rosenthal2021query,zhang2022quantum}
when there are $\mathcal{O}(2^n)$ ancilla qubits available.
Similarly, in many algorithms, the error also depends
on the number of qubits used in the output register, cf.~\cite{cleve1998quantum}.
Table~\ref{QAEvariants} illustrates the resulting trade-off between circuit depth and query complexity across several QAE variants.

We expand on each of these points in this section. In addition, noting that all of these are ultimately relevant to the practical, overriding consideration of total wall-clock time for implementation, we finish with some additional details on the wall-clock runtime in Section~\ref{sec:wallclock}.

\subsection{Constructing oracles}\label{subseq:oraclecontstruction}

The query complexity bounds of this review count applications of an oracle $U_f$ (or equivalently, of a rotation oracle, quantum query $Q_f$ or arbitrary quantum subroutine $\mathcal{A}$). An obvious objection arises: if constructing $U_f$ required evaluating $f$ at each point of its domain, then the quadratic quantum query complexity advantage would be negated by the cost of constructing $U_f$. The construction of oracles is one of the principal obstacles to the implementation of quantum Monte Carlo methods. Indeed, it is in some sense more fundamental than the challenges posed by circuit depth or the difficulty of robustly implementing phase estimation, which pose a challenge only due to the limitations of current hardware. However, in pratice, the function $f$ is never supplied as an arbitrary table of values, but through some compact description from which $U_f$ (or $\mathcal{R}_f$ or $Q_f$) can be constructed at a reasonable computational cost. We remark that quantum query complexity is conceptually the same as classical query complexity. The quantum circuit for the oracle can be simple or complicated, just as the classical circuit for evaluating $f$ at a point can be simple or complicated. Studying the classical or quantum query complexity of an algorithm taking $f$ as an input, (as opposed to the classical or quantum depth of the compiled circuit) allows us to disentangle the complexity of the algorithm from the complexity of evaluating the input $f$. 

We have already seen one instance of this in Section \ref{subseq:partitionfunctions}, where we have presented a quantum algorithm for estimating partition functions \cite{Montanaro_2015}. Here, the configuration space $\Omega$ is exponentially large, yet it is never enumerated: each Gibbs distribution $\pi_i$ is prepared as the stationary state of a rapidly mixing Markov chain, by a quantum walk~\cite{Temme_2011,Rall_2023,Montanaro_2015}, and the accompanying oracle $A_i$ applies a value $e^{-(\beta_{i+1}-\beta_i)H(x)}$. Both the state and the oracle are thus built from a succinct description of the problem. 

The simplest case is where $f$ is given by a closed-form expression or a small classical circuit. Frequently, such a description can be compiled into a quantum circuit. An example of this occurs in financial applications of \cref{sec:finance}, where the payoff of a derivative is an explicit, typically simple, function of the underlying and is implemented directly as an circuit~\cite{Woerner2019,Stamatopoulos2020optionpricingusing}.

Even when no explicit circuit for $f$ is available, structural information about $f$ may be leveraged. If $f$ is known to admit a sparse or rapidly convergent representation, such as a low-degree polynomial or a truncated Fourier series, the oracle can be built cheaply. This is precisely the mechanism of Herbert's Fourier-series decomposition of the integrand~\cite{Herbert2021}, which retains the full quadratic advantage at minimal depth by expanding the function whose mean is sought into a sum of elementary terms, replacing one intractable oracle by many simple ones.

Another way of constructing an oracle is by using the methods of Hamiltonian simulation. In particular, qubitization and quantum signal processing, realize a propagator $e^{-iHt}$ as a block encoding~\cite{low2017hamiltonian,gilyen2019quantum}. The recent work of Bravyi et al.~\cite{bravyi2026quantum} extends the methods of Hamiltonian simulation to classical noisy non-linear dynamics and is another potential method for constructing oracles.

\subsection{Circuit depth}

For a moment, let us assume that there is no complexity associated with state preparation (e.g., thanks to some oracles utilizing quantum random access memory),
and that there are no ancilla qubits available.
We remove these assumptions in Section~\ref{challenge:stateprep}.

The key challenge to realizing a QAE algorithm is the limited depth of circuits
that can be implemented on current NISQ devices.
While the details vary with the qubit technology, we can currently only execute circuits that are at best hundreds of gates deep~\cite{giurgica2022low2} (post state preparation, oracle and not considering QPE)
which severely limits either the quantum speedup or the error bound that can be achieved.
Moreover, there is a trade-off between the number of qubits available and the depth of the circuits that
can be reliably executed.

To capture this trade-off, IBM \cite{Moll_2018} suggested a randomized benchmarking procedure known as the ``quantum volume''.
A quantum volume of $v$ suggests that across an ensemble of random, dense quantum circuits of depth $\log_2 v$ on $\log_2 v$ qubits,
the device provides quantum states within a negligible margin from the target in terms of fidelity, with a probability close to 1.
For example, a quantum volume of $256$ suggests that a circuit of depth $8$ can be executed on an $8$-qubit register very reliably.
There are a number of important choices to make in setting up the benchmark. For instance, the quantum volume measured by the device manufacturer (e.g., 2048 for Honeywell's HQS-LTS2), with their specific choices,
need not match the quantum volume measured independently by third parties using their own settings (e.g., 256 for the same HQS-LTS2,
as measured by a team at the Los Alamos National Lab).

The recent work of \cite{giurgica2022low} has shown that there is a trade-off between the quantum speedup and the allowable depth.
Two recent variants of the QAE algorithm,
power-law AE and QoPrime AE \cite{giurgica2022low}, 
make it possible to bound the depth of the circuit from above,
at the price of lowering the quantum speedup. \cite{Herbert2021} shows how to obtain the full advantage at minimal depth using a Fourier series decomposition of the sum that approximates the integral of interest. In practice, of course, there is also a trade-off in accuracy and quantum volume, i.e., deeper circuits with more qubits exhibit greater noise. This can be partially mitigated through multiple runs of the circuit with the same input to obtain a sample average.

\subsection{The complexity of state preparation}
\label{challenge:stateprep}

Obtaining the quadratic speedup in quantum-assisted Monte Carlo requires the ability to construct $\mathcal{P}$, which encodes the probability density $p(x)$ onto the state $\ket{s_0}$, such that the resulting state $\mathcal{P}\ket{s_0}= \ket{s}$ is amplified, see Eq.~\eqref{circ:p}. Recent progress in the context of loading interesting probability distributions, especially for the purpose of Monte Carlo integration for derivative pricing, includes the re-parameterization technique of \cite{Chakrabarti2021}.
For distributions over (possibly high-dimensional) convex support, one may consider geometric random walks \cite{marecek2021cutting}.
We can also consider incorporating a circuit that itself generates the desired distribution, as in quantum generative adversarial networks (qGANs) \cite{zoufal2019quantum}. The latter, however, suffer from the usual problems of variational circuits (see, e.g., \cite{bittel2021training}), in addition to adding complexity to the circuit pipeline. Another approach to the problem is to first approximate the discretized function values by a
matrix product state (MPS), or quantized tensor train, and subsequently compile this
low-bond-dimension MPS into a shallow state-preparation circuit. Bohun
\emph{et al.}~\cite{bohun2026entanglement} show that, at least for smooth functions, the entanglement across MPS bonds associated with progressively finer binary scales decays exponentially and the subleading Schmidt coefficient satisfies $\Lambda_{k,1}\sim 2^{-k}\sqrt{g_1(f)/12}$, while the entanglement entropy obeys $S_k=\mathcal{O}(k/4^k)$. As such, this structure enables function-loading circuits
with $\mathcal{O}(N)$ gates for $N$ qubits. Using tensor-cross interpolation,
the MPS can be constructed without storing all $2^N$ function values, with $\mathcal{O}(N\chi^2)$ function evaluations for maximal bond dimension $\chi$.

Ultimately, the state preparation is a time-optimal control problem that is as hard as realizing any other quantum circuit in the same system, i.e., of any depth whatsoever.
In the time-optimal control, we seek a particular solution to the initial value problem for the Schr\"odinger equation
\begin{align}
\label{schroedinger}
\frac{\partial}{\partial t} \U (t) = \hat{A}(t) \U (t)
\end{align}
where $\hat{A}(t) = \hat{H}(t)/i \hbar$ can explicitly be written in terms of controls $u_j(t): [0, T] \to \mathbb{R}$ as
\begin{align}
\label{eq:hata}
    \hat{A}(t) = \sum_j u_j(t) ~\hat{H}_j/i\hbar.
\end{align}
In particular, we seek a solution that is optimal with respect to time horizon $T$, while using controls $\{u_j(t)\}$ constrained to some polynomially-representable set $\Upsilon$.
Formally, the quantum optimal control problem is expressed as follows:
\begin{equation}\label{eq:QOC}
    \begin{aligned}
    \min_{\U(t), \{u_j(t)\} \in \Upsilon} & \quad J(\hat{U}(t), \{ u_j(t)\}) \\
    \text{s.t.}\quad  & \frac{\partial}{\partial t} \U (t) =  \left[\sum_j u_j(t) ~\hat{H}_j/i\hbar\right] \U (t) \\
                & \U(0) = \hat{\pmb{1}} \\
                &  | \U (T) - \U^* | \le \varepsilon
    \end{aligned}
\end{equation}

\cite{marecek2020quantum} have shown that we can employ the Magnus expansion to obtain an asymptotically convergent procedure for the
time-optimal control problem for any unitary, but without bounds on the rate of convergence.
It is a textbook exercise \cite{kitaev2002classical} to show that the corresponding quantum circuit will require a number of gates that grows exponentially in the number of levels.
This applies to any algorithm \cite{rosenthal2021query,zhu2021generative,zhang2022quantum} including the work of \cite{grover2002creating}. 
Specifically, they ask if it is possible for a given probability distribution $\mathcal{D} \sim p(\vb*{x})$ to prepare a state of the form
\begin{align}\label{eq:distribution_state}
    \ket{\psi_{\mathcal{D}}} = \sum_{i=0}^{2^n-1} \sqrt{p_i}\ket{i},
\end{align}
where the index $i$ takes values in a discrete set $I \subset \mathbb{Z}_{\geq 0}$ of potentially very large cardinality,
and $p_i$ denotes the probability that $\mathcal{D}$ assigns to the $i$-th point.
However, note that Grover and Rudolph did not actually solve the problem of ``loading the distribution'' onto the quantum circuit. Rather, they showed that if $\mathcal{D}$ can be loaded efficiently, then we can produce the state of Eq.~\eqref{eq:distribution_state}. The idea of Grover and Rudolph is to start with a coarse discretization using $m < n$ qubits, where $n$ is the resolution of the discretization of $p(\vb*{x})$ we wish to achieve. That is, after initially loading the distribution onto the quantum circuit, we begin with the state
\begin{align}
    \ket{\psi_m} = \sum_{i=0}^{2^m-1} \sqrt{p_i^{(m)}}\ket{i},
\end{align}
where $p_{i}^{(m)}$ is the probability of drawing the sample $x_i$. Subsequently, in order to achieve the desired resolution, we need to add an ancilla qubit into the system such that the following evolution takes place
\begin{align}\label{eq:evolution}
    \sqrt{p_{i}^{(m)}}|i\rangle \xrightarrow[]{\mathcal{U}} \sqrt{\alpha_{i}}|i\rangle|0\rangle+\sqrt{\beta_{i}}|i\rangle|1\rangle,
\end{align}
for some operator $\mathcal{U}$, where $\alpha_i, \beta_i$ denote the probabilities for a new sample $x_j$ to be drawn from the sets defined by $L_i \coloneqq [x_{i-1}, x_i]$ or $R_i \coloneqq [x_i,x_{i+1}]$, respectively. The superscript $m$ labels the current resolution level: we start at level $m$, and each application of Eq.~\eqref{eq:evolution} introduces a new boundary, advancing to level $m+1$. We can repeat this process until $m=n$ and thus arrive at Eq.~\eqref{eq:distribution_state}. The question of the existence of such a $\mathcal{U}$ is discussed in \cite{grover2002creating}.
Nevertheless, Herbert showed in \cite{Herbert2021} that the Grover-Rudolph procedure can eliminate the potential algorithmic speedup even for log-concave distributions. In particular,
to achieve a root mean squared
error of $\hat{\varepsilon}$ using an unbiased
quantum Monte Carlo estimation method requires $\tilde{\Omega}(1/\hat \varepsilon^2)$
operations when the Grover-Rudolph method is used to prepare some log-concave distribution as a quantum state.
In the above, $\hat{\varepsilon} = \sqrt{\mathbb{E}[(\hat \mu - \mu)^2]}$, where $\mu$ is the mean of $p(x)$, $\hat\mu$ the estimate, and $\hat{\varepsilon}$ is the root mean square error.

Fortunately, Vazquez and Woerner \cite{vazquez2021efficient} have shown that this can be avoided by treating the state preparation problem and QAE in a unified way. To explain, consider the case of loading the uniform distribution
\begin{align}
    p_i = \frac{1}{2^n}, \quad \forall i.
\end{align}
Loading this probability distribution is quite easy using Hadamard gates that take an all-zero quantum state to the equal-superposition state. Then, the application of the Grover operator $G$, as described in Sec.~\ref{sec:quantum_alternatives}, to the composite system that implements QAE yields
\begin{align}
    \frac{1}{\sqrt{2^{n}}} \sum_{i=0}^{2^{n}-1}|i\rangle_{n}\left[\sqrt{1-p\left(x_{i}\right)}|0\rangle+\sqrt{p\left(x_{i}\right)}|1\rangle\right].
\end{align}
This results in measuring $\ket{1}$ on the ancilla qubit with probability $\tfrac{1}{2^n}\sum_{i=0}^{2^n-1}p(x_i)$ \cite{Montanaro_2015}. The efficient state preparation, as proposed in \cite{vazquez2021efficient}, considers an arbitrary function $f: \mathbb{R}^\ell \to [0,1]$ (with the distribution not necessarily log-concave, as opposed to \cite{grover2002creating}; we assume $\ell = 1$ for simplicity), for a system of $n$ qubits and two ancilla qubits (as opposed to the usual QAE). Furthermore, they consider the analogue of the controlled operator $\mathcal{R}$ of Eq.~\eqref{eq:R} acting on the last ancilla qubit:
\begin{align}
    \mathcal{R}_p\ket{i}^n\ket{j}\ket{0} = \ket{i}^n\ket{j}(\sqrt{1-p(x_i)}\ket{0}+ \sqrt{p(x_i)}\ket{1}),
\end{align}
as well as a rotation operator acting on the first ancilla qubit, whose task is to encode $f(x_i)$ into the amplitude of that ancilla:
\begin{align}
    \mathcal{R}_f \ket{i}^n\ket{0}\ket{k}=\ket{i}^n(\sqrt{1-f(x_i)}\ket{0}+ \sqrt{f(x_i)}\ket{1})\ket{k}.
\end{align}
Then, starting from $\ket{0}^n\ket{0}\ket{0}$ and bringing the first $n$ qubits to the equal superposition state, followed by $\mathcal{R}_f$ and $\mathcal{R}_p$, results in the state
\begin{align}
    \begin{aligned}
\ket{s}&=\frac{1}{\sqrt{2^{n}}} \sum_{i=0}^{2^{n}-1}|i\rangle_{n} \sqrt{1-f\left(x_{i}\right)} \sqrt{1-p\left(x_{i}\right)}|00\rangle \\
&+\frac{1}{\sqrt{2^{n}}} \sum_{i=0}^{2^{n}-1}|i\rangle_{n} \sqrt{1-f\left(x_{i}\right)} \sqrt{p\left(x_{i}\right)}|01\rangle \\
&+\frac{1}{\sqrt{2^{n}}} \sum_{i=0}^{2^{n}-1}|i\rangle_{n} \sqrt{f\left(x_{i}\right)} \sqrt{1-p\left(x_{i}\right)}|10\rangle \\
&+\frac{1}{\sqrt{2^{n}}} \sum_{i=0}^{2^{n}-1}|i\rangle_{n} \sqrt{f\left(x_{i}\right)} \sqrt{p\left(x_{i}\right)}|11\rangle,
\end{aligned}
\end{align}
where the probability of measuring the ancilla product state $\ket{11}$, that is
$$\tfrac{1}{2^n}\sum_{i=0}^{2^n-1}p(x_i)f(x_i),$$
approximates $\mathbb{E}_p[f(X)]$ for the random variable $X$. The analogue of the oracle shifting operator from Sec.~\ref{sec:quantum_alternatives} is
\begin{align}
    O^{(2)}_{f}=\left(\pmb{1}_{n+2}-2\left|w\right\rangle|11\rangle\left\langle w\right|\langle 11|\right),
\end{align}
where $\ket{11} \coloneqq \ket{w^\perp}$ in this case. Similarly to the previous approach, we can define the addition between functions \cite{vazquez2021efficient} by introducing another operator $\mathcal{R}_h$. In general, the three oracles $\mathcal{R}_p, \mathcal{R}_f$ and $\mathcal{R}_h$ allow one to consider arbitrary combinations of additions and multiplications of functions.

In the above, we focused on the $\ell=1$ case. However, it is not difficult to extend this to arbitrary $\ell$, i.e., to consider multivariate problems. This can be achieved by representing the dimensions with additional circuit registers (possibly with a different number
of qubits on each of them), and this can be implemented efficiently since the number of qubits required is $\mathcal{O}(\ell)$. This contrasts with the classical numerical integration schemes based on quadrature, where the dependence is usually exponential, except for Monte Carlo simulation (including quasi-Monte Carlo).


\subsection{The wall-clock time}
\label{sec:wallclock}

So far, we have considered the asymptotics of the query complexity. In practice, however,
it is important to consider the ``constant'' involved as well, or at least its order of magnitude.

In classical computers, the cycle time is determined by the clocking frequency, which is often
around 5 GHz. Therefore, the cycle time is on the order of nanoseconds (e.g., 0.2 ns = 2 $\times 10^{-10}$ s for 5 GHz).
Within a single cycle time, we can perform binary operations such as the addition of two numbers,
and set registers to a constant value such as zero.

Quantum hardware spans a range of qubit modalities whose gate, readout, and initialization times differ by orders of magnitude, so abstract gate-count complexities can translate into very different wall-clock costs. At the fast extreme, 2-qubit operations on charge-based semiconductor qubits can be as short as a few hundred picoseconds (e.g., 200 ps = 2 $\times 10^{-10}$ s) \cite{petta2005coherent}, albeit historically at low fidelity; modern high-fidelity silicon spin-qubit gates are slower, on the order of tens of nanoseconds \cite{zajac2018resonantly,huang2019fidelity}, while readout remains the slowest primitive, on the order of microseconds \cite{9563045}. A few tens of nanoseconds (e.g., 68 ns = 6.8 $\times 10^{-8}$ s on IBM Heron r3 \cite{ibmheron2026calibration}) is required for a 2-qubit operation on superconducting qubit technologies (incl. transmon at IBM, xmon at Google). On neutral-atom (Rydberg) platforms the entangling gate itself is now sub-microsecond, e.g. $\approx 270$ ns at $99.5\%$ fidelity in \cite{evered2023highfidelity}, although qubit imaging/readout is comparatively slow, on the order of milliseconds. Trapped-ion 2-qubit gates remain the slowest, typically tens to hundreds of microseconds (e.g., $\approx 226$ $\mu$s = 2.26 $\times 10^{-4}$ s for a $>99.99\%$-fidelity M{\o}lmer--S{\o}rensen gate \cite{hughes2025trappedion}).
This implies that for some of the qubit technologies, there can be up to a factor of a million
difference in the speed of a single operation compared to the classical computer. For superconducting qubit technologies, however, the raw gate time is rarely the limiting factor; the dominant overheads are coherence-limited circuit depth and the cost of quantum error correction discussed below.

\begin{table}[!tb]
    \centering
    \begin{tabular}{lcccc}\toprule[2pt]
Operation              & Semiconductor          & Superconducting       & Ion traps               & Neutral atoms   \\
                       & (spin)                 & (IBM Heron r3)        &                         & (Rydberg)       \\ \hline
2-qubit (CZ/CNOT)      & $\sim 50$              & $68$                  & $\sim 10^{5}$           & $\sim 270$      \\
1-qubit (X/H)          & $\sim 50$              & $32$                  & $\sim 10^{3}$--$10^{4}$ & $\sim 10^{3}$   \\
Init $\ket{0}$         & $\sim 10^{3}$          & $\sim 10^{3}$         & $\sim 10^{5}$           & $\sim 10^{3}$   \\
Readout                & $\sim 10^{3}$--$10^{4}$ & $2.18\times 10^{3}$  & $\sim 10^{5}$           & $\sim 10^{6}$   \\ \hline
    \end{tabular}
    \caption{Illustrating the Wall-Clock Challenge of Section \ref{sec:wallclock}: representative operation times for a number of qubit technologies (in ns), for reasonable gate fidelities. The structure of the table is based on \cite{Suchara2013}, with up-to-date numbers. Note that these are order-of-magnitude figures that vary substantially across devices; the entries combine the dominant entangling gate, single-qubit gate, initialization, and readout times. The semiconductor (silicon spin) numbers are taken from \cite{petta2005coherent,mills2019shuttling,zajac2018resonantly,huang2019fidelity,9563045}; charge qubits can reach $\sim 0.2$ ns 2-qubit gates \cite{petta2005coherent} but at much lower fidelity. The superconducting column reports median calibration data for an IBM Heron r3 processor (\texttt{ibm\_boston}): single-qubit gate length $32$ ns, median two-qubit gate length $68$ ns, and readout length $2{,}180$ ns \cite{ibmheron2026calibration}. The neutral-atom entangling-gate figure is a time-optimal Rydberg CZ gate of $\approx 270$ ns at $99.5\%$ fidelity \cite{evered2023highfidelity}, while atom imaging/readout is millisecond-scale. The ion-trap figures correspond to laser-driven M{\o}lmer--S{\o}rensen gates, with state-of-the-art high-fidelity ($>99.99\%$) gates around $226$ $\mu$s \cite{hughes2025trappedion} and commercial systems in the tens-to-hundreds-of-microseconds range; state preparation and detection are likewise $\sim 10^{5}$ ns.
For photonic qubits, the estimated gate times are on the order of 150 fs which effectively corresponds to the limit on the duration of the laser pulse implementing the gates \cite{Zeuner2018}.}
    \label{tab:gates}
\end{table}

Still, we need to consider:
\begin{itemize}
    \item overhead of quantum error correction, that is, mapping the protected qubits to sets of unprotected qubits,
    \item overhead of compiling the algorithms to the native gate set of the qubit technology,
    \item overhead of mapping the unprotected qubits to the physical qubits, while respecting the topology, in the sense of the pairs of qubits wherein 2-qubit gates can be applied.
\end{itemize}

To make this concrete, consider the credit-risk example of Egger \emph{et al.} \cite{egger2020credit}, who estimate the wall-clock time required to compute the value at risk (VaR) of a large loan portfolio on a fault-tolerant quantum computer built from IBM transmon qubits. The relevant quantum primitive is QAE which requires deep, coherent circuits and therefore full quantum error correction. For a portfolio of $K = 2^{20} \approx 10^{6}$ assets, the resulting circuit contains roughly $37$ million $T$/Toffoli gates, which form the dominant cost in a fault-tolerant implementation since these are the gates that must be distilled
and are far more expensive than Clifford operations. Assuming each
error-corrected $T$/Toffoli gate executes in $10^{-4}$~seconds, and exploiting a
variant that removes the quantum phase estimation step (halving the circuit depth), we arrive at an estimated run time of about $30$~minutes to evaluate the VaR of this one-million-asset portfolio.

\section{Discussion: Classical alternatives to Monte Carlo integration}
\label{sec:alt}

The preceding sections established that quantum amplitude estimation delivers a quadratic improvement over \emph{standard} Monte Carlo, reducing the $\mathcal{O}(N^{-1/2})$ error scaling to effectively $\mathcal{O}(N^{-1})$ in the number of queries. Standard Monte Carlo is, however, neither the only nor the fastest classical option: a range of classical techniques --quasi-Monte Carlo, multilevel Monte Carlo, and adaptive schemes such as importance sampling -- already improve on the canonical Monte Carlo rate by exploiting problem structure, albeit under additional assumptions. This raises a complementary question: rather than asking how quantum methods compare to vanilla Monte Carlo, can quantum techniques accelerate, or be combined with, these stronger classical baselines? In this section we survey such classical alternatives and the (currently limited) extent to which quantum algorithms address them.

The error of any Monte Carlo estimator is governed by its variance. Classical Monte Carlo exhibits a universal $\mathcal{O}(N^{-1})$ variance decay (see \eqref{eq:standard-mc-conv}) requiring no assumptions beyond finite variance, but the asymptotic convergence rate can be improved by modifications to the classical Monte Carlo procedure that exploit problem structure. However, these improvements come at the cost of stronger assumptions and, often, increased implementation complexity.

Quasi-Monte Carlo (QMC) methods \cite{niederreiter1992random, dick2013high} replace random samples with low-discrepancy sequences, achieving variance $\mathcal{O}(N^{-1}(\log N)^{2d-2})$ for smooth functions in dimension $d$, which translates to substantially improved sample complexity when the effective dimension is small. For problems with hierarchical structure, Multilevel Monte Carlo (MLMC) \cite{giles2008multilevel} exploits telescoping sums across discretization levels to achieve improved query complexity compared to standard MC on the finest level, provided a condition on the variance decay rate holds. Adaptive methods \cite{hammersley1956conditional, rubinstein2016simulation, chopin2002sequential} such as importance sampling and control variates maintain the $\mathcal{O}(N^{-1})$ variance scaling but reduce the asymptotic constant, sometimes dramatically. All of these complexity results are, however, conditioned on additional, nontrivial assumptions that one often does not have in practical applications.



\subsection{Adaptive Monte Carlo}
Importance sampling is a technique used in settings wherein some region of the sample space is in some sense more ``important'', either because it is of independent interest, or because preferential sampling will produce more accurate estimates due to the greater variance in the sampled region. By constructing an appropriate auxiliary distribution that places greater density in this region, and then reweighting accordingly after the fact, one ends up generating relatively more samples in the more oscillatory neighborhoods of the random field and lowering the overall variance. Thompson Sampling, in particular (see, e.g.~\cite{russo2020tutorial}), is a popular approach to adaptive sampling by weighting particularly promising regions of the parameter space.

This technique can play an important role in the simulation of quantum systems~\cite{ghanem2021population, de2021importance}. In particular, as a simulation of Schr\"odinger time evolution progresses, the wave functions steadily diffuse, and each sample becomes less informative, exhibiting a greater variance over time. However, there are significant discrepancies in the level of complexity associated with the Hamiltonian at different wave-state trajectories, whereby importance sampling in the form of modifying the Hamiltonian as $H'_{ij}= H_{ij}\frac{\langle D_i|\psi_G\rangle}{\langle D_j|\psi_G\rangle}$,
with a guiding wave function $\psi_G$, can reduce the overall variance of the simulation. 

 Still, we are not aware of any works in the literature on the use of quantum circuits and algorithms to implement an importance sampling procedure for Monte Carlo integration. This is a consequence of the fact that importance sampling is inherently \emph{adaptive}: it responds to the intermediate results of a sampler and refines it for greater accuracy, making it inherently sequential. Thus, it does not lend itself to parallelism, and consequently, it is not straightforwardly amenable to speedup with quantum circuits. We can potentially consider a variational algorithm wherein the quantum-realized results of a Monte Carlo run are assessed by a classical processor that then adaptively selects the points to reroute to the quantum circuits. As this is a generally open problem, the technical challenges for its realization are evidently extensive. In particular, while a quantum circuit may efficiently evaluate many points in parallel, and is thus a potential source of speedup, application of the observable collapses this into a summary statistic. In general, it is well known that an agglomeration of this sort can impede the algorithmic speedup of estimation tasks~\cite{montanari2015computational}. 

This points to a basic correspondence between how one reasons about algorithms involving quantum circuits and how one reasons about algorithms for parallel computing architectures. The latter has seen significant active development in the last two decades. Algorithms amenable to quantum speedup are those with a potentially large set of parallelizable tasks relative to sequences of iterative control flow. Similarly, the ability to use modular (quantum) chip architectures has the potential to achieve parallel speedups akin to those observed in classical computing \cite{Akhtar2023}.

In general, longer coherence times are well known to correspond to greater potential for speedup, since they permit deeper and longer quantum circuits. Control flow, meanwhile, is still conceptualized as a classical operation: an action whose specific nature is conditioned on a measurement outcome. To the extent that this remains a fundamental aspect of quantum computing, novel algorithms will hinge on the clever use of the available coherence time for a given circuit depth -- for near-term devices, typically through classical-quantum hybrid methods, and, in the theoretical quantum-circuit literature, by assuming prolonged coherence. 

\subsection{Multilevel Monte Carlo}
\emph{Multilevel Monte Carlo} (MLMC) is the principled application of the framework of multigrid methods, first developed for the finite element method for discretizing PDEs, and later extended to a general multilevel approach for any setting amenable to a hierarchy of discretizations (see, e.g.~\cite{borzi2005convergence}). In the specific context of stochastic sampling, MLMC can be thought of as a quadrature rule, a way of accurately estimating the integral that corresponds to the expectation of a random variable whose distribution function is known. 

With MLMC, we consider a sequence of unbiased random variables with progressively decreasing variance $\{ F_0, F_1, ...,F_L\}$, estimating $F=\int f(\mathbf{x})d\mathbf{x}$ and sampling the telescoping sum
\[
\mathbb{E}[F] = \mathbb{E}[F_L] = \sum\limits_{l=0}^L \mathbb{E}[F_l-F_{l-1}], \qquad (F_{-1}\equiv 0),
\]
by computing the sample average
\begin{align}
\label{sec:sldef}
\mathcal{S} = \sum\limits_{l=0}^L \mathcal{S}_l,\quad \mathcal{S}_l:=\frac{1}{N_l}\sum\limits_{i=1}^{N_l}\left(f(\mathbf x_l^{l,i})-f(\mathbf x_{l-1}^{l,i})\right).
\end{align}
Similar to classical multigrid results for finite element discretizations of PDEs, it can be shown that the multilevel strategy provides better accuracy relative to the number of sample points, by simultaneously reducing the error over multiple frequencies in the spectral decomposition of the function.

To the best of our knowledge, there is one work in the literature studying the use of quantum circuits in MLMC~\cite{an2021quantum}. They consider the application of MLMC for estimating a functional of a random variable governed by a stochastic differential equation (SDE), with applications to finance. In particular, they consider order-$r$ numerical discretization schemes for simulating SDEs whose solutions have no available analytical expression. They note that, to achieve accuracy $\varepsilon$, classical Monte Carlo with an order-$r$ scheme solves the SDE with complexity $\tilde{\mathcal{O}}(1/\varepsilon^{2+1/r})$, where $\tilde{\mathcal{O}}$ denotes the presence of an additional multiplicative logarithmic factor, whereas classical MLMC with a scheme of strong order $r>1$ improves this to $\tilde{\mathcal{O}}(1/\varepsilon^{2})$. In the context of the Black-Scholes model, if an analytical solution to the SDE exists, then quantum circuits can achieve $\tilde{\mathcal{O}}(1/\varepsilon)$ complexity. For numerically discretized SDEs, however, attaining the full quadratic speedup $\tilde{\mathcal{O}}(1/\varepsilon)$ with quantum-accelerated MLMC requires a numerical scheme of strong order $r>2$. Since $r>2$ can be unrealizable for many practical SDEs, and classical MLMC already attains $\tilde{\mathcal{O}}(1/\varepsilon^2)$ for $r>1$, the quantum advantage is limited in the regime $1<r\le 2$.

\subsection{Alternative Random Variable Quadrature Methods}\label{sec:alternativeB}
Quasi-Monte Carlo performs numerical quadrature by selecting a set of points $\{a_i\}$ from a \emph{low-discrepancy sequence}. A low-discrepancy sequence appears random in the sense that, for any reasonable subset of finite measure in the ambient space, the fraction of points falling in that subset is roughly proportional to its volume. In a classical experimental study~\cite{morokoff1995quasi} it was found that QMC typically computed integrals more accurately than Monte Carlo alternatives, especially for smooth and lower-dimensional functions. 

Typically, the convergence is reported as $\mathcal{O}(N^{-1}(\log N)^{s})$, where $s$ is the dimension. However, the dimensional dependence, which is still not too unfavorable compared to grid-based methods, can be improved for certain classes of problems, and the convergence rate with samples can also be $N^{-2}$ in some cases. The trick to achieving this rate is to consider weighted reproducing kernel Hilbert spaces
to balance the impact of dimension and samples. In particular, if certain $L^p$-like norms of the higher derivatives of the function of interest decay with the derivative order at an appropriate rate, then the integrability and limited oscillations enable tighter integration bounds. 
The natural question to ask is the possibility of a quantum speedup for quasi-Monte Carlo. The answer is not immediately obvious. 
In fact, despite the name, quasi-Monte Carlo is only minimally randomized. 
It relies on low-discrepancy sequences, such as those of  Halton \cite{halton1964radical} or Sobol \cite{SOBOL199055}.
Popular strategies can be seen as lattice rules \cite{wang2002historical}, where only the first point in the lattice is selected at random, while all the other points follow a precise sequence of displacement in the ambient space. On the other hand, the evaluation of a function on points defined on a lattice rule is highly parallelizable and potentially amenable to quantum parallelism. We can hypothesize hybrid MC-QMC strategies, aided by parallel computing. 
However, at this point in the literature we are not aware of any studies on the use of quantum circuits to perform QMC (or sparse grids). 

Sparse grids (for a tutorial, see~\cite{garcke2006sparse}), popularized for uncertainty quantification of partial differential equations~\cite{nobile2008sparse}, is a technique that generates quadrature points and weights so as to minimize the ratio of the quadrature error to the number of function evaluation points. While it scales exponentially with dimension, it performs well for approximating integrals with a small dimension $d$, in the sense of favorable scalability with sample size $N$. 

At this point, to the best of our knowledge, no quantum algorithms exist for implementing grid collocation. For a potential indication as to how such an algorithm could be developed, we can consider the approximation of the gradient of a function by the ``quantum gradient'' procedure (we refer to the most recent work in~\cite{gilyen2019optimizing} based on the work originally done in~\cite{jordan2005fast}). At a conceptual level, this procedure is based on the classical realization of the effectiveness of the central difference method to estimate a gradient, i.e.,
\[
\nabla_i f(x) \approx \frac{f(x+\Delta e_i)-f(x-\Delta e_i)}{2\Delta},
\]
for some appropriate $\Delta>0$, where $\nabla_i$ is the partial derivative with respect to the $i$-th coordinate, and $e_i$ is the vector of all zeros except for $1$ in the $i$-th component. Noting the superficial resemblance to the two-point trapezoidal rule for one-dimensional $f(x)$,
\[
\int_{x-\Delta}^{x+\Delta} f(y) dy \approx (f(x+\Delta)+f(x-\Delta))\Delta,
\]
we can clearly see the analogy for grid-based computing. 

The algorithm in~\cite{gilyen2019optimizing} considers a function of interest $f(x)$ to be defined as an oracle that computes
\[
O^f\ket{x_1}...\ket{x_n}:= e^{2\pi i f(x_1,...,x_n)}\ket{x_1}...\ket{x_n}.
\]
The key point is that the quantum gradient algorithm then starts with a uniform superposition
\[
\ket{\psi} = \frac{1}{\sqrt{|G^d_x|}}\sum_{\delta\in G^d_x}\ket{\delta},
\]
where $G^d_x$ is a grid of points at which the function is to be evaluated. Note that this set of points must be amenable to a binary representation (at least for computers working with qubits). 

The algorithm then applies the oracle $O^f$ to this superposition state and uses the inverse Fourier transform
\[
\ket{x}\to\frac{1}{\sqrt{2^n}}\sum\limits_{k\in G^d_x}e^{-2\pi i \, xk/2^n}\ket{k},
\]
and, subsequently, each input register will contain the measurement outcome for a corresponding dimension. 

We can see that the binary expansion corresponding to a multidimensional central-difference-type rule is key to its parallelizability, due to the inherent form of the Fourier transform. As such, nested collocation procedures with this structure are intuitively amenable to these techniques. The Smolyak sparse grids approach is a natural target. However, the quadrature weights as well as the grid points are derived by nontrivial calculations involving Chebyshev polynomials. This presents a challenge in implementing this procedure without extensive classical computation, thus limiting the potential for quantum speedup.

\section{Conclusion}
\label{sec:conclusions}

\subsection{Synthesis}
\label{sec:synthesis-summary}

The preceding sections have surveyed quantum amplitude estimation (QAE) and quantum approximate counting (QAC) and their modern variants and refinements. What distinguishes amplitude estimation from other quantum algorithms is the \emph{unconditional} character of its speedup. Unlike the conjectured exponential speedup of Shor's algorithm, which rests on unproven assumptions about the classical complexity of integer factorization, the quadratic improvement over classical Monte Carlo methods demonstrates a quantum advantage that is not conditional on any assumptions about the classical randomized complexity of integration. This is because classical Monte Carlo methods are optimal among randomized algorithms in the general setting. Moreover, the complexity of QAE and QAC match a provable lower bound on the (algorithm-agnostic) quantum query complexity~\cite{nayak1998quantum, NEURIPS2022_933e9533}.

We have provided a conceptual framework that classifies various quantum analogues of Monte Carlo methods. The end-to-end perspective of \cref{subsection:taxonomy} (see Figure~\ref{fig:hourglass}), shows how a variety of numerical problems can be recast as mean estimation problems. Upon discretization, these mean estimation problems give rise to the triple $((\pi_{i})_{i\in I},(U_{f_j})_{j\in J},(\mathcal{A}_{k})_{k\in K})$ comprising of states $\ket{\pi_i}$ whose amplitudes encode the relevant distributions along with oracles $U_{f_i}$ and subroutines $\mathcal{A}_k$ encoding function values or other problem-specific data. At this point, one can use QAE, QAC or one of their variants described in Section \ref{sec:quantum_opp} to obtain an approximate solution to the mean estimation problem. The first part of the procedure is essentially classical: as the dashed arrow of Figure~\ref{fig:hourglass} suggests. Once the mean estimation problem has been fixed, we could, in principle, proceed using a classical randomized algorithm. The second half of the procedure (estimating the mean given $((\pi_{i})_{i\in I},(U_{f_j})_{j\in J},(\mathcal{A}_{k})_{k\in K})$) is genuinely quantum. In general, the original problem at the top of Figure~\ref{fig:hourglass} possesses some structure which informs the optimal choice of $((\pi_{i})_{i\in I},(U_{f_j})_{j\in J},(\mathcal{A}_{k})_{k\in K})$. 

For example, by taking into account the \emph{regularity} of the integrand (or, more generally, the regularity of the problem specification), the optimal quantum complexity bounds can be refined. The query complexity of integration over H\"older and Sobolev classes has been characterized for deterministic, randomized, and quantum algorithms alike~\cite{novak2001quantum,heinrich2002quantum,heinrich2002optimal}. Moreover, we have shown that the regularity-based complexity results can be applied relatively straightforwardly to several problems of genuine interest such as the solution of elliptic PDE and path integration. 

Despite the numerous theoretical advantages of quantum Monte Carlo methods over their classical randomized counterparts, the implementation of these methods presents several difficulties: the construction of the requisite oracles, the circuit depth, the complexity of loading the input distribution, and the resulting wall-clock time. The asymptotic advantage is one not yet realized in practice. Many of the modern variants surveyed in \cref{sec:quantum_opp} adapt the core ideas of QAE to realize the advantage on NISQ devices.

\subsection{Outlook}
\label{sec:outlook}

We conclude this survey by proposing several directions for further research. 
One research program that we believe has been unduly neglected concerns the regularity-based query-complexity bounds of Heinrich and Novak~\cite{heinrich2002quantum,novak2001quantum}. The optimal quantum rates summarized in \cref{table:functionclasses} are obtained by a single mechanism, a classical quadrature that exploits the smoothness of the integrand, composed with a quantum subroutine applied to the residual. This mechanism is not specific to the H\"older and Sobolev spaces and should be extended to other function spaces. The most immediate candidate is the Besov space. The deterministic and randomized rates are known~\cite{li2022besovball,duan2022sphere}, but the computing the corresponding quantum rates remains an open problem. We suspect that this line of work has been underexplored not due to lack of interest but because the foundational papers that address Sobolev and Hölder classes are highly technical. However, we have already shown that the regularity-based query complexity results can be readily applied to many problems of practical interest to physicists.

Indeed, a complementary program consists of applying these bounds to new problems in physics and other fields. Heinrich's methods extend well beyond integration over a hypercube, and \cref{table:regularityapplications} collects regularity-based quantum query complexity bounds for path integration on Gaussian-measure spaces~\cite{traub2002path}, Feynman--Kac functionals of Brownian motion~\cite{kwas2006feynmankac}, parametric integration~\cite{wiegand2006parametric}, initial-value problems for ordinary differential equations~\cite{kacewicz2004ivp,kacewicz2005ivp}, the Sturm-Liouville eigenvalue problem~\cite{papageorgiou2005sturm}, elliptic boundary-value problems sampled on a submanifold~\cite{heinrich2006elliptic}, and $L_q$-approximation of Sobolev functions~\cite{heinrich2004approximation}. In all of these results, the methods is to reduce the problem to the estimation of the integral of a function whose regularity is known, and invoking the results of \cite{heinrich2002optimal}. Thus, a natural research direction would be to extend these results with further problems drawn from physics. Natural candidates include high-dimensional parabolic and backward Kolmogorov equations, whose Feynman--Kac representations tie this line directly to the stochastic differential equations of \cref{sec:finance}; spectral problems beyond the Sturm-Liouville setting; kinetic and transport equations; and the path integrals of statistical mechanics and lattice field theory.

We would also like to draw attention to the use of quantum algorithms other than QAE and QAC, which nonetheless generalise classical Monte Carlo methods. There has already been some promising research in this direction. For instance, Layden et al. \cite{layden2022quantum} proposed an alternative hybrid quantum-classical algorithm to perform Markov Chain Monte Carlo (MCMC) simulations for sampling from intractable distributions. The hybrid framework is a popular strategy for noisy near-term devices, since it relaxes the depth requirement for quantum circuits to compute potentially useful outputs. More generally, a standard component of classical Markov Chain Monte Carlo techniques to sample from and compute expectations of intractable distributions is to generate samples from some well-understood distribution and perform acceptance and rejection using the evaluation of the potential of the distribution of interest at the previous and new samples. Thus, rather than attempting to directly work with the oracle of the distribution as in QAE, we can consider translating this classical approach more directly, by generating the proposal distribution on a quantum device. There is also the question of exploring the ``quantization" of classical alternatives to Monte Carlo methods: although some of the approaches discussed in Section \ref{sec:alt} do not readily carry over to the quantum setting, some methods, such as multi-level Monte Carlo, could plausibly be implemented in a hybrid fashion, leveraging classical and quantum computing. We believe that this is another promising research direction. 

We conclude by proposing an ambitious but potentially very powerful line of work. Throughout this review, the triple $((\pi_{i})_{i\in I},(U_{f_j})_{j\in J},(\mathcal{A}_{k})_{k\in K})$ has been obtained via classical reduction and oracle construction/state preparation. However, one may also be interested in the case where the quantum data $((\pi_{i})_{i\in I},(U_{f_j})_{j\in J},(\mathcal{A}_{k})_{k\in K})$ is itself the output of a quantum algorithm, for example of a Hamiltonian simulation algorithm. In this case, amplitude estimation can act as a \emph{readout} subroutine, estimating the expectation of an observable $\bra{\psi}B\ket{\psi}$ with error $\varepsilon$ at cost $\mathcal{O}(\varepsilon^{-1})$ as opposed to the $\mathcal{O}(\varepsilon^{-2})$ of naive sampling~\cite{knill2007optimal,rall2020quantum}. Alternatively, quantum simulation may provide the oracle $U_f$ itself ~\cite{rall2020quantum,huggins2022nearly}. These Hamiltonian simulation techniques have recently been extended to the study of noisy classical nonlinear dynamics in ~\cite{bravyi2026quantum}. However, the integration of quantum Monte Carlo methods with Hamiltonian simulation techniques remains understudied.

\subsection*{Disclaimer}

This paper was prepared for information purposes, and is not a product of HSBC Bank Plc. or
its affiliates. Neither HSBC Bank Plc. nor any of its
affiliates make any explicit or implied representation or
warranty and none of them accept any liability in connection with this paper, including, but not limited to, the completeness, accuracy, reliability of information contained
herein and the potential legal, compliance, tax or accounting effects thereof. This document is not intended
as investment research or investment advice, or a recommendation, offer or solicitation for the purchase or sale
of any security, financial instrument, financial product or
service, or to be used in any way for evaluating the merits
of participating in any transaction.

\bibliographystyle{apsrmp4-2}
\bibliography{refs_final}

\end{document}